\documentclass[a4paper]{article}
\usepackage[a4paper, left=2.5cm, right=2.5cm, top=2.5cm, bottom=2.5cm]{geometry}
\usepackage{graphicx}
\usepackage{subcaption}    
\usepackage{multicol} 
\usepackage{multirow}
\usepackage{comment}
\usepackage{cite}
\usepackage{amsmath}  
\usepackage{booktabs}    
\usepackage{amssymb}    
\usepackage{float}      
\usepackage{array}
\usepackage{soul,xcolor,tcolorbox}
\usepackage{lineno}

\usepackage{hyperref}
\hypersetup{
	colorlinks=true,     
	linkcolor=blue,        
	urlcolor=blue,         
	filecolor=magenta,     
	citecolor=blue         
}

\date{}

\usepackage[font=small]{caption}  
\newcommand{\supercite}[1]{\!\mbox{\textsuperscript{\cite{#1}}}}
 
\def\ge{\geqslant}

\soulregister{\cite}{7} 
\soulregister{\supercite}{7} 
\soulregister{\ref}{7}

\newif\ifhighlightFirst
\highlightFirstfalse  

\newcommand{\modifyFirstYellow}[1]{\ifhighlightFirst \sethlcolor{yellow} \hl{#1} \else #1 \fi}

\newtcolorbox{highlightedGreen}{colback=green!30, colframe=green!80!black, sharp corners}
\newtcolorbox{highlightedYellow}{colback=yellow!30, colframe=yellow!80!black, sharp corners}

\begin{document} 
	
	\title{{\large \textbf{Multi-Partitioned Computing Quantum-Particle Approach: A Hybrid Quantum Framework for Fluid Flow}}}
	
	\author{
		\small{Yudong Li$^{1,2,}$\footnotemark[1],\quad Wenkui Shi$^{1}$,\quad Zhihao Qian$^{1,2}$,\quad Yan Li$^{3,4}$,\quad Chunfa Wang$^{3,4}$,}\\ \small{\quad Ling Tao$^{5}$,\quad Zhuojia Fu$^{6}$,\quad Moubin Liu$^{1,2,}$\footnotemark[1] ,\quad Zhiqiang Feng$^{3,4,7}$}
		\\[2mm]
		\footnotesize{1. School of Mechanics and Engineering Science, Peking University, Beijing 100871, PR China}
		\\
		\footnotesize{2. Nanchang Institute of Innovation, Peking University, Nanchang 330038, PR China}
		\\
		\footnotesize{3. School of Mechanics and Aerospace Engineering, Southwest Jiaotong University, Chengdu 611756, PR China}
		\\
		\footnotesize{4. Advanced Structural Material Mechanics and Service Safety Key Laboratory of Sichuan Province,}\\ \footnotesize{Chengdu 611756, PR China}
		\\
		\footnotesize{5. School of Architecture and Civil Engineering, Xihua University, Chengdu 610039, PR China}
		\\
		\footnotesize{6. College of Mechanics and Engineering Science, Hohai University, Nanjing 211100, PR China}
		\\
		\footnotesize{7. Laboratory of Mechanics and Energy, Univ-Evry, Paris-Saclay University, Evry 91190, France}
	}
	
	\renewcommand{\thefootnote}{\fnsymbol{footnote}}
	\footnotetext[1]{\hspace{0.5mm}Corresponding authors. \textit{E-mail address:} yudongli@pku.edu.cn (Y.D. Li), mbliu@pku.edu.cn (M.B. Liu).}
	\renewcommand{\thefootnote}{\arabic{footnote}}
	
	\maketitle	
	
	\thispagestyle{plain}
	
	\small
	\begin{abstract}
		\noindent \textbf{Abstract~~~}This study established a quantum-classical hybrid framework that integrates quantum computing paradigm with meshfree finite particle method. By harnessing quantum superposition and entanglement, it hybridized the critical computational kernels (termed as quantum finite particle method). A resource-efficient quantum computational strategy on multi-partitioned zones was proposed, which leverages a fixed small-scale quantum circuit as a fundamental processing unit to handle inner product for arbitrarily sized arrays. This approach employs iterative nesting of the quantum-core operation to accommodate varying input dimensions while maintaining hardware feasibility throughout. Motivated with developed quantum framework, the novel numerical discretization for hybrid quantum computational particle dynamics can be derived commonly and applied in fluid flows. Through a sequence of numerical experiments purposefully, the proposed numerical model was thoroughly validated and analyzed. Results demonstrate that integrating quantum computing to hybridize conventional linear combinations of particle dynamics serves as a novel computing paradigm. By further extending into the numerical investigation of viscoelastic, highly elastic and purely elastic fluids under high Weissenberg number conditions, the applicability of simulation framework is broadened. Despite bottlenecks in quantum hardware and computational efficiency on this process, these advances offer critical insights for transitioning quantum-enhanced fluid simulation to practical engineering applications.
		\\[2mm]
		\textbf{Key words~~~}Meshfree/particle methods; Finite particle methods; Quantum computing paradigm; Viscoelastic Weissenberg flows; Computational particle dynamics; Quantum processing unit
	\end{abstract}
	
	\section{Introduction}
	\normalsize \hspace{10pt}
	Quantum computing has rapidly emerged as a transformative paradigm in computational science, heralding a new era where quantum mechanical phenomena on superposition and entanglement is harnessed to perform calculations that are infeasible for classical computers. This unique property enables quantum computers to process vast amounts of data in parallel, potentially offering exponential speedups for certain computational tasks, including matrix inversion, optimization, and simulation of quantum systems\supercite{tennie2025quantum,DEFENU20241,Au-Yeung_2024}. Paralleling to the advancements in quantum superposition and entanglement, the field of computational fluid dynamics (CFD) has generated significant progress on quantum fluid dynamics\supercite{thomson2024unravelling,WOS:001341256800003,CHEN2024117428,Ye2024}. At present, some newish developments on meshfree computational particle dynamics and high performance computing (HPC) are being on the heels of quantum mechanical phenomena and crossed computing paradigms.
	
	Mathematical models of computational fluid dynamics usually have intractable non-linear partial differential equations with a large of variables in two or higher dimensions, especially where the prevalent neural networks are more likely to break out in higher-dimensional numerical problems\supercite{2022Physical}. Relatively comparing with classical numerical methods, as finite difference/element/volume methods and spectral Galerkin methods, quantum computing or quantum neural network has demonstrated astonishing computing potential\supercite{RIEFFEL2024598}. Innovative scientific exploration is proceeding, as that Harrow-Hassidim-Lloyd (HHL) algorithm improved the computational ability on solving Navier-Stokes (NS) equations and even non-linear partial differential equations\supercite{doi:10.1073/pnas.2311014120,2020Variational,2021Solving}, Todorova et al.\supercite{TODOROVA2020109347} built on the latter and presented the quantum lattice Boltzmann (LB) method between the LB streaming step and quantum walks\supercite{WOS:001650371600001}, and more efforts on Poisson equation solver by variational quantum algorithms (VQAs)\supercite{PhysRevA.104.052409} and quantum physics-informed neural networks (PINNs)\supercite{2024Potential} are developed.
	
	Likewise with regard to classical meshfree methods, improved numerical strategies have been motivated to promote the computational precision, consistency and efficiency. These approaches mainly depend on the discretization of  spatio-temporal variables and the numerical approximation on meshfree weak-strong forms\supercite{BELYTSCHKO19963,liu2003meshfree,LiuGR0014,NGUYEN2008763}.
	
	As a meshfree Lagrangian form, the smoothed particle hydrodynamics (SPH) is a particle-based method originally proposed by Gingold and Monaghan in 1977\supercite{Monaghan375,monaghan1992smoothed}. In early stages, the SPH technique did not receive significant attention compared with other well-established numerical methods, such as finite element, finite volume, and finite difference methods, that were already mature at the time. Although the SPH algorithm lacked the capability to conserve angular and linear momentum rigorously, it still produced results that were deemed acceptable for a variety of astrophysical applications. In contrast, when applied to problems in solid and fluid mechanics, the method encountered difficulties in accurately reproducing partial differential equations, particularly with respect to ensuring stability and achieving the desired level of accuracy. Subsequently, lacking and drawbacks were identified and improved by many researchers\supercite{belytschko1994element,swegle1995smoothed,liu2003constructing}. Improvements on the SPH method find its way to a wide range of applications\supercite{shadloo2016smoothed}. Afterwards, the element free Galerkin (EFG) method\supercite{belytschko1994element,lu1994new} was emerged as a solution in earlier meshless techniques. It was specifically devised to mitigate issues such as diminished boundary accuracy and various instabilities, which were notably prevalent in SPH implementation. This approach was also named as the diffuse element method introduced by Nayroles et al.\supercite{nayroles1992generalizing}. Kansa et al.\supercite{kansa1990multiquadrics} introduced the radial basis function (RBF) method as a means of addressing partial differential equations. Building upon this foundation, Fasshauer et al.\supercite{fasshauer1999solving} initially applied it to solve interpolation matrix problems and subsequently extended his research to enable the numerical resolution of partial differential equations. Certainly, considerable amount of meshless/meshfree methods was developed on different mathematical insights and intersectional processes. It spawned a lot of mathematical analysis and engineering applications\supercite{liu2003smoothed,wang2004extended,atluri1998new,aluru2000point,garg2018meshfree}.
	
	For the numerical approximation on meshfree methods, some means of linear combination, resolving matrices equation and time evolution are ubiquitous. With the enforcements on quantum superposition and quantum bits, these means can be of exponential acceleration by quantum computing. As a matter of course, the current consensus is that quantum physical devices will not simply supersede the classical high performance computing. Why is that? Firstly, the current noisy intermediate-scale quantum (NISQ)\supercite{RevModPhys.94.015004} device is not perfect enough to solve practical problems, but allows researchers to develop quantum algorithms and prepare for more advanced quantum hardware. Secondly, the specific quantum algorithms need to produce for different mathematics and engineering applications, existentially such as classical quantum phase estimation (QPE), Grover’s algorithm\supercite{1997A}, quantum walks\supercite{PhysRevA.67.052307}, Harrow-Hassidim-Lloyd (HHL) algorithm\supercite{PhysRevLett.103.150502}, variational quantum algorithms (VQAs)\supercite{Cerezo2021Variationalquantum}, etc. Thirdly, the present engineering building and application are both constructed based on central processing unit (CPU) chipset, accompanying with specified commands and special execution on graphics processing unit (GPU) chipset. Similarly, the interface on quantum processing unit (QPU) chipset is also facing great challenges, such as clock speed mismatch\supercite{Madsen2022Quantum}, engineering quantum data encoding\supercite{PhysRevA.109.052423} and data interconnections\supercite{Kenton2023Optimising}. As a fact, the present computing strategy is quantum-hybrid computer, both many parts of classical computation and key components by quantum processor.
	
	\modifyFirstYellow{Among the growing repertoire of quantum algorithms applicable to common computational fluid dynamics, the aforementioned HHL algorithm and its variants for solving linear systems of equations, as well as the optimized VQAs for non-linear problems, have attracted considerable attentions\supercite{CHEN2024117428}. The HHL computing with a Hermitian matrix $\boldsymbol{A}$ and a normalized vector $\boldsymbol{b}$ encoded as a quantum state $|b\rangle$, proceeds via three stages: QPE on the unitary $e^{i\boldsymbol{A} t}$ to extract eigenvalues $\lambda_j$ of $\boldsymbol{A}$ into an ancillary register, controlled rotations that invert the eigenvalues (mapping $|b\rangle=\sum_j \beta_j |u_j\rangle$ to $\sum_j \beta_j \lambda_j^{-1}|u_j\rangle$ up to a known factor), and inverse QPE to uncompute the eigenvalue register. The output is the quantum state $|x\rangle$ proportional to the solution. The circuit width scales as $\mathcal{O}(\log N + \log(1/\varepsilon))$ qubits, where $N$ is the system dimension and $\varepsilon$ is the allowable error in eigenvalue estimation. The depth is dominated by controlled rotations and QPE. Using qubitization or sparse Hamiltonian simulation, the overall gate complexity is $\widetilde{\mathcal{O}}(\log(N)\cdot s^2\kappa^2/\varepsilon)$, with $\kappa$ the condition number of $\boldsymbol{A}$ and $s$ its sparsity. For the ill-conditioned matrices frequently encountered in meshfree/particle discretisations, $\kappa$ can be large, rendering HHL implementations prohibitively expensive on current hardware. The VQAs rely on a parameterized quantum circuit of width $n$ (number of qubits) and depth $d$ (number of sequential gate layers) to prepare a trial state $|\psi(\boldsymbol{\theta})\rangle$. The hybrid procedure repeatedly evaluates the expectation value on a quantum processor and updates the parameters $\boldsymbol{\theta}$ using a classical optimizer to minimize the cost function, with the depth typically kept moderate to mitigate noise while the width determines the expressivity of ansatz. However, VQAs introduce substantial classical overhead: the optimisation landscape is often plagued by barren plateaus, and the number of circuit evaluations required for gradient estimation scales unfavourably with the number of parameters. Moreover, the convergence guaranteeing to true solution remains an active subject of theoretical investigation. The swap test adopted herein operates on a quantum register of width $\mathcal{O}({\log _2}N + 1 + t)$ and depth $\mathcal{O}(({2^{t + 1}} + 1) \cdot 2N)$, offering exponential compression in circuit width ($N$ system dimension, $t$ the number of ancilla qubits). While the linear encoding depth currently bounds the end-to-end speedup, the multi-partitioned strategy adopted in this work caps this depth at a hardware-tractable constant, rendering the overall framework viable on near-term devices described in the following main contents.}
	
	As an improved weak-form collocation method, the finite particle method (FPM) has been extensively developed in fields of meshfree computational particle dynamics\supercite{2019A,2018A,10.1063/5.0197088,LI2023112213}, which was proposed based on the integral approximated scheme and the conception on multivariate mathematical Taylor series expansion\supercite{2005Modeling}. Building upon the foundational residuals formulation in integral domains similarly as SPH methodology, several enhanced approaches have been developed to refine the FPM framework, supported by comprehensive numerical investigations\supercite{2010Smoothed,2019Smoothed,2023On,LI2023104532}. By constructing residual bounds through moving least squares (MLS) and multivariate Taylor series expansion, the generalized finite difference method (GFDM) has also been adapted for applications in computational solid and fluid mechanics\supercite{BENITO20011039,lxxb2021-439}. During the evolved procedure on numerical computation, the linear combination, resolving matrices equation and explicit/implicit integration usually take up most of the time and play the crucial role on general numerical performance\supercite{AUYEUNG2024108909,DEFENU20241,10.1063/5.0268240}. Yet, the computational burden of FPM escalates quickly as particle numbers and interaction complexity rise, making high-fidelity simulations of turbulent or multiphase flows particularly demanding\supercite{LI2025119962}. The principles of quantum interference and entanglement further augment these capabilities. This paper investigates a quantum-classical hybrid framework that integrates quantum computing methodologies into the meshfree FPM simulation paradigm. By harnessing quantum superposition and entanglement, we improved critical computational kernels, overcoming the scalability constraints of conventional particle dynamics.
	
	On the other hand, the extra elastic stress tensor in complex fluids, as a fundamental physical quantity in viscoelastic constitutive equations, evolves from the objective physical laws governing the modeled systems. Consequently, numerous nonlinear constitutive models were developed in early studies, including classical formulations such as Oldroyd-B, upper convected Maxwell (UCM), finite extensible nonlinear elastic (FENE), Phan-Thien-Tanner (PTT), and Giesekus models\supercite{John_Poole_Kowalski_Fonte_2024,KING2021104556}. Notably, the nonlinear parameters presented in other models can be generally derived as extensions of the Oldroyd-B model, making it a benchmark for testing limits on numerical scheme stability. The UCM model represents a limiting case of Oldroyd-B, where their nonlinear elastic effects are governed by specific parameters $(\alpha, \beta, \gamma)$. Through parameter variation $(\alpha \rightarrow 0, \beta \rightarrow 0, \gamma \rightarrow 0)$, these characteristic models can degenerate into the Oldroyd-B formulation.
	
	In the previous work\supercite{LI2023112213}, we identified computational artifacts when directly applying time-dependent numerical discretization methods to the elastic UCM model. Two primary mechanisms underlie these instabilities: (a) Similar to dissipation in viscoelastic constitutive equations, the strain rate tensor diminishes with increasing relaxation time; (b) The conformation tensor inducing elastic stress cannot maintain symmetric positive definiteness (SPD) under arbitrary velocity gradient modifications. As established by Fattal, Kupferman et al.\supercite{FATTAL2004281,FATTAL200523}, this numerical instability stems from an imbalance between exponential stress growth (originally from deformation) and convective transport, a phenomenon termed as the high Weissenberg number (HWN) problem\supercite{KING2021104556}. Fattal et al.\supercite{HULSEN200527} pioneered the logarithmic conformation representation within finite element methods to resolve HWN challenges in viscoelastic flow simulations. Subsequent adaptations extended this approach to finite volume\supercite{LOPEZHERRERA2019144} and finite difference methods\supercite{COMMINAL201537}, achieving stable simulations at Weissenberg numbers $\mathrm{We} \ge 10$. In this paper, numerical analysis on high Weissenberg viscoelastic fluid flows is further investigated based on present hybrid quantum computing procedure. It is also to rigorously evaluate the HWN behaviors and improved meshfree numerical performance on accuracy and convergence. \modifyFirstYellow{Through this paradigm, we demonstrated both the methodological novelty and the computational potential of developed FPM framework and quantum-hybridized implementation.}
	
	This paper is organized as follows: The fundamental quantum theory on swap test is described in Section 2; The proposed hybrid quantum framework on multi-partitioned meshfree quantum finite particle method is introduced in Section 3; The hybrid quantum computational particle dynamics involving common numerical discretization and viscoelastic Weissenberg fluid flows with Log conformation is described in Section 4; Some benchmarks of present hybrid quantum verification, valid time-dependent evolutionary flow states and numerical analysis on high Weissenberg viscoelastic fluid flows are proceeded orderly in Section 5; Some conclusions and remarks on present hybrid quantum resolution are given in Section 6. \modifyFirstYellow{The partial evaluation algorithms for this manuscript are openly available in https://github.com/lyd12311/Quantum-computing-FPM.}
	
	\section{Quantum theory: Swap test}
	\normalsize \hspace{10pt}
	To introduce a unified understanding of the distinctions between von Neumann computer and universal quantum computer, the operational principles $\left | 0  \right \rangle$, $\left | 1  \right \rangle$ are described similarly from the expression on binary system. In classical computer architecture, the single bit $0$, $1$ can definitely describe the machine process of instruction order and data storage. While in universal quantum computer, the single quantum bit (qubit) is usually in a superposition state $\left | \psi  \right \rangle$. The uncertain state can disappear (collapse) until that a definite result (observation) is executed. The property on uncertain superposition state can embody the potential parallel computing capability and data storage efficiency even combining with the entanglement effect of multiple quantum bits.
	
\begin{figure}[H]
		\centering
		\includegraphics[scale=0.35]{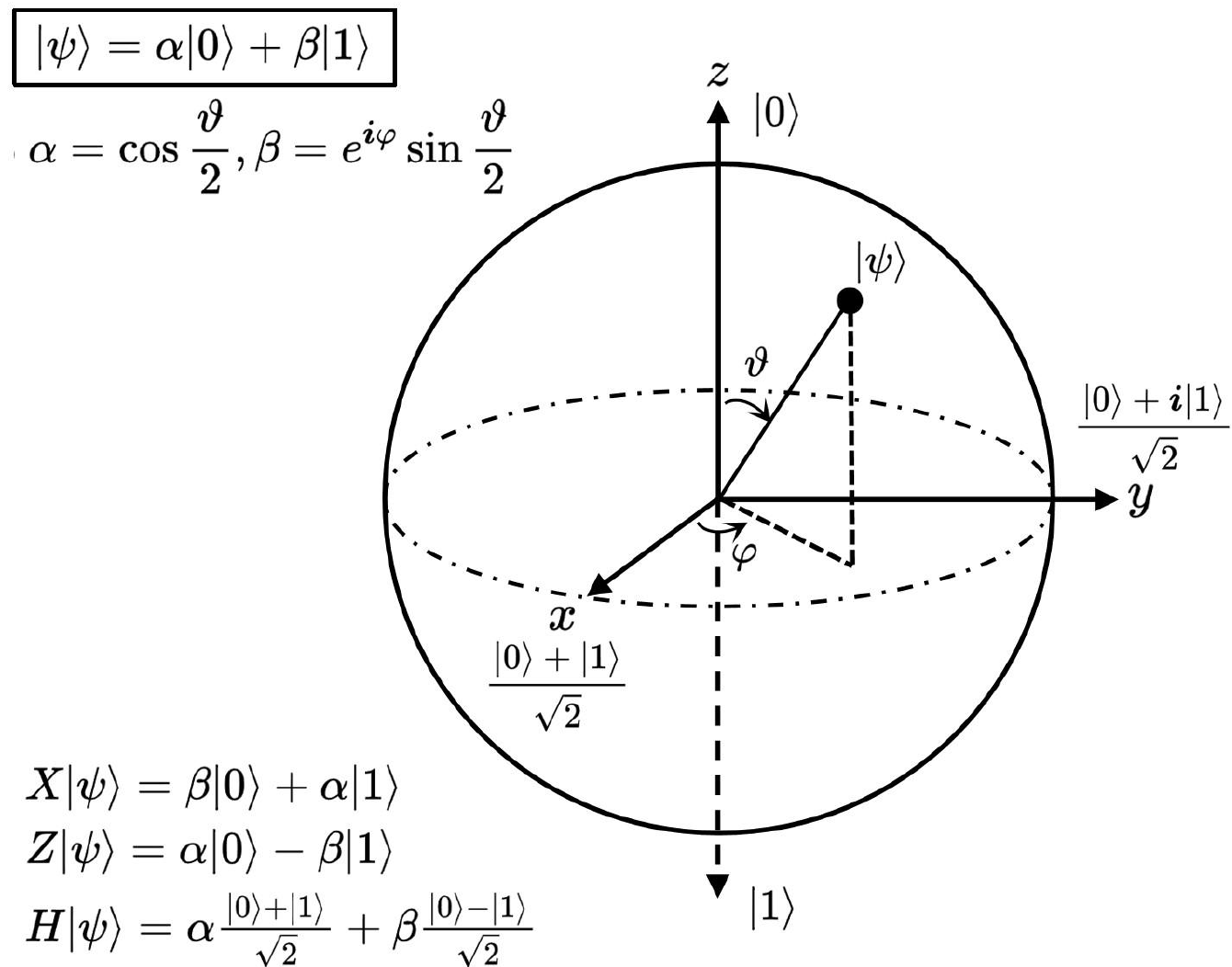}
		\caption{\small Bloch sphere of quantum state $\left | \psi  \right \rangle$ with basic gate operations.}
		\label{fig_Quantum:bloch}
	\end{figure}

	In order to facilitate the understanding on quantum system, the superposition state of single qubit can be described in Eq.(\ref{eq_Quantum:1}) based on the Hilbert space. As shown in Fig.(\ref{fig_Quantum:bloch}), the quantum superposition state can be quantized in the form of Bloch sphere. The state position is decided from the numerical amplitude and rotation angle, for which it notes that the state is still in an uncertain superposition state of probability amplitude. More mathematical details of quantum theory can be referenced\supercite{Au-Yeung_2024,CHEN2024117428,Ye2024}.
	
	\begin{equation}\label{eq_Quantum:1}
		\left | \psi  \right \rangle =\alpha \left | 0  \right \rangle+\beta \left | 1  \right \rangle
	\end{equation}
	where $\alpha, \beta$ denote the quantum amplitude $\alpha, \beta \in \mathbb{C}$ and $\left | \alpha  \right | ^{2} +\left | \beta   \right | ^{2} =1$. $\left | 0  \right \rangle$, $\left | 1  \right \rangle$ denote a set of base space in $\mathbb{C}^{2}$, $\left | 0  \right \rangle=\left [ 1,0 \right ] ^{\mathrm{T} }$ and $\left | 1  \right \rangle=\left [ 0,1 \right ] ^{\mathrm{T} }$.
	
	\textbf{Quantum encoding:} Quantum state encoding maps classical data into quantum systems for computation. The primary methods\supercite{2021Encoding,2023Configurable} include:
	\\ (a) Basis encoding by storing binary data directly in qubit states. It is simple but limited to discrete values.
	\\ (b) Angle encoding by embedding data in qubit rotation angles. It is efficient for near-term quantum machine learning.
	\\ (c) Amplitude encoding by encoding data into quantum amplitudes. It is highly compacted but requires complex preparation.
	\\ (d) Hamiltonian encoding by encoding data into quantum system interactions. It is used in quantum simulation but sensitive to noise.
	
\begin{figure}[H]
		\centering
		\includegraphics[scale=0.3]{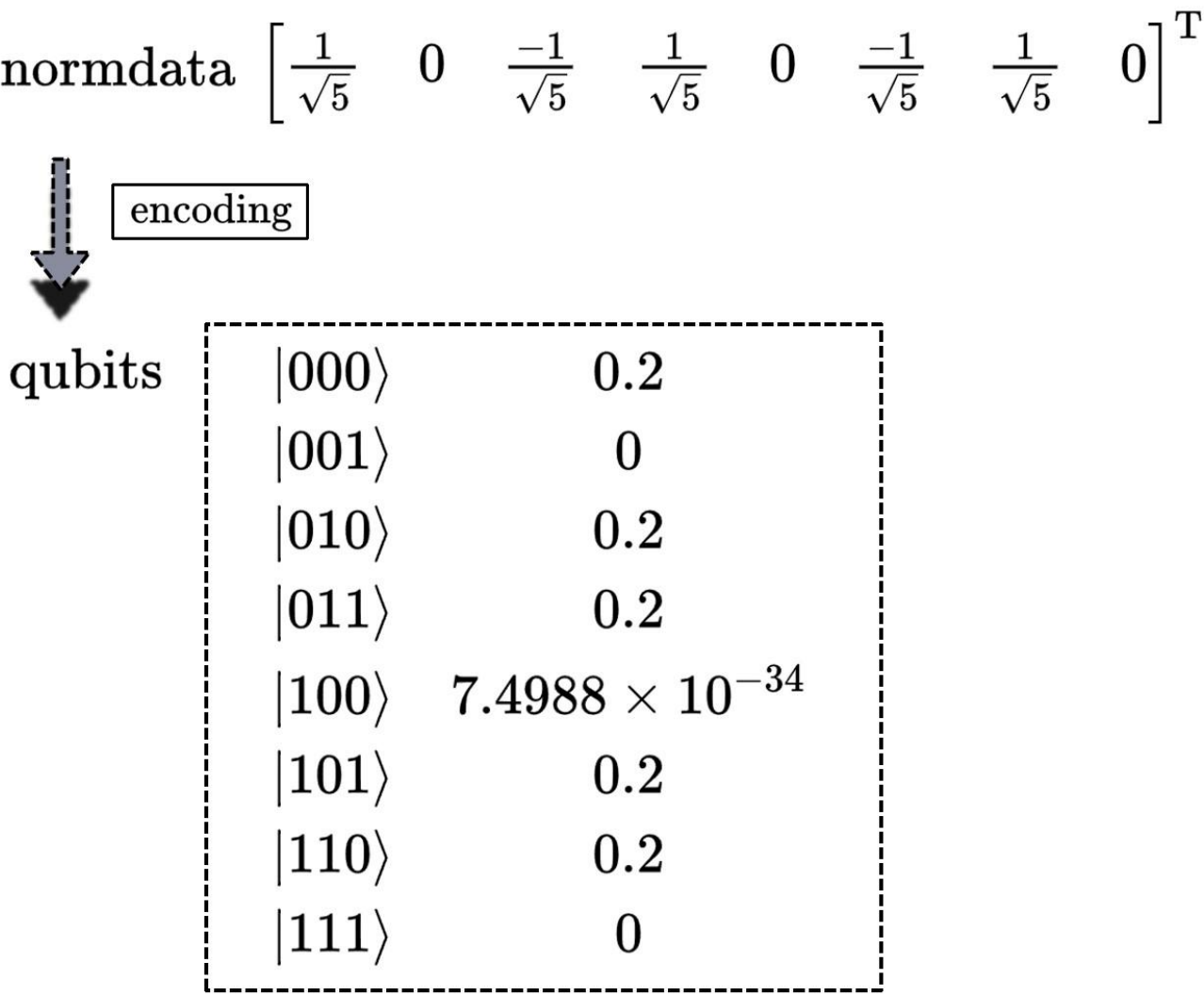}
		\caption{\small Amplitude encoding from normalized data array to quantum state.}
		\label{fig_Quantum:encoding}
	\end{figure}

	This study adopts the top-down amplitude encoding\supercite{2023Configurable}, whose protocol operates through two phases: preprocessing the input vector to construct an angle tree representation, and then sequential encoding initiated from the root node. \modifyFirstYellow{The implementation demonstrates the linear scaling in both circuit width $O\left( {{\log }_2}N \right)$ and circuit depth $O\left( N \right)$. The encoding example is displayed in Fig.(\ref{fig_Quantum:encoding}).}
	
\begin{figure}[H]
		\centering
		\includegraphics[scale=0.3]{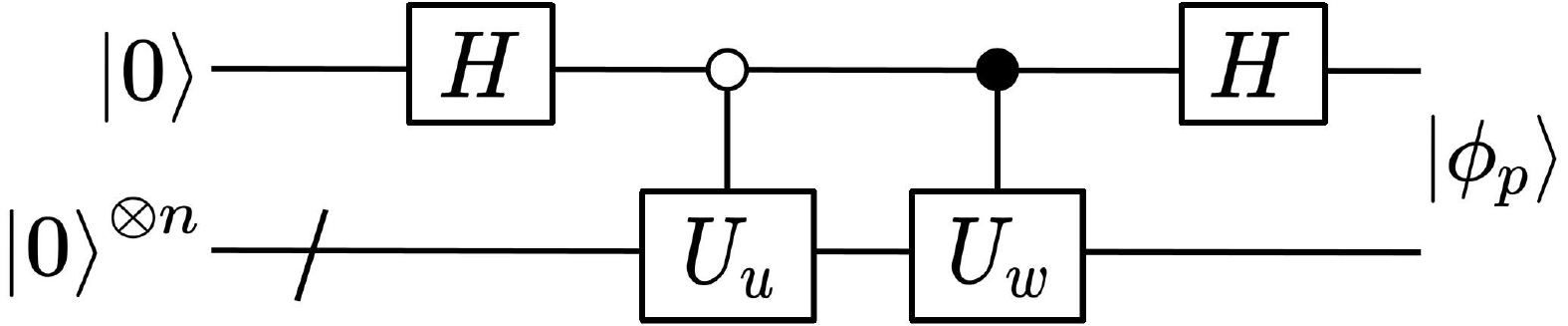}
		\caption{\small Quantum circuit to prepare the specified state $\left | \phi_{p}   \right \rangle$.}
		\label{fig_Quantum:statepre}
	\end{figure}

	The swap test is a quantum algorithm that estimates the overlap between two quantum states $\left| u \right\rangle $ and $\left| w \right\rangle $ by using an ancillary control qubit. The protocol involves preparing the ancilla in a superposition state, performing a controlled-swap operation between the target states, and measuring the ancilla to determine the state overlap from output probabilities. This method provides an efficient way to compare quantum states and evaluate their similarity, with applications ranging from quantum state characterization to machine learning, while requiring minimal quantum resources. However, its accuracy is inevitably limited by current noisy quantum hardware. The swap test method has been extensively applied into fields on quantum computing\supercite{CHEN2025107442,HUANG2024129614}, which can promote the procedure on quantum inner product. Its quantum circuits usually include:
	
	\textbf{Prepare the state:} In order to estimate the overlap between two quantum states $\left| u \right\rangle $ and $\left| w \right\rangle $ by swap test, the specified quantum state $\left | \phi_{p}   \right \rangle$ is prepared shown in Fig.(\ref{fig_Quantum:statepre}). The ${U_u}$ and ${U_w}$ denote unitary transformation on quantum state encoding for $\left| u \right\rangle $ and $\left| w \right\rangle $.
	\begin{equation}\label{eq_Quantum:statepre}
		\left | \phi_{p}   \right \rangle =\frac{\sqrt{2}}{2}\left ( \left | +  \right \rangle \left | u  \right \rangle +\left | -  \right \rangle \left | w  \right \rangle  \right ) 
	\end{equation}
	
	The state $\left | \phi_{p}   \right \rangle$ of Eq.(\ref{eq_Quantum:statepre}) can be rewritten as
	\begin{equation}\label{eq_Quantum:2}
		\left| {{\phi _p}} \right\rangle  = \frac{1}{2}\left( {\left| 0 \right\rangle \left( {\left| u \right\rangle  + \left| w \right\rangle } \right) + \left| 1 \right\rangle \left( {\left| u \right\rangle  - \left| w \right\rangle } \right)} \right)
	\end{equation}
	where the amplitude of $\left| 0 \right\rangle $ is${{\sqrt {1 + {\mathop{\rm Re}\nolimits} \left\langle {u}
				\mathrel{\left | {\vphantom {u w}}
					\right. \kern-\nulldelimiterspace}
				{w} \right\rangle } } \mathord{\left/
			{\vphantom {{\sqrt {1 + {\mathop{\rm Re}\nolimits} \left\langle {u}
							\mathrel{\left | {\vphantom {u w}}
								\right. \kern-\nulldelimiterspace}
							{w} \right\rangle } } {\sqrt 2 }}} \right.
			\kern-\nulldelimiterspace} {\sqrt 2 }}$, and the amplitude of $\left| 1 \right\rangle$ is ${{\sqrt {1 - {\mathop{\rm Re}\nolimits} \left\langle {u}
				\mathrel{\left | {\vphantom {u w}}
					\right. \kern-\nulldelimiterspace}
				{w} \right\rangle } } \mathord{\left/
			{\vphantom {{\sqrt {1 - {\mathop{\rm Re}\nolimits} \left\langle {u}
							\mathrel{\left | {\vphantom {u w}}
								\right. \kern-\nulldelimiterspace}
							{w} \right\rangle } } {\sqrt 2 }}} \right.
			\kern-\nulldelimiterspace} {\sqrt 2 }}$.
	
	Further, we conveniently represent $\left| x \right\rangle$ and $\left| y \right\rangle$ as the normalized states of $\left| u \right\rangle  + \left| w \right\rangle$ and $\left| u \right\rangle  - \left| w \right\rangle$, respectively. Then there is a real number ${\theta _p} \in \left[ {0,{\pi  \mathord{\left/
				{\vphantom {\pi  2}} \right.
				\kern-\nulldelimiterspace} 2}} \right]$ such that
	\begin{equation}\label{eq_Quantum:transform}
		\left| {{\phi _p}} \right\rangle  = \sin {\theta _p}\left| 0 \right\rangle \left| x \right\rangle  + \cos {\theta _p}\left| 1 \right\rangle \left| y \right\rangle 
	\end{equation}
	\begin{equation}\label{eq_Quantum:transform2}
		\left\{ \begin{array}{l}
			\cos {\theta _p} = {{\sqrt {1 - {\mathop{\rm Re}\nolimits} \left\langle {u}
						\mathrel{\left | {\vphantom {u w}}
							\right. \kern-\nulldelimiterspace}
						{w} \right\rangle } } \mathord{\left/
					{\vphantom {{\sqrt {1 - {\mathop{\rm Re}\nolimits} \left\langle {u}
									\mathrel{\left | {\vphantom {u w}}
										\right. \kern-\nulldelimiterspace}
									{w} \right\rangle } } {\sqrt 2 }}} \right.
					\kern-\nulldelimiterspace} {\sqrt 2 }}\\
			\sin {\theta _p} = {{\sqrt {1 + {\mathop{\rm Re}\nolimits} \left\langle {u}
						\mathrel{\left | {\vphantom {u w}}
							\right. \kern-\nulldelimiterspace}
						{w} \right\rangle } } \mathord{\left/
					{\vphantom {{\sqrt {1 + {\mathop{\rm Re}\nolimits} \left\langle {u}
									\mathrel{\left | {\vphantom {u w}}
										\right. \kern-\nulldelimiterspace}
									{w} \right\rangle } } {\sqrt 2 }}} \right.
					\kern-\nulldelimiterspace} {\sqrt 2 }}
		\end{array} \right.
	\end{equation}
	\begin{equation}\label{eq_Quantum:transform3}
		{\mathop{\rm Re}\nolimits} \left\langle {u}
		\mathrel{\left | {\vphantom {u w}}
			\right. \kern-\nulldelimiterspace}
		{w} \right\rangle  =  - \cos 2{\theta _p}
	\end{equation}
	
\begin{figure}[H]
		\centering
		\includegraphics[scale=0.4]{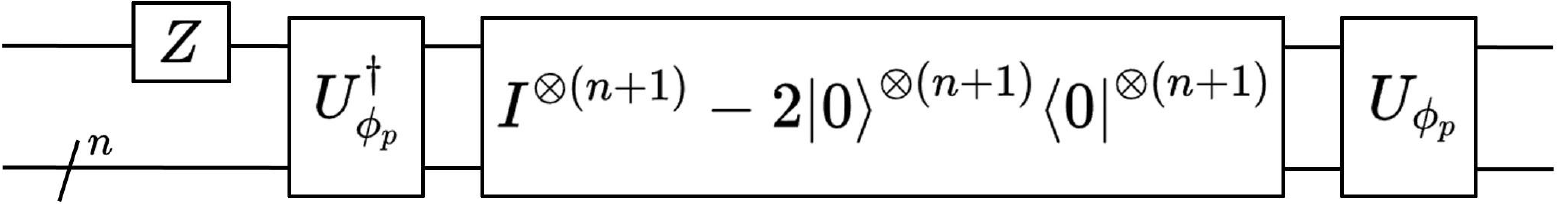}
		\caption{\small Construct the unitary transformation ${Q_p}$.}
		\label{fig_Quantum:construct}
	\end{figure}

	\textbf{Construct the unitary transformation:} Through preparing the specified state and the entanglement of two quantum states, the inner product can be embodied by their amplitude angle in Eq.(\ref{eq_Quantum:transform3}). The present purpose is to calculate the ${\theta _p}$ of Eq.(\ref{eq_Quantum:transform3}) by quantum phase estimation algorithm. As a result, the specified constructed procedure on unitary transformation ${Q_p}$ is implemented in Fig.(\ref{fig_Quantum:construct}).
	\begin{equation}\label{eq_Quantum:construct}
		\begin{aligned}
			{Q_p}& = \left( {{I^{ \otimes \left( {n + 1} \right)}} - 2\left| {{\phi _p}} \right\rangle \left\langle {{\phi _p}} \right|} \right)\left( {Z \otimes {I^{ \otimes n}}} \right)\\&
			= {U_{{\phi _p}}}\left( {{I^{ \otimes \left( {n + 1} \right)}} - 2{{\left| 0 \right\rangle }^{ \otimes \left( {n + 1} \right)}}{{\left\langle 0 \right|}^{ \otimes \left( {n + 1} \right)}}} \right)U_{{\phi _p}}^\dag \left( {Z \otimes {I^{ \otimes n}}} \right)
		\end{aligned}
	\end{equation}
	where $Z = \left| 0 \right\rangle \left\langle 0 \right| - \left| 1 \right\rangle \left\langle 1 \right|$ is the Pauli Z matrix.
	
\begin{figure}[H]
		\centering
		\includegraphics[scale=0.2]{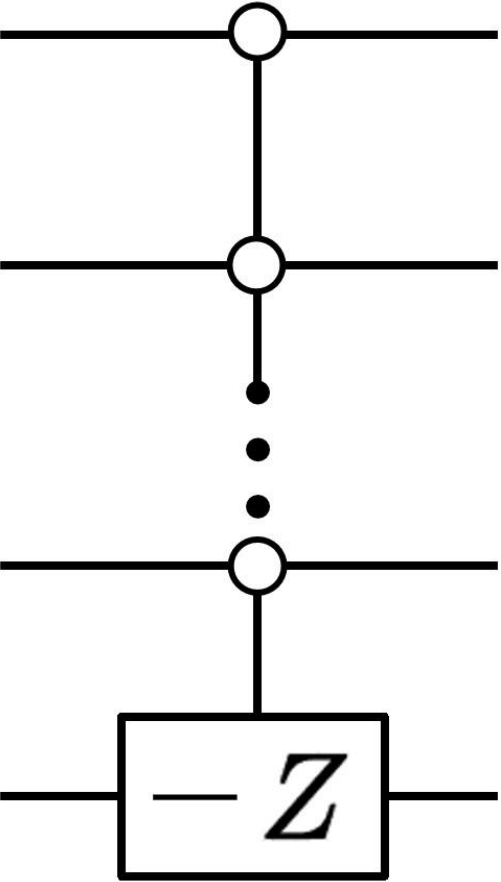}
		\caption{\small Quantum circuit of unitary operator ${I^{ \otimes \left( {n + 1} \right)}} - 2{\left| 0 \right\rangle ^{ \otimes \left( {n + 1} \right)}}\langle 0{|^{ \otimes \left( {n + 1} \right)}}$.}
		\label{fig_Quantum:zGate}
	\end{figure}

	As for the unitary operator ${I^{ \otimes \left( {n + 1} \right)}} - 2{\left| 0 \right\rangle ^{ \otimes \left( {n + 1} \right)}}\langle 0{|^{ \otimes \left( {n + 1} \right)}}$, it flips the sign of the all-zero quantum state while leaving other states unchanged, acting as a key phase-based oracle in algorithms like Grover's search\supercite{1997A}.  It can be implemented efficiently using controlled quantum gates despite operating on an exponentially large state space. We can run it in the circuit shown in Fig.(\ref{fig_Quantum:zGate})\supercite{WOS:000477879500005}.
	
	Subsequently, Eq.(\ref{eq_Quantum:transform}) can be transformed by Schmidt decomposition method\supercite{WOS:000291924100002} as
	\begin{equation}
		\begin{aligned}
			\left| {{\phi _p}} \right\rangle  &= \frac{{ - {\boldsymbol{i}}}}{{\sqrt 2 }}\left( {{e^{{\theta _p}{\boldsymbol{i}}}}\left| {{\zeta _ + }} \right\rangle  - {e^{ - {\theta _p}{\boldsymbol{i}}}}\left| {{\zeta _ - }} \right\rangle } \right)\\
			\left| {{\zeta _ \pm }} \right\rangle & = \frac{1}{{\sqrt 2 }}\left( {\left| 0 \right\rangle \left| x \right\rangle  \pm {\boldsymbol{i}}\left| 1 \right\rangle \left| y \right\rangle } \right)
		\end{aligned}
	\end{equation}
	for the decomposed prepared state $\left| {{\phi _p}} \right\rangle $, the unitary transformation ${Q_p}$ has the following characteristic equation as
	\begin{equation}
		{Q_p}\left| {{\zeta _ \pm }} \right\rangle  = {e^{ \pm 2{\theta _p}{\boldsymbol{i}}}}\left| {{\zeta _ \pm }} \right\rangle 
	\end{equation}
	where $\left| {{\zeta _ \pm }} \right\rangle$ denotes the eigenstates of ${Q_p}$, the phase ${\theta _p}$ is skillfully contained in the eigenvalues.
	
\begin{figure}[H]
		\centering
		\includegraphics[scale=0.32]{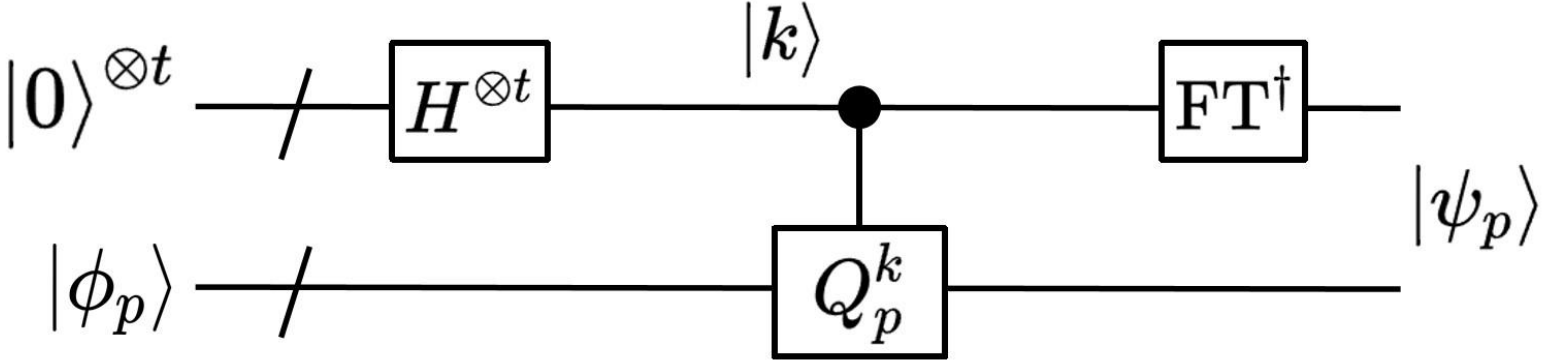}
		\caption{\small Estimate the proper phase ${\theta _p}$ of unitary transformation ${Q_p}$.}
		\label{fig_Quantum:qpe}
	\end{figure}

	\textbf{Estimate the phase:} Currently adopt the quantum phase estimation (QPE) algorithm to estimate the phase ${\theta _p}$. The quantum circuit is shown in Fig.(\ref{fig_Quantum:qpe}). During the procedure on quantum phase estimation, the circuit of auxiliary $t$ qubits is constructed to estimate the phase ${\theta _p}$ with higher precision, whose number of $t$ qubits controls its calculation accuracy.
	
	In Fig.(\ref{fig_Quantum:qpe}), the ${\rm{FT}}$ indicates the quantum Fourier transform\supercite{Shor1997POLYNOMIAL}. The control gate $Q_p^k$ is composed with a series of controlled gates $Q_p^{{2^m}}$ by ergodic ${m^{{\rm{th}}}}$ qubit in the first register as control qubit, where $m = 0, \ldots ,t - 1$.
	
	The output state $\left| {{\psi _p}} \right\rangle $ of quantum circuit by present algorithm can be described as
	\begin{equation}
		\left| {{\psi _p}} \right\rangle  = \frac{{ - {\boldsymbol{i}}}}{{\sqrt 2 }}\left( {{e^{{\theta _p}{\boldsymbol{i}}}}\left| {{\lambda _e}} \right\rangle \left| {{\zeta _ + }} \right\rangle  - {e^{ - {\theta _p}{\boldsymbol{i}}}}\left| {{2^t} - {\lambda _e}} \right\rangle \left| {{\zeta _ - }} \right\rangle } \right)
	\end{equation}
	where ${\left| {{\lambda _e}} \right\rangle }$ denotes the auxiliary estimation state on $t$ qubits, the binary value ${\lambda _e}$ of maximum amplitude state has ${\lambda _e} \in \left[ {0,{2^{t - 1}}} \right]$ and then ${{{\lambda _e}\pi } \mathord{\left/
			{\vphantom {{{\lambda _e}\pi } {{2^t}}}} \right.
			\kern-\nulldelimiterspace} {{2^t}}}$ is the approximation on phase ${\theta _p}$.
	
	Finally, combining with Eq.(\ref{eq_Quantum:transform3}), the inner product has
	\begin{equation}\label{eq_Quantum:finallInnerProduct}
		{\rm{Re}}\left\langle {u}
		\mathrel{\left | {\vphantom {u w}}
			\right. \kern-\nulldelimiterspace}
		{w} \right\rangle  \cong  - \cos \left( {{{{\lambda _e}\pi } \mathord{\left/
					{\vphantom {{{\lambda _e}\pi } {{2^{t - 1}}}}} \right.
					\kern-\nulldelimiterspace} {{2^{t - 1}}}}} \right)
	\end{equation}
	
	Note that the imaginary part ${\mathop{\rm Im}\nolimits} \left\langle {u}
	\mathrel{\left | {\vphantom {u w}}
		\right. \kern-\nulldelimiterspace}
	{w} \right\rangle $ can be also estimated similarly in the quantum procedure of the real part of inner product. Given that the present study does not investigate the imaginary component solution of quantum computing, we will not elaborate further on this aspect\supercite{WOS:000477879500005}.
	
	\section{Numerical approximation}
	\subsection{Meshfree finite particle method}
	\normalsize \hspace{10pt}
	As a comprehensive numerical scheme based on mathematical multivariate Taylor series expansion in Eq.(\ref{eq_FPM:1}), the meshfree finite particle method (FPM) considers a weighted discretization of weakly integral form described in Eq.(\ref{eq_FPM:2}).
	\begin{equation}\label{eq_FPM:1}
		f\left( {\bf{x}} \right) = {f_i} + {f_{i,\alpha }}\left( {{{\bf{x}}^\alpha } - {\bf{x}}_i^\alpha } \right) + O\left( {{{\left( {{\bf{x}} - {{\bf{x}}_i}} \right)}^2}} \right)
	\end{equation}
	\begin{equation}\label{eq_FPM:2}
		\int_{\Omega }{f\left( \mathbf{x} \right)\Phi \left( \mathbf{x}-{{\mathbf{x}}_{i}} \right)d\mathbf{x}}={{f}_{i}}\int_{\Omega }{\Phi \left( \mathbf{x}-{{\mathbf{x}}_{i}} \right)d\mathbf{x}}+{{f}_{i,\alpha }}\int_{\Omega }{\left( {{\mathbf{x}}^{\alpha }}-\mathbf{x}_{i}^{\alpha } \right)\Phi \left( \mathbf{x}-{{\mathbf{x}}_{i}} \right)d\mathbf{x}}+O\left( {{\left( \mathbf{x}-{{\mathbf{x}}_{i}} \right)}^{2}} \right)
	\end{equation}
	\begin{equation}\label{eq_FPM:3}
		{{f}_{i}}=f\left( {{\mathbf{x}}_{i}} \right)
	\end{equation}
	\begin{equation}\label{eq_FPM:4}
		{{f}_{i,\alpha }}={{f}_{\alpha }}\left( {{\mathbf{x}}_{i}} \right)={{\left( {\partial f}/{\partial {{\mathbf{x}}^{\alpha }}}\; \right)}_{i}}
	\end{equation}
	\begin{equation}\label{eq_FPM:5}
		\left[ \begin{matrix}
			\int_{\Omega }{\omega \left( \mathbf{x}-{{\mathbf{x}}_{i}} \right)d\mathbf{x}} & \int_{\Omega }{\left( {{\mathbf{x}}^{\alpha }}-\mathbf{x}_{i}^{\alpha } \right)\omega \left( \mathbf{x}-{{\mathbf{x}}_{i}} \right)d\mathbf{x}}  \\
			\int_{\Omega }{\nabla \omega \left( \mathbf{x}-{{\mathbf{x}}_{i}} \right)d\mathbf{x}} & \int_{\Omega }{\left( {{\mathbf{x}}^{\alpha }}-\mathbf{x}_{i}^{\alpha } \right)\nabla \omega \left( \mathbf{x}-{{\mathbf{x}}_{i}} \right)d\mathbf{x}}  \\
		\end{matrix} \right]\left[ \begin{matrix}
			{{f}_{i}}  \\
			{{f}_{i,\alpha }}  \\
		\end{matrix} \right]\cong \left[ \begin{matrix}
			\int_{\Omega }{f\left( \mathbf{x} \right)\omega \left( \mathbf{x}-{{\mathbf{x}}_{i}} \right)d\mathbf{x}}  \\
			\int_{\Omega }{f\left( \mathbf{x} \right)\nabla \omega \left( \mathbf{x}-{{\mathbf{x}}_{i}} \right)d\mathbf{x}}  \\
		\end{matrix} \right]
	\end{equation}
	\begin{equation}\label{eq_FPM:6}
		\mathbf{L}=\left[ \begin{matrix}
			\int_{\Omega }{\omega \left( \mathbf{x}-{{\mathbf{x}}_{i}} \right)d\mathbf{x}} & \int_{\Omega }{\left( {{\mathbf{x}}^{\alpha }}-\mathbf{x}_{i}^{\alpha } \right)\omega \left( \mathbf{x}-{{\mathbf{x}}_{i}} \right)d\mathbf{x}}  \\
			\int_{\Omega }{\nabla \omega \left( \mathbf{x}-{{\mathbf{x}}_{i}} \right)d\mathbf{x}} & \int_{\Omega }{\left( {{\mathbf{x}}^{\alpha }}-\mathbf{x}_{i}^{\alpha } \right)\nabla \omega \left( \mathbf{x}-{{\mathbf{x}}_{i}} \right)d\mathbf{x}}  \\
		\end{matrix} \right],\mathbf{F}=\left[ \begin{matrix}
			{{f}_{i}}  \\
			{{f}_{i,\alpha }}  \\
		\end{matrix} \right],\mathbf{B}=\left[ \begin{matrix}
			\int_{\Omega }{f\left( \mathbf{x} \right)\omega \left( \mathbf{x}-{{\mathbf{x}}_{i}} \right)d\mathbf{x}}  \\
			\int_{\Omega }{f\left( \mathbf{x} \right)\nabla \omega \left( \mathbf{x}-{{\mathbf{x}}_{i}} \right)d\mathbf{x}}  \\
		\end{matrix} \right]
	\end{equation}\\
	where $\Phi \left( \mathbf{x} \right)$ represents a basis function defined within the support domain $\Omega$. Providing a set of basis functions $\Phi (\mathbf{x})=\left\{ {{\omega }^{0}}(\mathbf{x})\text{, }{{\omega }^{1}}(\mathbf{x})\text{, }{{\omega }^{2}}(\mathbf{x}),... \right\}$ satisfies the corresponding consistency conditions on both sides of Eq.(\ref{eq_FPM:1}), a linear matrix system can be constructed using integrated kernel forms and purely physical discretized nodes. The choice of basis functions ${{\omega }^{0}}(\mathbf{x}), {{\omega }^{1}}(\mathbf{x}), {{\omega }^{2}}(\mathbf{x}), \ldots$ follows the approach proposed by Liu et al.\supercite{2005Modeling}. Consequently, the smoothing function $\omega \left( \mathbf{x} \right)$ and its derivatives $\nabla \omega \left( \mathbf{x} \right)$ may serve as appropriate members of such a basis set. By employing distinct basis functions and omitting higher-order Taylor expansion terms, the linear matrix equation corresponding to Eq.(\ref{eq_FPM:5}) is derived.
	
	By expressing the solution to the matrix Eq.(\ref{eq_FPM:5}) as $\mathbf{F} \cong {{\mathbf{L}}^{-1}}\mathbf{B}$, the finite particle method (FPM) is formally introduced. Neglecting the inversion of approximated functional value ${{f}_{i}}$ leads to the kernel gradient correction (KGC) technique presented in Eq.(\ref{eq_FPM:7}), which can be also obtained through transformation of Eq.(\ref{eq_FPM:2}). In contrast to KGC, the FPM incorporates the resolved functional value ${{f}_{i}}$ into the reorganized matrix system to achieve improved numerical accuracy.
	\begin{equation}\label{eq_FPM:7}
		\left[ \int_{\Omega }{\left( {{\mathbf{x}}^{\alpha }}-\mathbf{x}_{i}^{\alpha } \right)\nabla \omega \left( \mathbf{x}-{{\mathbf{x}}_{i}} \right)d\mathbf{x}} \right]\left[ {{f}_{i,\alpha }} \right]\cong \left[ \int_{\Omega }{\left( f\left( \mathbf{x} \right)-{{f}_{i}} \right)\nabla \omega \left( \mathbf{x}-{{\mathbf{x}}_{i}} \right)d\mathbf{x}} \right]
	\end{equation}
	
	In the case of smoothed particle hydrodynamics (SPH), the resolved matrix $\mathbf{L}$ is commonly approximated as a unit matrix and neglected, allowing the physical variable to be directly approximated via Eq.(\ref{eq_FPM:8}) as a standard interpolation procedure. Although traditional SPH exhibits relatively low accuracy, it retains considerable numerical robustness, making it suitable for a wide range of practical engineering simulations\supercite{WOS:001501337900001}.
	\begin{equation}\label{eq_FPM:8}
		\left[ \begin{matrix}
			{{f}_{i}}  \\
			{{f}_{i,\alpha }}  \\
		\end{matrix} \right]\cong \left[ \begin{matrix}
			\int_{\Omega }{f\left( \mathbf{x} \right)\omega \left( \mathbf{x}-{{\mathbf{x}}_{i}} \right)d\mathbf{x}}  \\
			\int_{\Omega }{f\left( \mathbf{x} \right)\nabla \omega \left( \mathbf{x}-{{\mathbf{x}}_{i}} \right)d\mathbf{x}}  \\
		\end{matrix} \right]
	\end{equation}
	
	Similarly, within the SPH framework, the approximations presented in Eq.(\ref{eq_FPM:5}) satisfy the properties required for constant functions, as specified in Eq.(\ref{eq_FPM:9}). By incorporating this condition, symmetric formulations of the aforementioned methods can be obtained for particle–particle interactions, as shown in Eqs.(\ref{eq_FPM:10}–\ref{eq_FPM:12}). These symmetric forms exhibit improved conservation properties in interactive dynamics.
	\begin{equation}\label{eq_FPM:9}
		\left[ \begin{matrix}
			\int_{\Omega }{\omega \left( \mathbf{x}-{{\mathbf{x}}_{i}} \right)d\mathbf{x}} & \int_{\Omega }{\left( {{\mathbf{x}}^{\alpha }}-\mathbf{x}_{i}^{\alpha } \right)\omega \left( \mathbf{x}-{{\mathbf{x}}_{i}} \right)d\mathbf{x}}  \\
			\int_{\Omega }{\nabla \omega \left( \mathbf{x}-{{\mathbf{x}}_{i}} \right)d\mathbf{x}} & \int_{\Omega }{\left( {{\mathbf{x}}^{\alpha }}-\mathbf{x}_{i}^{\alpha } \right)\nabla \omega \left( \mathbf{x}-{{\mathbf{x}}_{i}} \right)d\mathbf{x}}  \\
		\end{matrix} \right]\left[ \begin{matrix}
			\text{Const}  \\
			0  \\
		\end{matrix} \right]\cong \left[ \begin{matrix}
			\text{Const}\int_{\Omega }{\omega \left( \mathbf{x}-{{\mathbf{x}}_{i}} \right)d\mathbf{x}}  \\
			\text{Const}\int_{\Omega }{\nabla \omega \left( \mathbf{x}-{{\mathbf{x}}_{i}} \right)d\mathbf{x}}  \\
		\end{matrix} \right]
	\end{equation}
	\begin{equation}\label{eq_FPM:10}
		\text{FPM:}\begin{aligned}
			& \left[ \begin{matrix}
				{{f}_{i}}  \\
				-  \\
			\end{matrix} \right]\cong {{\left[ \begin{matrix}
						\sum{{{\omega }_{ij}}\Delta {{V}_{j}}} & \sum{\mathbf{x}_{ji}^{\beta }{{\omega }_{ij}}\Delta {{V}_{j}}}  \\
						\sum{{{\omega }_{ij,\alpha }}\Delta {{V}_{j}}} & \sum{\mathbf{x}_{ji}^{\alpha }{{\omega }_{ij,\beta }}\Delta {{V}_{j}}}  \\
					\end{matrix} \right]}^{-1}}\left[ \begin{matrix}
				\sum{{{f}_{j}}{{\omega }_{ij}}\Delta {{V}_{j}}}  \\
				\sum{{{f}_{j}}{{\omega }_{ij,\beta }}\Delta {{V}_{j}}}  \\
			\end{matrix} \right] \\ 
			& \left[ \begin{matrix}
				-  \\
				{{f}_{i,\alpha }}  \\
			\end{matrix} \right]\cong {{\left[ \begin{matrix}
						\sum{{{\omega }_{ij}}\Delta {{V}_{j}}} & \sum{\mathbf{x}_{ji}^{\beta }{{\omega }_{ij}}\Delta {{V}_{j}}}  \\
						\sum{{{\omega }_{ij,\alpha }}\Delta {{V}_{j}}} & \sum{\mathbf{x}_{ji}^{\alpha }{{\omega }_{ij,\beta }}\Delta {{V}_{j}}}  \\
					\end{matrix} \right]}^{-1}}\left[ \begin{matrix}
				\sum{\left( {{f}_{j}}-{{f}_{i}} \right){{\omega }_{ij}}\Delta {{V}_{j}}}  \\
				\sum{\left( {{f}_{j}}-{{f}_{i}} \right){{\omega }_{ij,\beta }}\Delta {{V}_{j}}}  \\
			\end{matrix} \right] \\ 
		\end{aligned}
	\end{equation}
	\begin{equation}\label{eq_FPM:11}
		\text{KGC:}\left[ {{f}_{i,\alpha }} \right]\cong {{\left[ \sum{\mathbf{x}_{ji}^{\alpha }{{\omega }_{ij,\beta }}\Delta {{V}_{j}}} \right]}^{-1}}\left[ \sum{\left( {{f}_{j}}-{{f}_{i}} \right){{\omega }_{ij,\beta }}\Delta {{V}_{j}}} \right]
	\end{equation}
	\begin{equation}\label{eq_FPM:12}
		\text{SPH:}{{f}_{i}}\cong \sum{{{f}_{j}}{{\omega }_{ij}}\Delta {{V}_{j}}},\text{      }{{f}_{i,\alpha }}\cong \sum{\left( {{f}_{j}}-{{f}_{i}} \right){{\omega }_{ij,\alpha }}\Delta {{V}_{j}}}
	\end{equation}\\
	where the particle-particle interactional form is precisely as ${{A}_{ij}}={{A}_{i}}-{{A}_{j}}$, ${{\omega }_{ij,\beta }}$ denotes the partial derivative of basis functional value and $\Delta {{V}_{j}}$ denotes the volume of particle $j$.
	
	\subsection{Quantum finite particle method with multi-partitioned zones}
\begin{figure}[H]
		\centering
		\includegraphics[scale=0.5]{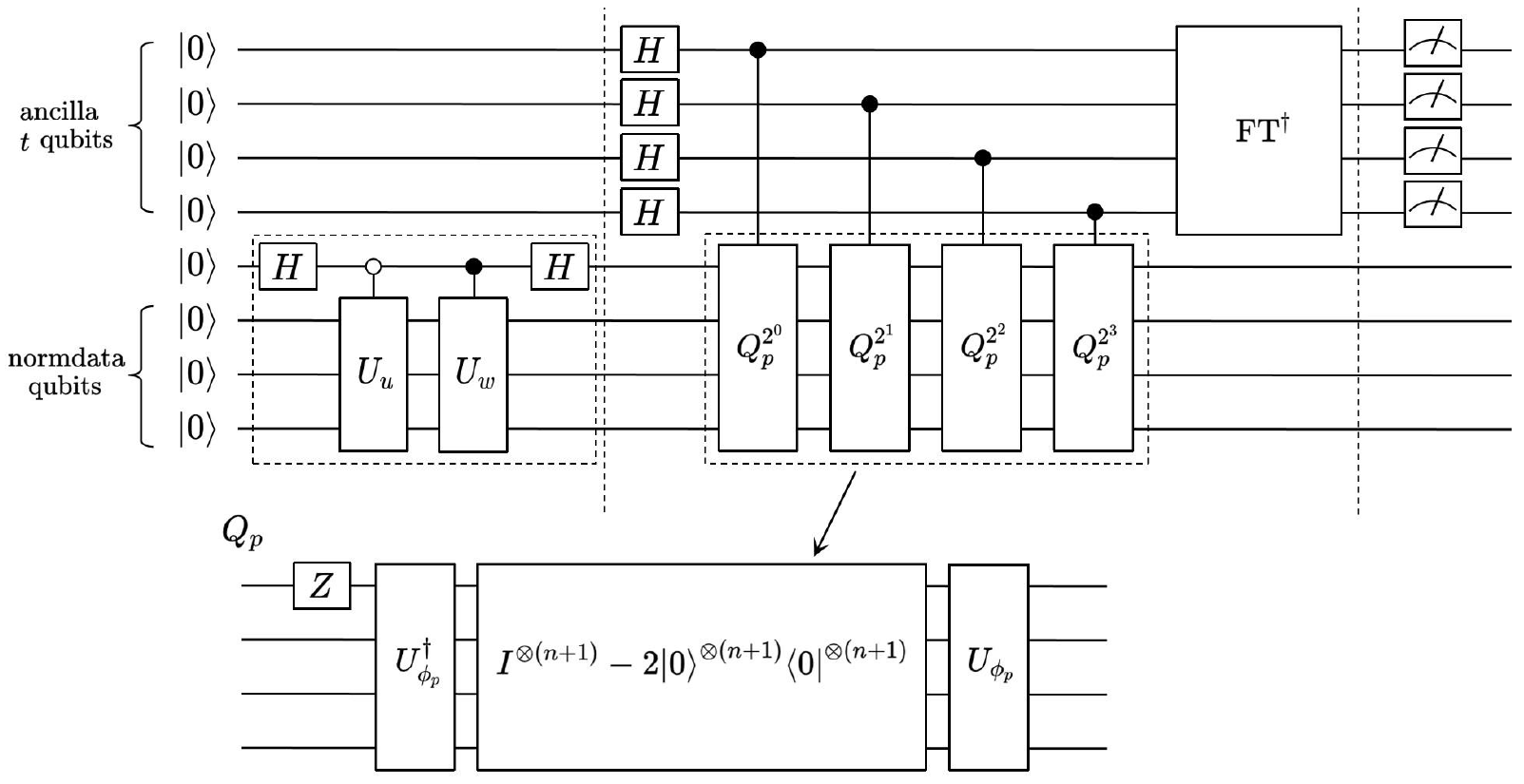}
		\caption{\small Complete quantum circuit diagram in quantum finite particle method.}
		\label{fig_quantumFPM:partition}
	\end{figure}

	\normalsize \hspace{10pt}
	According to the aforementioned quantum descriptions and numerical approximation on meshfree finite particle method, the hybrid quantum finite particle method (Q-FPM) is presented in this subsection. Based on the quantum phase estimation (QPE) on ancilla qubits, the complete quantum circuit diagram is described in Fig.(\ref{fig_quantumFPM:partition}), in which three qubits of norm data (the size of array is ${2^n}$, ${n}$ is the size of qubits of norm data) and four ancilla qubits ($t=4$) for observation are requested. Quantum results on QPE measurement are counted and concluded by a large number of quantum observations (quantum collapse). The procedure is to find the specific quantum superposition state that has been observed at the most times.
	
	\begin{figure}[H]
		\centering
		\includegraphics[scale=0.45]{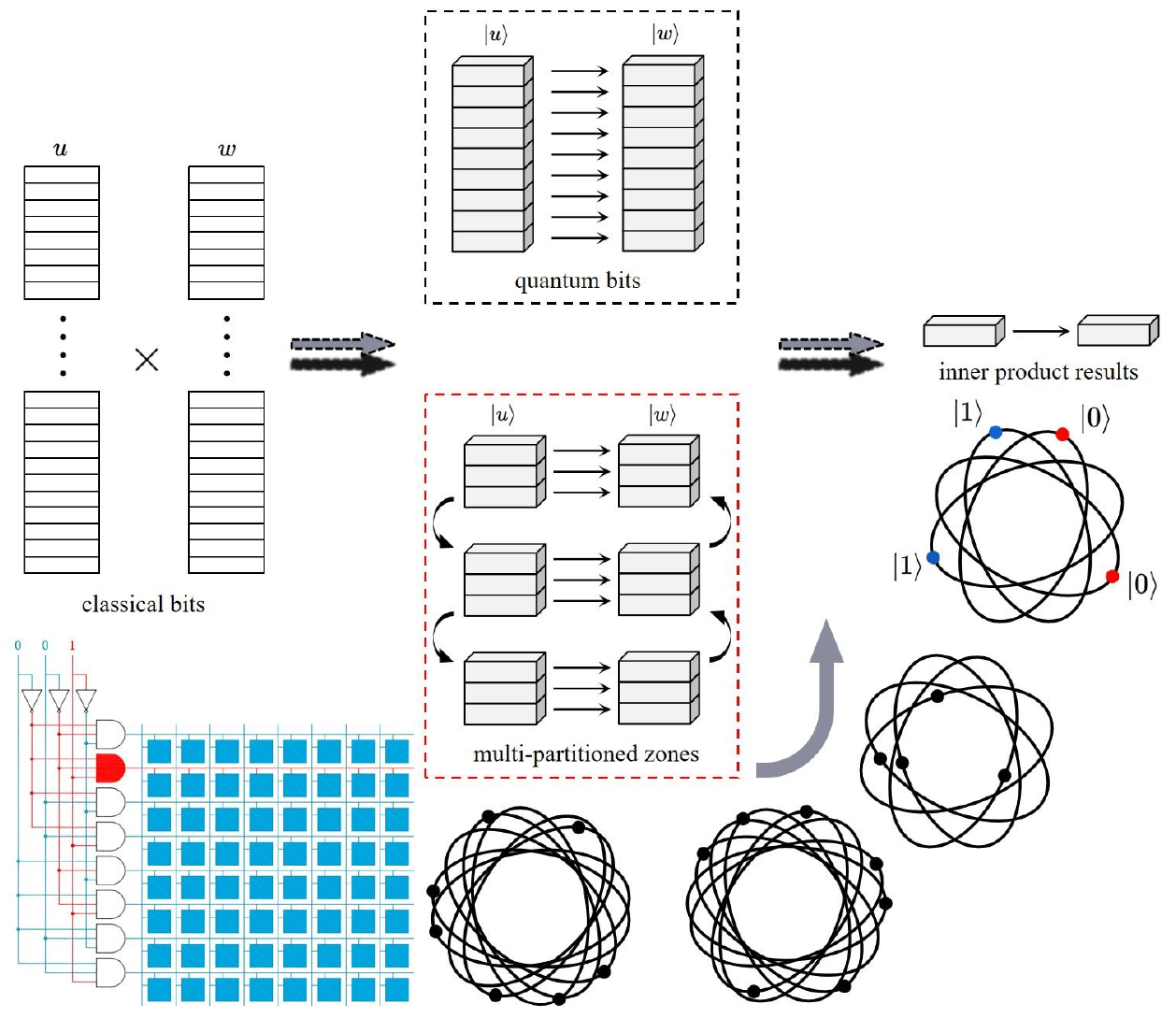}
		\caption{\small The fixed small-scale quantum procedure on advantaged multi-partitioned zones.}
		\label{fig_quantumFPM:partitionedzones}
	\end{figure}
	
	In Eqs.(\ref{eq_FPM:10}-\ref{eq_FPM:12}) of meshfree numerical approximation, the superimposed form array of adjacent summation is usually the changeable structure and of a large scale for general computational particle methods. Further considering the indefinite requests on computing size of qubits and the current limitation on NISQ device, this study proposed the strategy of multi-partitioned zones to optimize the requests on qubits and improve the efficiency on quantum computing. As shown in Fig.(\ref{fig_quantumFPM:partitionedzones}), the classical $N$ bits of data arrays are requested on the form of binary system and the quantum bits $n = \log \left( N \right)$ are requested accordingly. The resource-efficient quantum computational strategy on multi-partitioned zones leverages a fixed small-scale quantum circuit as a fundamental processing unit, which is iteratively nested to handle inner product calculations for arbitrarily sized arrays described in Eq.(\ref{eq_quantumFPM:1}). Hence the quantum finite particle method with multi-partitioned zones is more suitable to perform an inner product process of indefinite scales in the current quantum device.
	
	Through the hybrid quantum finite particle method and the representation on corresponding quantum circuits, the quantum approximation can be transformed in Eqs.(\ref{eq_quantumFPM:2}-\ref{eq_quantumFPM:4}) based on the meshfree particle approximation on Eqs.(\ref{eq_FPM:10}-\ref{eq_FPM:12}). Following the Eq.(\ref{eq_Quantum:finallInnerProduct}) and Eq.(\ref{eq_quantumFPM:1}), ${\left\langle {{u_j}\left| {{w_j}} \right.} \right\rangle _{j = 1,2,...}}$ denotes the quantum inner product calculations for arbitrarily sized arrays ($j = 1,2,...$). In this form, the meshfree numerical approximation can be represented as the quantum structure of neighboring particle cluster. Therein, the special smoothing kernel ${\omega _{ij}^{}}$ and its derivatives ${\omega _{ij,\alpha }^{}}$ are locally weighted and constructed as the right product state. The superscript $c$ denotes the kernel correction on FPM scheme and the definition on spatial dimension $\left( {\alpha ,\beta  = 0,x,y,z} \right)$ contributes the approximation on original function value and spatial directional discretization, where the KGC neglected the inversion on approximated functional value ${{f}_{i}}$ and is only related with the discretization on spatial direction. In this study, smoothing kernel functions are requested and described detailedly in the previous work\supercite{WOS:001501337900001,10.1063/5.0197088}.
	
	\begin{equation}\label{eq_quantumFPM:1}
		u \cdot w = {u_j}{w_j} \cong {\left\langle {u\left| w \right.} \right\rangle _{j = 1,2,...}}
	\end{equation}
	
	\begin{equation}\label{eq_quantumFPM:2}
		\text{FPM:}\left\{\begin{aligned}
			{f_i} \cong {\left\langle {{f_j}\Delta {V_j}\left| {\omega _{ij,\alpha  = 0}^c} \right.} \right\rangle _{j = 1,2,...}}\\
			{f_{i,\alpha }} \cong {\left\langle {\left( {{f_j} - {f_i}} \right)\Delta {V_j}\left| {\omega _{ij,\alpha }^c} \right.} \right\rangle _{j = 1,2,...}}\\
			\omega _{ij,\alpha }^c \cong {\left\langle {{{\left[ {L_i^{\alpha \beta }} \right]}^{ - 1}}\left| {{\omega _{ij,\beta }}} \right.} \right\rangle _{\beta  = 0,x,y,z}}\\
			\left[ {L_i^{\alpha \beta }} \right] \cong {\left\langle {{x}_{ji}^\alpha \Delta {V_j}\left| {{\omega _{ij,\beta }}} \right.} \right\rangle _{j = 1,2,...}}
		\end{aligned}\right.
	\end{equation}
	\begin{equation}\label{eq_quantumFPM:3}
		\text{KGC:}\left\{\begin{aligned}
			{f_{i,\alpha }} \cong {\left\langle {\left( {{f_j} - {f_i}} \right)\Delta {V_j}\left| {\omega _{ij,\alpha }^c} \right.} \right\rangle _{j = 1,2,...}}\\
			\omega _{ij,\alpha }^c \cong {\left\langle {{{\left[ {L_i^{\alpha \beta }} \right]}^{ - 1}}\left| {{\omega _{ij,\beta }}} \right.} \right\rangle _{\beta  = x,y,z}}\\
			\left[ {L_i^{\alpha \beta }} \right] \cong {\left\langle {{x}_{ji}^\alpha \Delta {V_j}\left| {{\omega _{ij,\beta }}} \right.} \right\rangle _{j = 1,2,...}}
		\end{aligned}\right.
	\end{equation}
	\begin{equation}\label{eq_quantumFPM:4}
		\text{SPH:}\left\{\begin{aligned}
			{f_i} \cong {\left\langle {{f_j}\Delta {V_j}\left| {\omega _{ij}^{}} \right.} \right\rangle _{j = 1,2,...}}\\
			{f_{i,\alpha }} \cong {\left\langle {\left( {{f_j} - {f_i}} \right)\Delta {V_j}\left| {\omega _{ij,\alpha }^{}} \right.} \right\rangle _{j = 1,2,...}}
		\end{aligned}\right.
	\end{equation}

\subsection{Remarks on asymptotic complexity}\label{sec:complexity}
\normalsize \hspace{10pt}
\modifyFirstYellow{The swap test theory totally operates with a relatively lean quantum footprint. For an inner product between two $N$-dimensional vectors, the quantum register containing top-down amplitude encoding requires the $\mathcal{O}({\log _2}N + 1 + t)$ circuit width and $t$ ancilla qubits for phase estimation procedure observed in Fig.(\ref{fig_quantumFPM:partition}). Crucially, the circuit depth scales $\mathcal{O}(({2^{t + 1}} + 1) \cdot 2N)$ complexity with $t$ ancilla qubits, additionally with the corresponding fixed depth of other unitary transformation shown in Eq.(\ref{eq_quantumFPM:333}). This complexity profile reveals a fundamental trade-off. The logarithmic scaling of the circuit width $\widetilde{\mathcal{O}}({\log _2}N)$ constitutes an exponential compression of classical data, representing a decisive advantage in storage efficiency. Conversely, the linear growth of circuit depth $\widetilde{\mathcal{O}}({2^{t + 2}}N)$ erodes the end-to-end speedup, and the susceptibility to decoherence during repeated measurements imposes further fidelity limitations under current hardware constraints. However, these obstacles are tied to present-day coherence times and the absence of fast quantum memory, and are expected to recede as quantum hardware matures. More immediately, the multi-partitioned strategy adopted in this work decomposes arbitrarily large datasets into small blocks, capping the effective $N$ and thus the encoding depth at a hardware-tractable constant. From this perspective, the shallow-circuit design of swap test core renders the overall hybrid framework viable on near-term devices.}

\begin{equation}\label{eq_quantumFPM:333}
	\mathcal{O}\left( {\underbrace {2N}_{{\rm{state-preparing}}} + \underbrace {4N \times {2^t}}_{{\rm{phase-estimating}}}} \right)=\mathcal{O}\left( {\left( {{2^{t + 1}} + 1} \right) \cdot 2N} \right)
\end{equation}

\modifyFirstYellow{In a further step, the asymptotic complexity on quantum finite particle method with multi-partitioned zones is analyzed. Considering the extensive asymptotic process with $N$-dimensional arrays in Eq.(\ref{eq_quantumFPM:1}) and a fixed small-scale block of $k$ particles in Fig.(\ref{fig_quantumFPM:partitionedzones}), the space complexity can be derived in Eq.(\ref{eq_quantumFPM:space}), and the time complexity can be derived in Eq.(\ref{eq_quantumFPM:time}). In current quantum computing architectures, sequentially constructing and executing quantum circuits in a serial manner requires only a modest quantum circuit width, yet incurs a time complexity that scales linearly with both the arrays dimension and the number of measurement shots $\mathcal{O}(N \cdot M)$. Conversely, if multiple quantum circuits can be constructed and executed in parallel forms (analogous to multi-GPU computing in classical high-performance architectures), the total time cost can be substantially reduced to a function of only the common measurement count $M$ and a fixed small-scale block $\mathcal{O}(M \cdot k)$. With continued advances in quantum hardware technology, these parallelization strategies (multi-QPU) may gradually become realizable and attain practical significance under certain conditions.}

\begin{equation}\label{eq_quantumFPM:space}
	\mathcal{O}\underbrace {\left( {{{\log }_2}k + 1 + t} \right)}_{{\rm{sequential-execution}}} \Rightarrow \mathcal{O}\underbrace {\left( {\frac{N}{k}\left( {{{\log }_2}k + 1 + t} \right)} \right)}_{{\rm{parallel-execution}}}
\end{equation}

\begin{equation}\label{eq_quantumFPM:time}
	\mathcal{O}\underbrace {\left( {\frac{{NM}}{k}\left[ {\left( {{2^{t + 1}} + 1} \right) \cdot 2k} \right]} \right)}_{{\rm{sequential - execution}}} \Rightarrow \mathcal{O}\underbrace {\left( {M\left[ {\left( {{2^{t + 1}} + 1} \right) \cdot 2k} \right]} \right)}_{{\rm{parallel - execution}}}
\end{equation}

\modifyFirstYellow{To further comprehend the asymptotic behaviors between current work and classical bit-based computing, we briefly summarized the complexity of classical computing paradigm below. From the perspective of elementary bit operations, the dot product of two floating‑point arrays $\boldsymbol{a}, \boldsymbol{b} \in \mathbb{R}^N$,
	$S = \sum_{i=1}^{N} a_i b_i$, is evaluated by $N$ multiplications and $N-1$ additions
	using the IEEE~754 representation\supercite{WOS:000247914800005}. For a format with $b$ total bits (e.g., $b=32$ for single precision) and an effective mantissa width $m$ (e.g., $m=24$ for single precision), each floating‑point multiplication requires $m^2$ AND operations to generate the partial products plus $\mathcal{O}(m \log m)$ full‑adder steps to sum them, while each addition needs $\mathcal{O}(m)$ full‑adder operations for exponent alignment, mantissa addition and normalization. Consequently, the total number of bit operations for the dot product is}\[{T_{{\rm{bit}}}} = N \cdot \mathcal{O}\left( {{m^2}} \right) + \left( {N - 1} \right) \cdot \mathcal{O}\left( m \right) = \widetilde{\mathcal{O}}\left( {N \cdot {m^2}} \right),\] \modifyFirstYellow{which is $\mathcal{O}(N)$ for fixed $m$ but with a constant factor proportional to $m^2$.
	The space complexity in bits includes the storage of two input arrays ($2N\cdot b$ bits) and a constant number of temporary registers (partial‑product registers, shifters, accumulator) requiring $\mathcal{O}(b)$ bits, thus}\[{S_{{\rm{bit}}}} = 2N \cdot b + \mathcal{O}\left( b \right) = \widetilde{\mathcal{O}}(N \cdot b).\] \modifyFirstYellow{Hence, at the bit‑operation level, the dot product scales linearly in $N$ for both time and space complexity, but the time constant grows quadratically with the mantissa precision $m$.}

\begin{table}[H]
	\caption{\small\centering Computing complexity between current work and classical bit-based computing.}
	\label{tab_Case2:compare}
	\centering \footnotesize 
	\begin{tabular}{>{\centering\arraybackslash}p{1.7cm} >{\centering\arraybackslash}p{3.0cm} >{\centering\arraybackslash}p{4.5cm} >{\centering\arraybackslash}p{4.5cm}}
		\toprule
		\textbf{Complexity}& \textbf{\parbox{3.0cm}{Classical IEEE~754}} & \textbf{\parbox{4.5cm}{Sequential quantum circuits}} & \textbf{\parbox{4.5cm}{Parallel quantum circuits}}\\
		\midrule
		\textbf{Time}& $\widetilde{\mathcal{O}}\left( {N \cdot {m^2}} \right)$ &$\mathcal{O}\left( {\frac{{NM}}{k}\left[ {\left( {{2^{t + 1}} + 1} \right) \cdot 2k} \right]} \right)$ &$\mathcal{O}\left( {M\left[ {\left( {{2^{t + 1}} + 1} \right) \cdot 2k} \right]} \right)$  \\[1.8mm]
		\textbf{Space} & $\widetilde{\mathcal{O}}(N \cdot b)$ & $\mathcal{O}\left( {{{\log }_2}k + 1 + t} \right)$& $\mathcal{O}\left( {\frac{N}{k}\left( {{{\log }_2}k + 1 + t} \right)} \right)$ \\
		\bottomrule
	\end{tabular}
\end{table}

\begin{table}[H]
	\caption{\small\centering Asymptotic complexity with extensive $N$-dimensional arrays and specific measurements $M=100,000$ ($m=24, b=32, k=4, t=6$).}
	\label{tab_Case2:compare2}
	\centering \footnotesize 
	\begin{tabular}{>{\centering\arraybackslash}p{1.7cm} >{\centering\arraybackslash}p{3.0cm} >{\centering\arraybackslash}p{4.5cm} >{\centering\arraybackslash}p{4.5cm}}
		\toprule
		\textbf{Complexity}& \textbf{\parbox{3.0cm}{Classical IEEE~754}} & \textbf{\parbox{4.5cm}{Sequential quantum circuits}} & \textbf{\parbox{4.5cm}{Parallel quantum circuits}}\\
		\midrule
		\textbf{Time}& $\widetilde{\mathcal{O}}\left( {N} \right)$ &$\widetilde{\mathcal{O}}\left( {N \cdot M} \right)$ &$\widetilde{\mathcal{O}}\left( {M} \right)$  \\[1.8mm]
		\textbf{Space} & $\widetilde{\mathcal{O}}(N)$ & $\widetilde{\mathcal{O}}\left( {1} \right)$& $\widetilde{\mathcal{O}}(N)$ \\
		\bottomrule
	\end{tabular}
\end{table}

\modifyFirstYellow{Finally, Table \ref{tab_Case2:compare} summarizes the asymptotic complexity of entire quantum FPM pipeline, encompassing all three quantum stages: amplitude encoding, quantum inner product based on swap test, and quantum state measurement. In the QPE procedure, the $k$ particles in fixed small-scale block is proportional to the number of ancilla qubits $t$, which requires a larger $t$ to maintain sufficient estimation accuracy, as evidenced in the numerical results discussed below. Hence, considering the small constant parameters involved, Table \ref{tab_Case2:compare2} further isolates the asymptotic dependence on the data dimension $N$ alone. A comparison with classical bit-based computing reveals that, setting aside current hardware constraints such as limited coherence times and the absence of multi quantum random access memory, the proposed framework theoretically possesses a degree of quantum advantage. This theoretical advantage can be moderately enhanced when the aforementioned parallel quantum circuit strategy is adopted and when the data scale becomes sufficiently extreme. However, it should be noted that, in practice, the overall computation may be bottlenecked by the data transfer across the classical-quantum interface (the classical pre-processing required for state preparation and the I/O on retrieving measurement outcomes). More generally, for the numerical discretization of computational particle dynamics discussed subsequently, the extension of asymptotic complexity relative to the classical bit-operation complexity also follows an analogous rationale, with the essential distinction residing solely in the aforementioned quantum inner product complexity and the associated classical-quantum I/O overhead.}
	
	\section{Hybrid quantum computational particle dynamics}
	\normalsize \hspace{10pt}
	In computational particle dynamics, the numerical discretization of physical variables can be formulated within meshfree frameworks. Building upon the developed quantum inner product strategy, this study introduces a novel discretization approach tailored for hybrid quantum computational particle dynamics. Specifically, Eq.(\ref{eq_BASICdiscretization:1}) discretizes the divergence of fluid velocity, Eq.(\ref{eq_BASICdiscretization:2}) provides a gradient discretization accounting for variable density, and Eq.(\ref{eq_BASICdiscretization:3}) approximates the Laplace operator via finite differences likely for second-order derivatives\supercite{WOS:000287553400034}, which may be also represented by nested summation. Furthermore, Eq.(\ref{eq_BASICdiscretization:4}) presents a finite-difference-type formulation integrated with the proposed hybrid quantum procedure.
	\begin{equation}\label{eq_BASICdiscretization:1}
		\nabla  \cdot {{\boldsymbol{v}}_i} \cong {\left\langle {\frac{{{m_j}}}{{{\rho _j}}}\left| {{{\left\langle {v_{ji}^\beta \left| {\omega _{ij,\beta }^c} \right.} \right\rangle }_{\beta  = x,y,z}}} \right.} \right\rangle _{j = 1,2,...,N}}
	\end{equation}
	\begin{equation}\label{eq_BASICdiscretization:2}
		\frac{1}{{{\rho _i}}}\nabla {f_i} \cong {\left\langle {{m_j}\left( {\frac{{{f_i}}}{{\rho _i^2}} + \frac{{{f_j}}}{{\rho _j^2}}} \right)\left| {\nabla \omega _{ij}^c} \right.} \right\rangle _{j = 1,2,...,N}}
	\end{equation}
	\begin{equation}\label{eq_BASICdiscretization:3}
		{\nabla ^2}{f_i} \cong {\left\langle {\frac{{4{f_{ij}}{m_j}}}{{{{\left( {{\rho _i} + {\rho _j}} \right)}^{}}\left( {r_{ij}^2} \right)}}\left| {{{\left\langle {r_{ij}^\beta \left| {\omega _{ij,\beta }^c} \right.} \right\rangle }_{\beta  = x,y,z}}} \right.} \right\rangle _{j = 1,2,...,N}}
	\end{equation}
	\begin{equation}\label{eq_BASICdiscretization:4}
		\left\{ {\begin{array}{*{20}{l}}
				{{f_{i,\beta }} \cong {{\left\langle {\frac{{2{f_{ji}}{m_j}}}{{{{\left( {{\rho _i} + {\rho _j}} \right)}^{}}}}\left| {\omega _{ij,\beta }^c} \right.} \right\rangle }_{j = 1,2,...,N}}}\\
				{{{f'}_{i,\beta }} \cong \frac{{{{f'}_j} - {{f'}_i}}}{{r_{ji}^\beta }} = \frac{{{{f'}_{ji}} - {{f'}_{ij}}}}{{r_{ji}^\beta }} = \frac{{2{{f'}_{ij}}}}{{r_{ij}^\beta }}}\\
				{{\nabla ^2}{f_i} = \nabla  \cdot \nabla {f_i} \cong {{\left\langle {2{{f'}_{ij}}\left| {\frac{1}{{r_{ij}^\beta }}} \right.} \right\rangle }_{j = 1,2,...,N}} \cong {{\left\langle {\frac{{4{f_{ji}}{m_j}}}{{{{\left( {{\rho _i} + {\rho _j}} \right)}^{}}}}\left| {{{\left\langle {\frac{1}{{r_{ij}^\beta }}\left| {\omega _{ij,\beta }^c} \right.} \right\rangle }_{\beta  = x,y,z}}} \right.} \right\rangle }_{j = 1,2,...,N}}}\\
				{ \cong  - {{\left\langle {\frac{{4{f_{ji}}{m_j}}}{{{{\left( {{\rho _i} + {\rho _j}} \right)}^{}}}}\left| {{{\left\langle {\frac{{r_{ij}^\beta }}{{\left( {r_{ij}^2} \right)}}\left| {\omega _{ij,\beta }^c} \right.} \right\rangle }_{\beta  = x,y,z}}} \right.} \right\rangle }_{j = 1,2,...,N}} \cong {{\left\langle {\frac{{4{f_{ij}}{m_j}}}{{{{\left( {{\rho _i} + {\rho _j}} \right)}^{}}\left( {r_{ij}^2} \right)}}\left| {{{\left\langle {r_{ij}^\beta \left| {\omega _{ij,\beta }^c} \right.} \right\rangle }_{\beta  = x,y,z}}} \right.} \right\rangle }_{j = 1,2,...,N}}}
		\end{array}} \right.
	\end{equation}
	where $N$ denotes the number of surrounding neighboring particles $j$ at source particle $i$. ${{m}_{j}}$, ${{\rho }_{j}}$ denote the mass and density of particle $j$ respectively, ${{f}_{i}}$ denotes the approximated physical variable, ${{\boldsymbol{v}}_{i}}$ denotes the general velocity and $\omega _{ij,\beta }^{c}$ denotes the first-order derivative on approximated kernels derived by aforementioned quantum schemes. Note that ${\left \langle | \right \rangle _{j = 1,2,...}}$ denotes the aforementioned quantum-hybrid numerical implementation.
	
	\subsection{Numerical discretization on computational fluid equations}
	\normalsize \hspace{10pt}
	Based on the Navier Stokes framework, the complete governing equations for viscoelastic fluid flow are presented in Eqs.(\ref{eq_NSdiscretization:1}–\ref{eq_NSdiscretization:13}). These include the continuity Eq.(\ref{eq_NSdiscretization:1}), the momentum Eq.(\ref{eq_NSdiscretization:3}), and an independent constitutive relation for Oldroyd-B viscoelasticity in Eq.(\ref{eq_NSdiscretization:11}).
	\begin{equation}\label{eq_NSdiscretization:1}
		\frac{{d\rho }}{{dt}} =  - \rho \nabla  \cdot \boldsymbol{v}
	\end{equation}
	\begin{equation}\label{eq_NSdiscretization:3}
		\frac{{d\boldsymbol{v}}}{{dt}} = \frac{1}{\rho }\nabla  \cdot  \boldsymbol{\hat \sigma}  + \boldsymbol{g}
	\end{equation}
	\begin{equation}\label{eq_NSdiscretization:8}
		\boldsymbol{\hat \sigma}  =  - p\boldsymbol{\hat I} + \boldsymbol{\hat \tau _s} + \vartheta \boldsymbol{\hat \tau _p}
	\end{equation}
	\begin{equation}\label{eq_NSdiscretization:11}
		\boldsymbol{{\hat \tau }_p} + {\lambda _1}\mathop {\boldsymbol{{\hat \tau }_p}}\limits^\nabla   = {\mu _p}\left( {\nabla \boldsymbol{v} + {{\left( {\nabla \boldsymbol{v}} \right)}^{\rm{T}}}} \right)
	\end{equation}
	\begin{equation}\label{eq_NSdiscretization:12}
		\mathop {\boldsymbol{{\hat \tau }_p}}\limits^\nabla    = \frac{{d\boldsymbol{{\hat \tau }_p}}}{{dt}} - \boldsymbol{\hat \tau _p} \cdot \left( {\nabla \boldsymbol{v}} \right) - {\left( {\nabla \boldsymbol{v}} \right)^{\rm{T}}} \cdot \boldsymbol{\hat \tau _p}
	\end{equation}
	\begin{equation}\label{eq_NSdiscretization:13}
		\frac{{d\boldsymbol{{\hat \tau }_p}}}{{dt}} = \boldsymbol{\hat \tau _p} \cdot \left( {\nabla \boldsymbol{v}} \right) + {\left( {\nabla \boldsymbol{v}} \right)^{\rm{T}}} \cdot \boldsymbol{\hat \tau _p} - \frac{1}{{{\lambda _1}}}\boldsymbol{\hat \tau _p} + \frac{{{\mu _p}}}{{{\lambda _1}}}\left( {\nabla \boldsymbol{v} + {{\left( {\nabla \boldsymbol{v}} \right)}^{\rm{T}}}} \right)
	\end{equation}
	
	Building upon the numerical discretization established for the quantum computational particle dynamics approach and integrating the weakly compressible smoothed particle hydrodynamics (WCSPH) framework\supercite{2019Smoothed}, the corresponding particle–particle interaction formulations can be derived as follows:
	\begin{equation}\label{eq_NSdiscretization:2}
		{\left( {\frac{{d\rho }}{{dt}}} \right)_i} \cong {\left\langle {\frac{{{m_j}{\rho _i}}}{{{\rho _j}}}\left| {{{\left\langle {v_{ij}^\beta \left| {\omega _{ij,\beta }^c} \right.} \right\rangle }_{\beta  = x,y,z}}} \right.} \right\rangle _{j = 1,2,...,N}}
	\end{equation}
	\begin{equation}\label{eq_NSdiscretization:4}
		{\left( {\frac{{d{v^\alpha }}}{{dt}}} \right)_i} \cong {\left\langle {{m_j}\left| {{{\left\langle {\left( {\frac{{\sigma _i^{\alpha \beta }}}{{\rho _i^2}} + \frac{{\sigma _j^{\alpha \beta }}}{{\rho _j^2}}} \right)\left| {\omega _{ij,\beta }^c} \right.} \right\rangle }_{\beta  = x,y,z}}} \right.} \right\rangle _{j = 1,2,...,N}} - {\left\langle {{m_j}\left| {{{\left\langle {{\Pi _{ij}}{\delta ^{\alpha \beta }}\left| {\omega _{ij,\beta }^c} \right.} \right\rangle }_{\beta  = x,y,z}}} \right.} \right\rangle _{j = 1,2,...,N}} + g_i^\alpha 
	\end{equation}
	\begin{equation}\label{eq_NSdiscretization:9}
		\sigma _i^{\alpha \beta } =  - {p_i}{\delta ^{\alpha \beta }} + {\mu _s}\left( {k_i^{\alpha \beta } + k_i^{\beta \alpha }} \right) + \vartheta  \cdot \tau _{pi}^{\alpha \beta }
	\end{equation}
	\begin{equation}\label{eq_NSdiscretization:7}
		{{p}_{i}}=\frac{{{\rho }_{0}}{{c}^{2}}}{\gamma }\left( {{\left( \frac{{{\rho }_{i}}}{{{\rho }_{0}}} \right)}^{\gamma }}-1 \right)
	\end{equation}
	\begin{equation}\label{eq_NSdiscretization:14}
		{\left( {\frac{{d\tau _p^{\alpha \beta }}}{{dt}}} \right)_i} \cong k_i^{\alpha \gamma }\tau _{pi}^{\gamma \beta } + k_i^{\beta \gamma }\tau _{pi}^{\gamma \alpha } - \frac{1}{{{\lambda _1}}}\tau _{pi}^{\alpha \beta } + \frac{{{\mu _p}}}{{{\lambda _1}}}\left( {k_i^{\alpha \beta } + k_i^{\beta \alpha }} \right)
	\end{equation}
	\begin{equation}\label{eq_NSdiscretization:10}
		k_i^{\alpha \beta } = {\left( {\frac{{\partial {v^\alpha }}}{{\partial {x^\beta }}}} \right)_i} \cong {\left\langle {v_{ji}^\alpha \frac{{{m_j}}}{{{\rho _j}}}\left| {\omega _{ij,\beta }^c} \right.} \right\rangle _{j = 1,2,...,N}}
	\end{equation}
	
	The fluid stress $\sigma _i^{\alpha \beta }$ comprises from isotropic pressure $p_i^{}$, solvent viscous stress $\tau _{si}^{\alpha \beta }$, and polymeric elastic stress $\tau _{pi}^{\alpha \beta }$. The constitutive model is adjusted by the numerical factor $\vartheta$. Solvent and polymeric contributions are characterized by viscosities ${\mu _s}$ and ${\mu _p}$, respectively, while ${\lambda _1}$ represents the relaxation time of viscoelastic fluid. Gravitational acceleration is incorporated via the spatial vector $g_i^\alpha $. The equation of state in Eq.(\ref{eq_NSdiscretization:7}) models the isotropic pressure, where ${\rho _0}$ is the reference density, $\gamma = 7$ is a constant, and $c$ denotes the speed of sound, typically an order of magnitude greater than the characteristic velocity.
	
	In hydrodynamic simulations, the absence of dissipative terms in the momentum equation often leads to significant numerical oscillations. To mitigate such instabilities particularly in highly deformable flows and shock-induced dissipation, the artificial viscosity models were introduced by Monaghan et al.\supercite{monaghan1992smoothed}. In contrast to low-Reynolds-number viscous flows, the artificial viscosity formulation in Eq.(\ref{eq_NS_AVdiscretization:1}) employs parameters ${\alpha _\Pi } = 0.01$, ${\beta _\Pi } = 2.0$ for Newtonian fluids and ${\alpha _\Pi } = 1.0$, ${\beta _\Pi } = 2.0$ for viscoelastic non-Newtonian fluids, accounting for physical viscosity through nested summation. A widely adopted form integrated with hybrid quantum computing is implemented as follows:
	\begin{equation}\label{eq_NS_AVdiscretization:1}
		{\Pi _{ij}} = \left\{ {\begin{array}{*{20}{c}}
				{\frac{{ - {\alpha _\Pi }{{\bar c}_{ij}}{\phi _{ij}} + {\beta _\Pi }\phi _{ij}^2}}{{{{\bar \rho }_{ij}}}},{{\left\langle {v_{ij}^\alpha \left| {r_{ij}^\alpha } \right.} \right\rangle }_{\alpha  = x,y,z}} < 0}\\
				{0,{{\left\langle {v_{ij}^\alpha \left| {r_{ij}^\alpha } \right.} \right\rangle }_{\alpha  = x,y,z}} \ge 0}
		\end{array}} \right.
	\end{equation}
	\begin{equation}\label{eq_NS_AVdiscretization:2}
		{\phi _{ij}} = \frac{{h{{\left\langle {v_{ij}^\alpha \left| {r_{ij}^\alpha } \right.} \right\rangle }_{\alpha  = x,y,z}}}}{{r_{ij}^2 + 0.01{h^2}}}
	\end{equation}
	\begin{equation}\label{eq_NS_AVdiscretization:3}
		{\bar c_{ij}} = ({c_i} + {c_j})/2,{\bar \rho _{ij}} = ({\rho _i} + {\rho _j})/2
	\end{equation}
	
	Employing the weakly compressible SPH framework and assuming approximate incompressibility ($\nabla \cdot \boldsymbol{v} \approx 0$), the momentum equation given in Eq.(\ref{eq_NSdiscretization:3}) can be reformulated as Eq.(\ref{eq_NSdiscretization:5}). This reformulation corresponds to an alternative numerical approach in which the viscous term is discretized in a finite-difference-like form within the SPH context\supercite{WOS:000287553400034}. This technique effectively suppresses numerical oscillations while preserving the physical fidelity of solvent viscosity in both Newtonian and viscoelastic non-Newtonian fluid flows.
	\begin{equation}\label{eq_NSdiscretization:5}
		\frac{{d{\boldsymbol{v}}}}{{dt}} =  - \frac{1}{\rho }\nabla p + \frac{{\mu _s^{}}}{\rho }{\nabla ^2}{\boldsymbol{v}} + \frac{\vartheta }{\rho }\nabla  \cdot \boldsymbol{\hat \tau _p} + \boldsymbol{g}
	\end{equation}
	\begin{equation}\label{eq_NSdiscretization:6}
		\begin{aligned}
			{\left( {\frac{{d{v^\alpha }}}{{dt}}} \right)_i} \cong  &- {\left\langle {{m_j}\left( {\frac{{{p_i}}}{{\rho _i^2}} + \frac{{{p_j}}}{{\rho _j^2}}} \right)\left| {\omega _{ij,\alpha }^c} \right.} \right\rangle _{j = 1,2,...,N}} + {\left\langle {\frac{{4\left( {\mu _{si}^{} + \mu _{sj}^{}} \right){m_j}v_{ij}^\alpha }}{{{{\left( {{\rho _i} + {\rho _j}} \right)}^2}\left( {r_{ij}^2 + 0.01{h^2}} \right)}}\left| {{{\left\langle {r_{ij}^\beta \left| {\omega _{ij,\beta }^c} \right.} \right\rangle }_{\beta  = x,y,z}}} \right.} \right\rangle _{j = 1,2,...,N}}&\\
			&+ {\left\langle {\vartheta {m_j}\left| {{{\left\langle {\left( {\frac{{\tau _{pi}^{\alpha \beta }}}{{\rho _i^2}} + \frac{{\tau _{pj}^{\alpha \beta }}}{{\rho _j^2}}} \right)\left| {\omega _{ij,\beta }^c} \right.} \right\rangle }_{\beta  = x,y,z}}} \right.} \right\rangle _{j = 1,2,...,N}} + g_i^\alpha 
		\end{aligned}
	\end{equation}	
	where Eq.(\ref{eq_NSdiscretization:6}) denotes the hybrid quantum FPM form on finite-difference-like formula for viscosity diffusion term. $r_{ij}^{{}}$ denotes the spatial distance between particle $i$ and particle $j$, and $h$ denotes the smoothing kernel length. For the simulations performed in this study, the density variation can be controlled to be within 0.01 to maintain a nearly constant density.
	
	\subsection{High Weissenberg number and Log conformation}
	\normalsize \hspace{10pt}
	In the numerical implementation of time-dependent constitutive equations particularly for purely elastic upper convected Maxwell (UCM) models, explicit numerical integration in Eq.(\ref{eq_NSdiscretization:14}) often encounters the high Weissenberg number (HWN) problem under critical conditions. In reality, this instability arises from the loss of symmetric positive definite (SPD) properties in the conformation tensor during the computational procedure on high Weissenberg number. The explicit integration process cannot effectively preserve the SPD properties. This fundamental limitation explains why that traditional stabilization techniques ultimately fail beyond certain Weissenberg thresholds, necessitating alternative approaches like Log-conformation representations to maintain both stability and physical fidelity in viscoelastic flow simulations.
	
	As previously discussed, the viscoelastic stress is governed by the strain function and conformation tensor in Eq.(\ref{eq_HWNLog:1}). The conformation tensor ${\boldsymbol{\hat A}}$ provides a macroscopic representation of the average orientation of polymer chains and inherently satisfies SPD property. By substituting Eq.(\ref{eq_HWNLog:1}) into the constitutive model of Eq.(\ref{eq_NSdiscretization:11}), the time evolution of conformation tensor is expressed as Eq.(\ref{eq_HWNLog:2}), where Eq.(\ref{eq_HWNLog:3}) defines its upper-convected derivative.
	
	\begin{equation}\label{eq_HWNLog:1}
		{\boldsymbol{{\hat \tau}_{p}}} = \frac{{{\mu _p}}}{{{\lambda _1}}}({\boldsymbol{\hat A}} - {\boldsymbol{\hat I}})
	\end{equation}
	\begin{equation}\label{eq_HWNLog:2}
		\mathop {{\boldsymbol{\hat A}}}\limits^\nabla   =  - \frac{1}{{{\lambda _1}}}({\boldsymbol{\hat A}} - {\boldsymbol{\hat I}})
	\end{equation}
	\begin{equation}\label{eq_HWNLog:3}
		\mathop {{\boldsymbol{\hat A}}}\limits^\nabla   = \frac{{{\rm{d}}{\boldsymbol{\hat A}}}}{{{\rm{d}}t}} - {\boldsymbol{\hat A}} \cdot {\left( {\nabla {\boldsymbol{v}}} \right)^{\rm{T}}} - \left( {\nabla {\boldsymbol{v}}} \right) \cdot {\boldsymbol{\hat A}}
	\end{equation}
	
	Here, the upper-convected derivative ensures the frame invariance for both material deformation and rotational effects. By extracting the symmetric positive definite (SPD) conformation tensor from the constitutive formulation and coupling it with an improved particle discretization scheme, the SPD property can be rigorously maintained, thereby resolving the high Weissenberg number problem. The logarithmic representation of conformation tensor ${\boldsymbol{\hat \Psi }} = \log ({\boldsymbol{\hat A}})$ is introduced and mathematically expressed as follows:
	
	\begin{equation}\label{eq_HWNLog:4}
		{\boldsymbol{\hat A}} = {\boldsymbol{\hat {\rm E}\hat \Lambda }}{{\boldsymbol{\hat {\rm E}}}^{\mathop{\rm T}\nolimits} }
	\end{equation}
	\begin{equation}\label{eq_HWNLog:5}
		{\boldsymbol{\hat \Psi }} = {\boldsymbol{\hat {\rm E}}}{\rm{log}}({\boldsymbol{\hat \Lambda }}){{\boldsymbol{\hat {\rm E}}}^{\rm T}}
	\end{equation}
	where ${\boldsymbol{\hat {\rm E}}}$ and ${\boldsymbol{\hat \Lambda }}$ denote the eigenvector matrix and eigenvalue matrix, respectively. The SPD tensor can be completely represented by its spectral decomposition and the logarithmic transformation must be constrained to non-zero components for preventing singularities. 
	
	The velocity gradient tensor $\nabla {\boldsymbol{v}}$ can be decomposed into three fundamental components: a traceless symmetric extension component, a commutation tensor and an antisymmetric pure rotation tensor, as follows:
	\begin{equation}\label{eq_HWNLog:6}
		\nabla {\boldsymbol{v}} = {\boldsymbol{\hat \Omega }} + {\boldsymbol{\hat B}} + {\boldsymbol{\hat N}}{{\boldsymbol{\hat A}}^{ - 1}}
	\end{equation}
	\begin{equation}\label{eq_HWNLog:7}
		{{{\boldsymbol{\hat {\rm E}}}}^{\mathop{\rm T}\nolimits} }\nabla {\boldsymbol{v\hat {\rm E}}} = {{{\boldsymbol{\hat {\rm E}}}}^{\mathop{\rm T}\nolimits} }{\boldsymbol{\hat \Omega \hat {\rm E}}} + {{{\boldsymbol{\hat {\rm E}}}}^{\mathop{\rm T}\nolimits} }{\boldsymbol{\hat B\hat {\rm E}}} + {\boldsymbol{\hat N}}{{{\boldsymbol{\hat \Lambda }}}^{ - 1}}
	\end{equation}
	\begin{equation}\label{eq_HWNLog:8}
		\left[ {\begin{array}{*{20}{c}}
				{{m_{11}}}&{{m_{12}}}&{{m_{13}}}\\
				{{m_{21}}}&{{m_{22}}}&{{m_{23}}}\\
				{{m_{31}}}&{{m_{32}}}&{{m_{33}}}
		\end{array}} \right] = {{{\boldsymbol{\hat {\rm E}}}}^{\mathop{\rm T}\nolimits} }\nabla {\boldsymbol{v\hat {\rm E}}},{\rm{  }}{\boldsymbol{\hat \Omega }} = {\boldsymbol{\hat {\rm E}}}\left[ {\begin{array}{*{20}{c}}
				0&{{\omega _{12}}}&{{\omega _{13}}}\\
				{ - {\omega _{12}}}&0&{{\omega _{23}}}\\
				{ - {\omega _{13}}}&{ - {\omega _{23}}}&0
		\end{array}} \right]{{{\boldsymbol{\hat {\rm E}}}}^{\rm T}},{\rm{  }}{\boldsymbol{\hat B}} = {\boldsymbol{\hat {\rm E}}}\left[ {\begin{array}{*{20}{c}}
				{{m_{11}}}&0&0\\
				0&{{m_{22}}}&0\\
				0&0&{{m_{33}}}
		\end{array}} \right]{{{\boldsymbol{\hat {\rm E}}}}^{\mathop{\rm T}\nolimits} }
	\end{equation}
	\begin{equation}\label{eq_HWNLog:9}
		{\omega _{12}} = \frac{{{\lambda _2}{m_{12}} + {\lambda _1}{m_{21}}}}{{{\lambda _2} - {\lambda _1}}},{\rm{  }}{\omega _{13}} = \frac{{{\lambda _3}{m_{13}} + {\lambda _1}{m_{31}}}}{{{\lambda _3} - {\lambda _1}}},{\rm{  }}{\omega _{23}} = \frac{{{\lambda _3}{m_{23}} + {\lambda _2}{m_{32}}}}{{{\lambda _3} - {\lambda _2}}}
	\end{equation}
	where the tensors ${\boldsymbol{\hat B}}$, ${\boldsymbol{\hat \Omega }}$, and ${\boldsymbol{\hat N}}$ represent a symmetric matrix and two antisymmetric matrices, respectively. This numerical relationship can be characterized by the definition in Eq.(\ref{eq_HWNLog:6}). Note that the matrix ${\boldsymbol{\hat N}}$ vanishes in subsequent derivative derivations and is therefore omitted. For the computed eigenvalues of SPD tensor ${\boldsymbol{\hat A}}$, uniqueness and distinctness are guaranteed in three-dimensional space.
	
	During the subsequent time integration process, the key objective is to decompose the velocity gradient to preserve the SPD property. By substituting Eq.(\ref{eq_HWNLog:6}) into Eq.(\ref{eq_HWNLog:2}) and Eq.(\ref{eq_HWNLog:3}), we obtained the reformulated expression given in Eq.(\ref{eq_HWNLog:10}). Further introducing the logarithmic transformation of conformation tensor, the log-conformation representation can be described in Eq.(\ref{eq_HWNLog:11}).
	\begin{equation}\label{eq_HWNLog:10}
		\frac{{{\rm{d}}{\boldsymbol{\hat A}}}}{{{\rm{d}}t}} = {\boldsymbol{\hat \Omega \hat A}} - {\boldsymbol{\hat A\hat \Omega }} + {\boldsymbol{\hat B\hat A}} + {\boldsymbol{\hat A\hat B}} - \frac{1}{{{\lambda _1}}}{f_R}({\boldsymbol{\hat A}})
	\end{equation}
	\begin{equation}\label{eq_HWNLog:11}
		\frac{{{\rm{d}}{\boldsymbol{\hat \Psi }}}}{{{\rm{d}}t}} = {\boldsymbol{\hat \Omega \hat \Psi }} - {\boldsymbol{\hat \Psi \hat \Omega }} + 2{\boldsymbol{\hat B}} - \frac{1}{{{\lambda _1}}}{e^{ - {\boldsymbol{\hat \Psi }}}}{f_R}({e^{{\boldsymbol{\hat \Psi }}}})
	\end{equation}
	
	Therefore, Eq.(\ref{eq_HWNLog:11}) describes the evolution of the SPD property in the Log-conformation formulation. Compared to direct computation via Eq.(\ref{eq_NSdiscretization:13}), this transformation leverages the intrinsic properties of symmetric matrices to rigorously maintain SPD characteristics throughout iterative computations.
	
	To distinguish advection, source terms, and maintain stable SPD properties during numerical integration, a time-splitting numerical scheme\supercite{KING2021104556} is adopted for polymer stress tensor integration, as follows:
	\begin{equation}\label{eq_HWNLog:12}
		\left( {\frac{{{\rm{d}}A}}{{{\rm{d}}t}}} \right)_i^{\alpha \beta } = \Omega _i^{\alpha \gamma }A_i^{\gamma \beta } - A_i^{\alpha \gamma }\Omega _i^{\gamma \beta } + B_i^{\alpha \gamma }A_i^{\gamma \beta } + A_i^{\alpha \gamma }B_i^{\gamma \beta }
	\end{equation}
	\begin{equation}\label{eq_HWNLog:13}
		\left( {\frac{{{\rm{d}}A}}{{{\rm{d}}t}}} \right)_i^{\alpha \beta } =  - \frac{1}{{{\lambda _1}}}{f_R}(A_i^{\alpha \beta })
	\end{equation}
	\begin{equation}\label{eq_HWNLog:14}
		\left( {\frac{{{\rm{d}}{{\Psi }}}}{{{\rm{d}}t}}} \right)_i^{\alpha \beta } = {{\Omega }}_i^{\alpha \gamma }{{\Psi }}_i^{\gamma \beta } - {{\Psi }}_i^{\alpha \gamma }{{\Omega }}_i^{\gamma \beta } + {\rm{2}}{{B}}_i^{\alpha \beta }
	\end{equation}
	\begin{equation}\label{eq_HWNLog:15}
		\left( {\frac{{{\rm{d}}{{\Psi }}}}{{{\rm{d}}t}}} \right)_i^{\alpha \beta } =  - \frac{1}{{{\lambda _1}}}{e^{ - {{\Psi }}_i^{\alpha \beta }}}{f_R}({e^{{{\Psi }}_i^{\alpha \beta }}})
	\end{equation}
	
	\subsection{Time integration and other technologies}
\begin{figure}[H]
		\centering
		\includegraphics[scale=0.31]{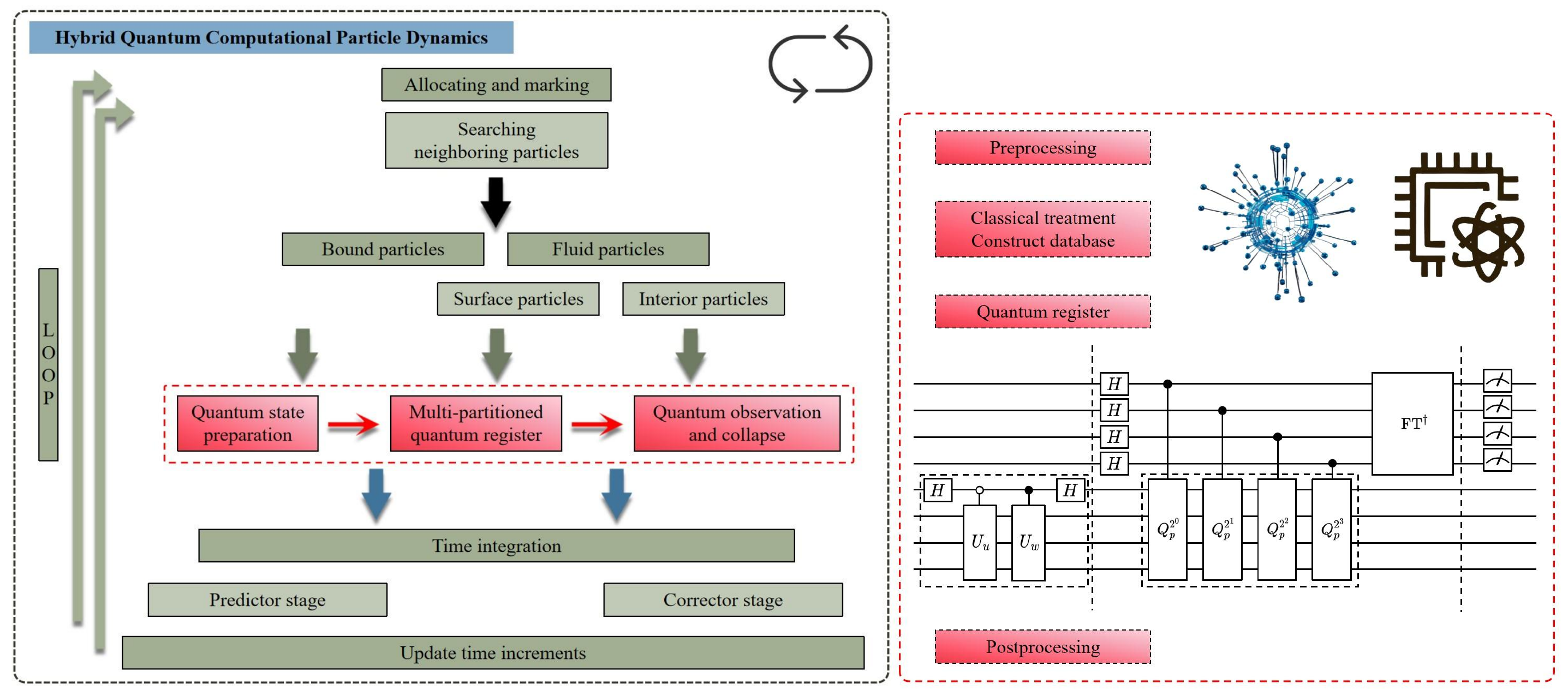}
		\caption{\small Thematic concept on hybrid quantum computational particle dynamics.}
		\label{fig:quantum_frame}
	\end{figure}

	\normalsize \hspace{10pt}
	During this hybrid procedure on meshfree computational particle dynamics and quantum computing, the classical time integration on explicit propulsion and other technologies of boundary treatment, etc. are adopted in the present study to ensure the effectiveness and continuity of dynamic computing process shown in Fig.(\ref{fig:quantum_frame}). Therein, the time evolutional treatment, predictor-corrector Symplectic scheme is introduced, whose time step has the second-order accuracy to calculate independent physical variables. Specific details on some necessary numerical technologies have been introduced in literatures\supercite{2010Smoothed,2019Smoothed,LI2023112213,10.1063/5.0197088,LI2025119962}. As this aspect falls outside the main scope of present work, it will not be elaborated further apologetically. \modifyFirstYellow{Furthermore, a detailed discussion of error propagation associated with the quantum inner product nested within the time-stepping predictor-corrector scheme, as well as its implications for the conservation of invariants, is deferred to Section \ref{Case2} Case 2.}
	
	\section{Numerical simulation}
	\normalsize \hspace{10pt}
	Considering the current limitation on noisy intermediate-scale quantum (NISQ) device, we mainly adopted the quantum algorithm based on classical computer \modifyFirstYellow{such as QPanda from Origin Quantum,}whose quantum assembly instruction can be same and suitable with the quantum hardware. \modifyFirstYellow{A limited set of computations was executed on the physical quantum processor from the “OriginQ Wukong”, chosen according to the specific objective of each test and practical considerations of economic cost.}It is entirely demonstrated that the distinction of final results merely stems from hardware devices between von Neumann computer and universal quantum computer with noise\supercite{CHEN2024117428}. \modifyFirstYellow{Case 1 was conducted entirely on a quantum simulator to validate the effectiveness of algorithm prototype. Case 2 was partially executed on the quantum registers via the Origin Quantum Cloud, establishing a quantum-classical hybrid computing platform. Therein, two synthetic noise profiles (Noise I and II) were additionally analyzed using the Origin Quantum simulator with corresponding noise models. Involving the high Weissenberg number analysis, Case 3 was performed on the quantum simulator to accommodate the economic cost and large number of circuit executions required by time-stepping computations.}The relative norm error evaluations are adopted in these numerical investigations shown in Eq.(\ref{eq:L2norm}).
	\begin{equation}\label{eq:L2norm}
		{{E}_{2}}={\sqrt{\sum\limits_{i}^{N}{{{(f_{i}^{numerical}-f_{i}^{analytical})}^{2}}}}}/{\sqrt{\sum\limits_{i}^{N}{{{(f_{i}^{analytical})}^{2}}}}}\;
	\end{equation}
	
	\subsection*{Case 1: Hybrid quantum verification on statical exact solution}
	\normalsize \hspace{10pt}
	In this subsection, the valid verification on present hybrid quantum FPM with statical exact solution is proceeded purposefully. We introduced a simple function of Eq.(\ref{eq_Case1:1}) and gradually a complicated function of Eq.(\ref{eq_Case1:2}) with multi-curvature changing. It adopts orthogonal particle distributions and the first-order partial derivative of gradient $\nabla F$ and second-order partial derivative of Laplace ${{\nabla }^{2}}F$ are numerically investigated.
	
	\begin{equation}\label{eq_Case1:1}
		F(x,y)={{x}^{2}}+{{y}^{2}} 
	\end{equation}
	\begin{equation}\label{eq_Case1:2}
		F(x,y)=\sin (10x)\cos (10y)
	\end{equation}
	
	\begin{figure}[H]
		\centering
		\begin{subfigure}[t]{0.32\textwidth}
			\centering
			\includegraphics[scale=0.3]{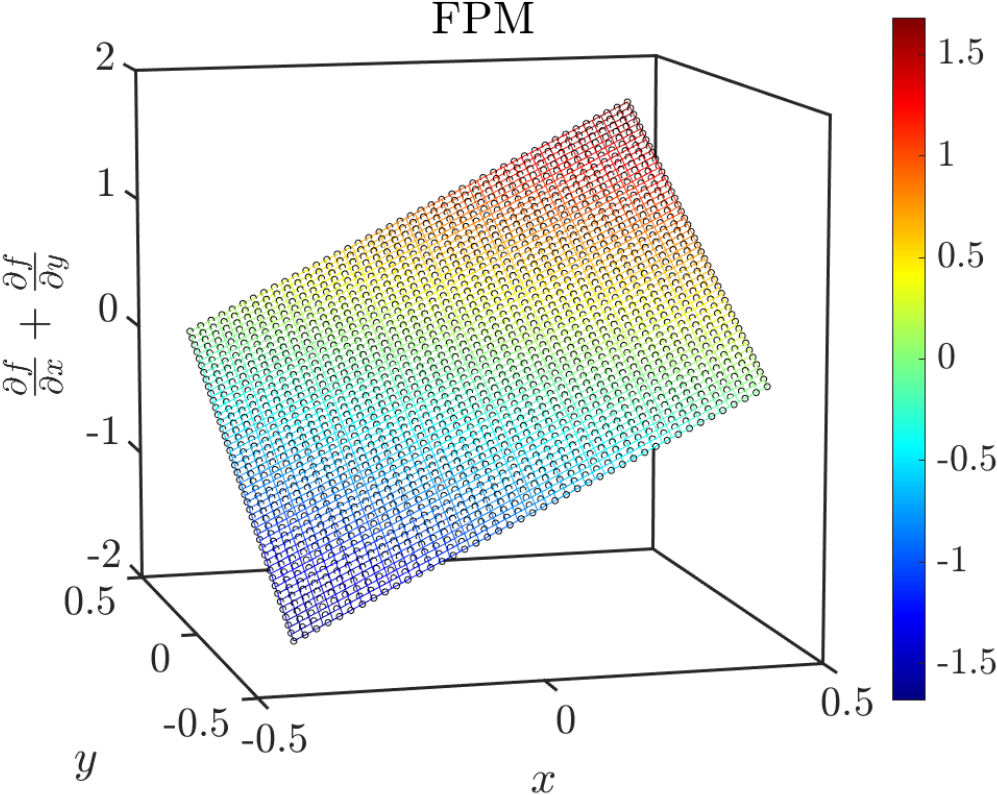}
			\caption{\centering\footnotesize Entire FPM computing.}
		\end{subfigure}
		\begin{subfigure}[t]{0.32\textwidth}
			\centering
			\includegraphics[scale=0.3]{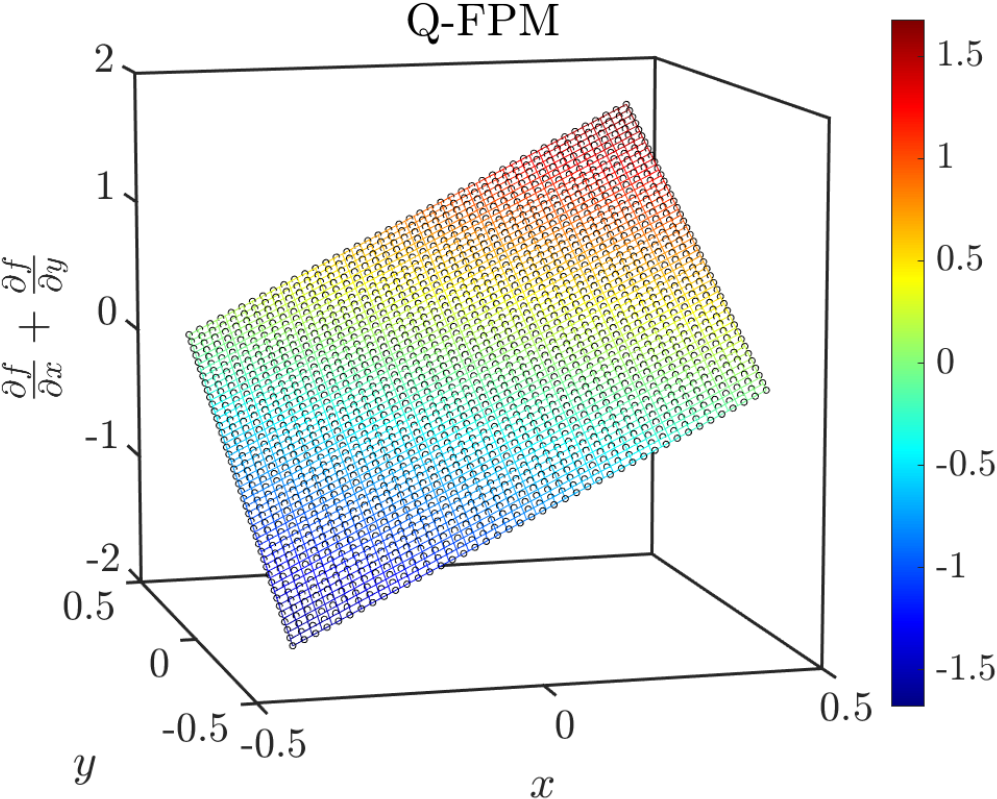}
			\caption{\centering\footnotesize Hybrid quantum FPM (Q-FPM).}  
		\end{subfigure}
		\begin{subfigure}[t]{0.32\textwidth}
			\centering
			\includegraphics[scale=0.3]{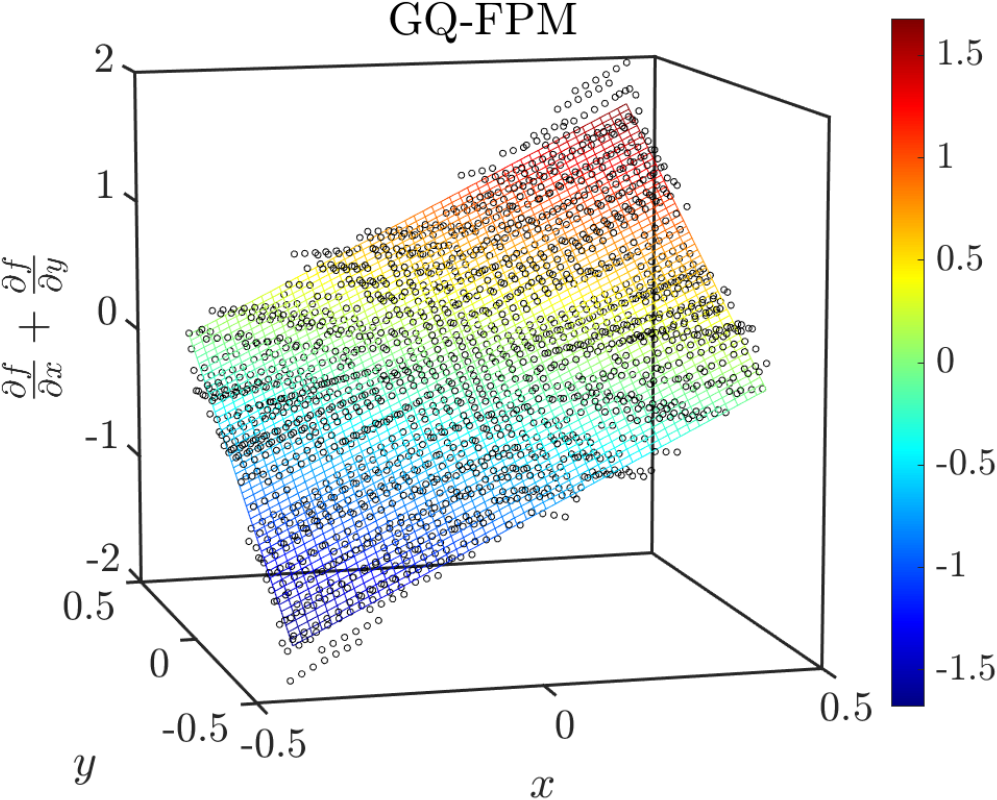}
			\caption{\centering\footnotesize Global quantum FPM (GQ-FPM) with multi-partitioned zones.}  
		\end{subfigure}
		\caption{\small Numerical approximation on first-order partial derivative.}
		\label{fig_Case1:1}
	\end{figure}
	
	Based on the numerical approximation on Eq.(\ref{eq_Case1:1}), Fig.(\ref{fig_Case1:1}) described the approximation on first-order partial derivative by the entire meshfree particle computing, hybrid quantum computing and global quantum computing with multi-partitioned zones. The corresponding relative errors of quantified nodes are calculated shown in Table \ref{tab_Case1:1}. The integrated FPM approximation is superior to traditional SPH on computational accuracy, whose conclusion had been demonstrated comprehensively\supercite{2010Smoothed,2019Smoothed,WOS:001501337900001}. However, quantum-hybrid meshfree algorithms offer theoretical advantages and exploratory prospects. Their current implementation fidelity imposes unavoidable accuracy reductions of approximately $10^{-4}\sim10^{-3}$ compared to classical meshfree solutions (error $10^{-7}\sim10^{-6}$). The fidelity estimation is also subject to errors induced by matrix inversion. The Q-FPM represents the hybrid quantum kernel procedure shown in Eq.(\ref{eq_quantumFPM:2}), which usually involves small size computation of inner product at each particle. Further the quantum procedure on particle dynamics is constructed based on the linear combination of surrounding neighboring particles with arbitrarily sized arrays and GQ-FPM represents the global hybrid quantum procedure with automatic multi-partitioned zones shown in Eqs.(\ref{eq_BASICdiscretization:1}-\ref{eq_BASICdiscretization:3}) and Figs.(\ref{fig_quantumFPM:partition}-\ref{fig_quantumFPM:partitionedzones}). Inevitably the accuracy of fidelity computation is reduced approximately $10^{-1}$ from quantum kernel to quantum particle approximation within more inner product. On the other hand, the discrete forms on second-order partial derivative exist different numerical performances in Table \ref{tab_Case1:1}, which possess the similar conclusion between FPM and SPH\supercite{WOS:001501337900001}. The capacity of current hybrid particle approximation mainly depends on the computational fidelity of quantum procedure.
	
\begin{figure}[H]
		\centering
		\begin{subfigure}[t]{0.32\textwidth}
			\centering
			\includegraphics[scale=0.3]{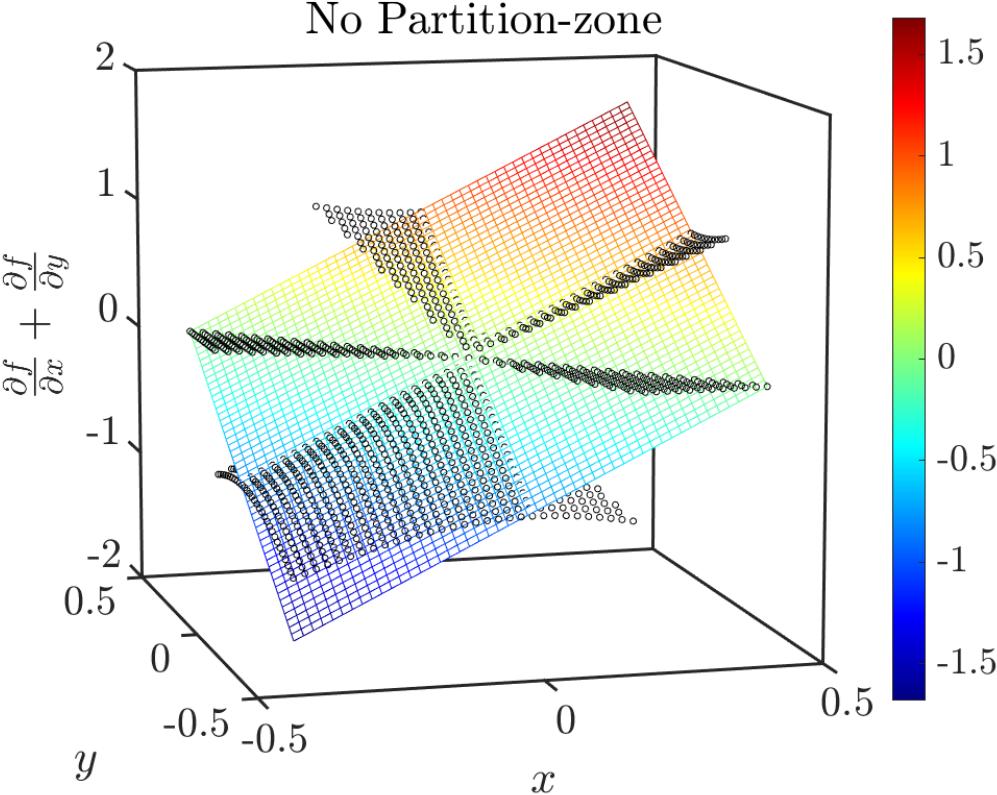}
			\caption{\centering\footnotesize Ancilla qubits ($t=4$) without partitioned zones.}
			\label{fig_Case1:2a}
		\end{subfigure}
		\begin{subfigure}[t]{0.32\textwidth}
			\centering
			\includegraphics[scale=0.3]{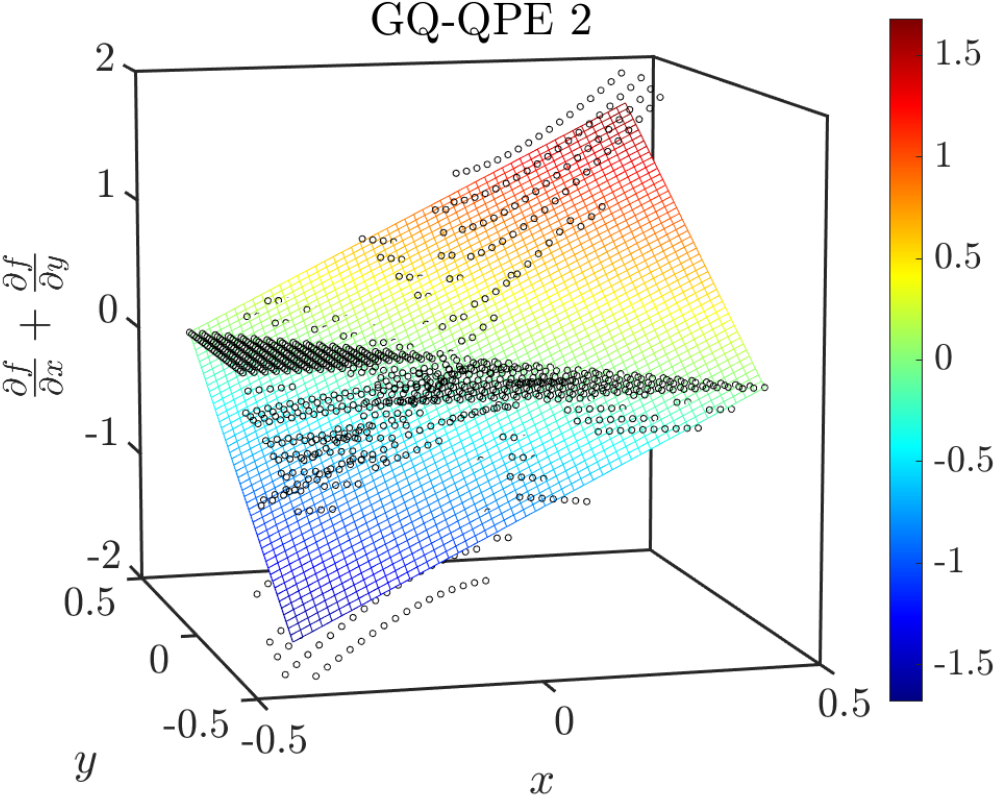}
			\caption{\centering\footnotesize Ancilla qubits ($t=2$).}  
			\label{fig:6b}
		\end{subfigure}
		\begin{subfigure}[t]{0.32\textwidth}
			\centering
			\includegraphics[scale=0.3]{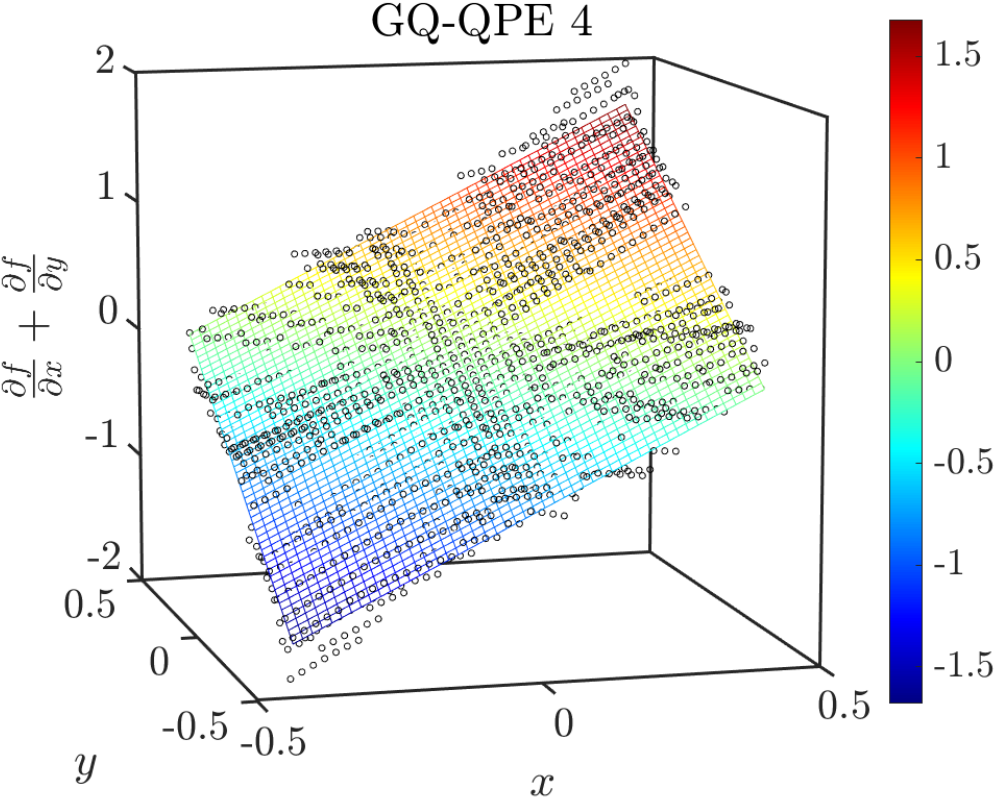}
			\caption{\centering\footnotesize Ancilla qubits ($t=4$).}  
			\label{fig_Case1:2c}
		\end{subfigure}
		\begin{subfigure}[t]{0.32\textwidth}
			\centering
			\includegraphics[scale=0.3]{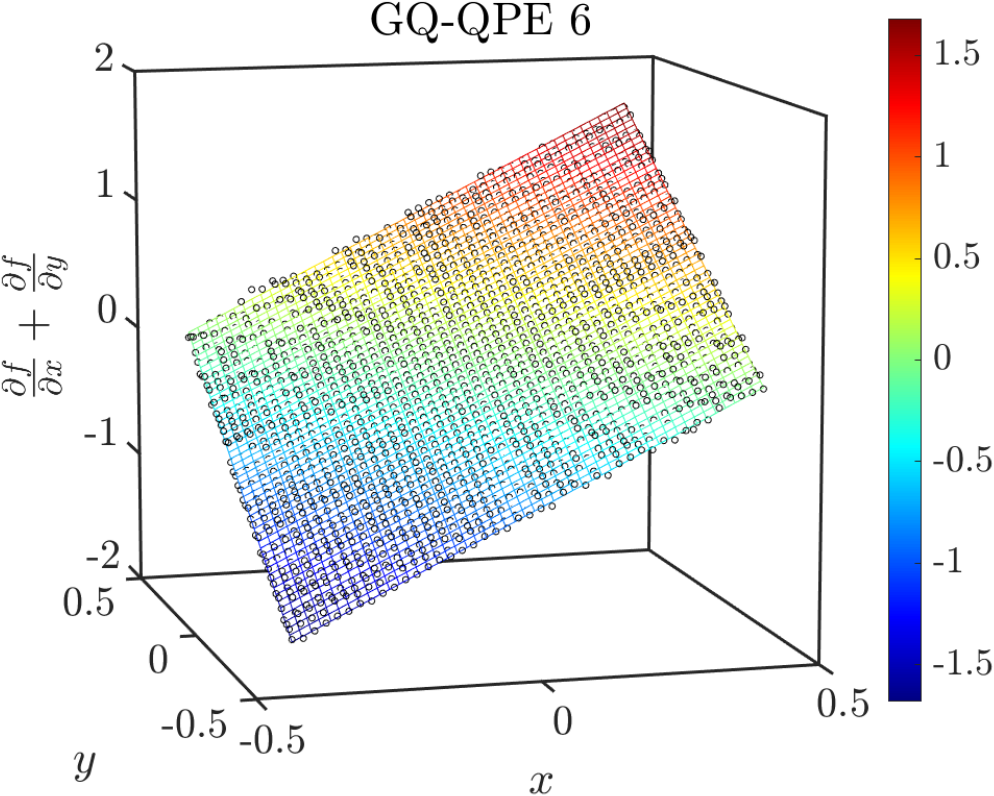}
			\caption{\centering\footnotesize Ancilla qubits ($t=6$).}
			\label{fig:6d}
		\end{subfigure}
		\begin{subfigure}[t]{0.32\textwidth}
			\centering
			\includegraphics[scale=0.3]{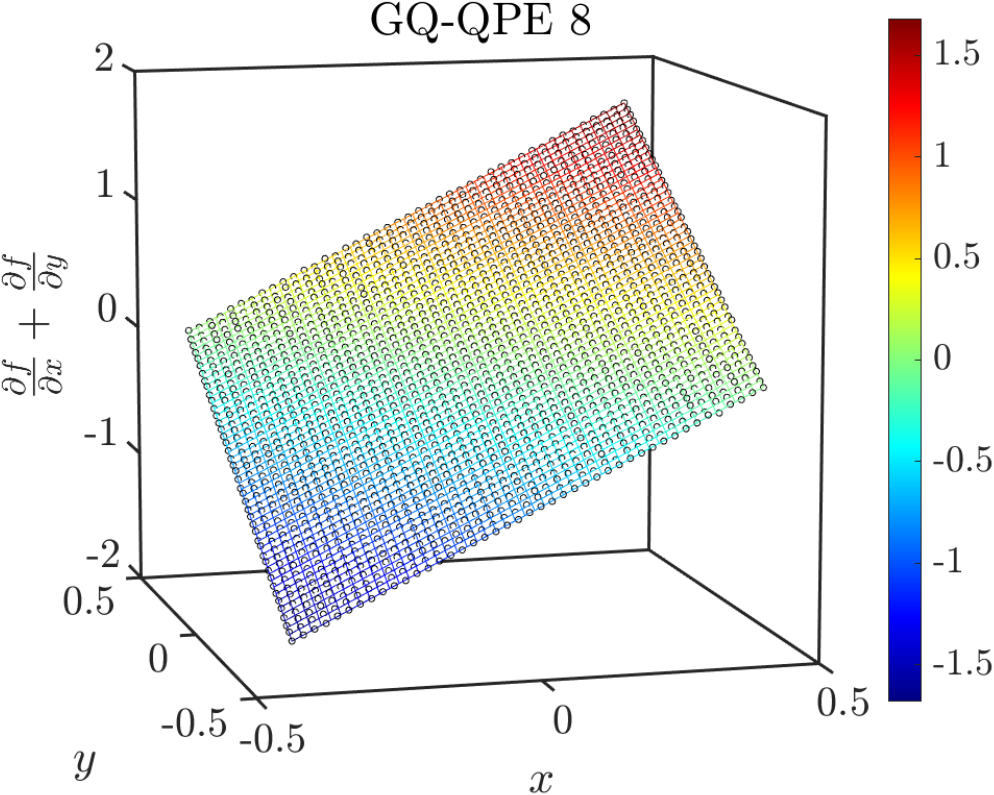}
			\caption{\centering\footnotesize Ancilla qubits ($t=8$).}  
			\label{fig:6e}
		\end{subfigure}
		\caption{\small Numerical approximation (meshfree FPM) on different ancilla qubits with multi-partitioned zones.}
		\label{fig_Case1:2}
	\end{figure}

	These numerical performances on multi-partitioned zones and ancilla qubits for QPE estimation are recorded in Table \ref{tab_Case1:2} and Fig.(\ref{fig_Case1:2}). It is obvious that the accuracy on QPE estimation depends on ancilla qubits. The small size computation of inner product by multi-partitioned zones is advantaged than that without partitioned procedure shown in Figs.(\ref{fig_Case1:2a},\ref{fig_Case1:2c}). This multi-partitioned strategy can also improve the utilization efficiency on quantum register and satisfy the current stored limitation on NISQ device especially in 3-D particle dynamics.
	
	\begin{table}[H]
		\caption{\small\centering Relative ${{\rm{L}}_{\rm{2}}}$ norm error evaluations on different computing modules.}
		\label{tab_Case1:1}
		\centering \footnotesize 
		\begin{tabular}{
				>{\centering\arraybackslash}p{3.6cm} >{\centering\arraybackslash}p{1.8cm} >{\centering\arraybackslash}p{1.8cm}
				>{\centering\arraybackslash}p{1.8cm}}
			\toprule
			\textbf{Derivative} & \textbf{FPM} & \textbf{Q-FPM} & \textbf{GQ-FPM}\\
			\midrule
			\multirow{1}{*}
			{First-order} & $1.0917\times10^{-7}$ & $2.7992\times10^{-4}$ & $1.3934\times10^{-1}$ \\
			Second-order (Like-diff) & $9.7665\times10^{-6}$ & $2.7984\times10^{-4}$ & $1.6638\times10^{-0}$ \\
			Second-order (Nested sum) & $1.4013\times10^{-6}$ & $5.6018\times10^{-4}$ & $3.2616\times10^{-1}$ \\
			\bottomrule
		\end{tabular}
	\end{table}
	
	\begin{table}[H]
		\caption{\small\centering Relative ${{\rm{L}}_{\rm{2}}}$ norm error evaluations on different ancilla qubits with multi-partitioned zones.}
		\label{tab_Case1:2}
		\centering \footnotesize 
		\begin{tabular}{
				>{\centering\arraybackslash}p{3.6cm} >{\centering\arraybackslash}p{1.9cm} >{\centering\arraybackslash}p{1.9cm} >{\centering\arraybackslash}p{1.9cm} >{\centering\arraybackslash}p{1.9cm} >{\centering\arraybackslash}p{1.9cm}}
			\toprule
			\textbf{Derivative} & \textbf{No Partition} & \textbf{QPE 2} & \textbf{QPE 4} & \textbf{QPE 6} & \textbf{QPE 8} \\
			\midrule
			\multirow{1}{*}
			{First-order} & $2.9506\times10^{-1}$ & $7.0863\times10^{-1}$ & $1.3934\times10^{-1}$ & $2.9278\times10^{-2}$ & $7.1423\times10^{-3}$ \\
			Second-order (Like-diff) & $9.9891\times10^{-1}$ & $3.5114\times10^{-0}$ & $1.6638\times10^{-0}$ & $4.4617\times10^{-1}$ & $1.0890\times10^{-1}$ \\
			Second-order (Nested sum) & $7.3994\times10^{-1}$ & $2.1146\times10^{-0}$ & $3.2616\times10^{-1}$ & $8.6567\times10^{-2}$ & $1.9793\times10^{-2}$ \\
			\bottomrule
		\end{tabular}
	\end{table}
	
\begin{figure}[H]
		\centering
		\begin{subfigure}[t]{0.35\textwidth}
			\centering
			\includegraphics[scale=0.3]{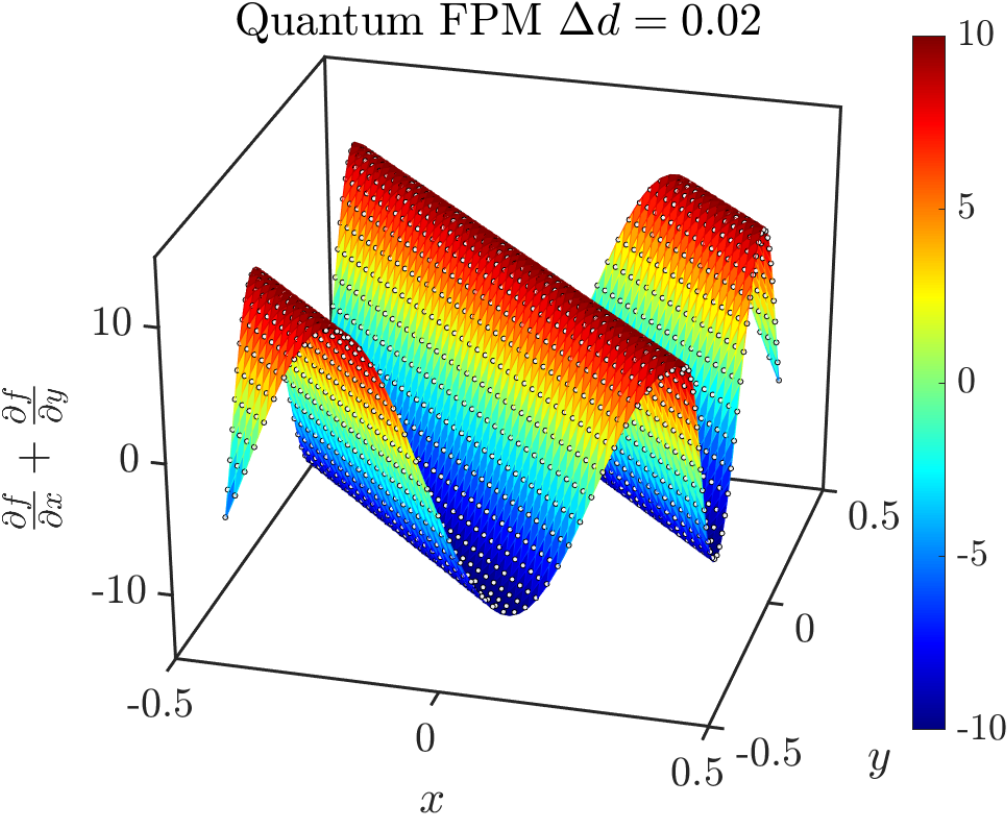}
			\caption{\centering\footnotesize First-order partial derivative (2601 nodes).}
			\label{fig:6a}
		\end{subfigure}
		\begin{subfigure}[t]{0.35\textwidth}
			\centering
			\includegraphics[scale=0.3]{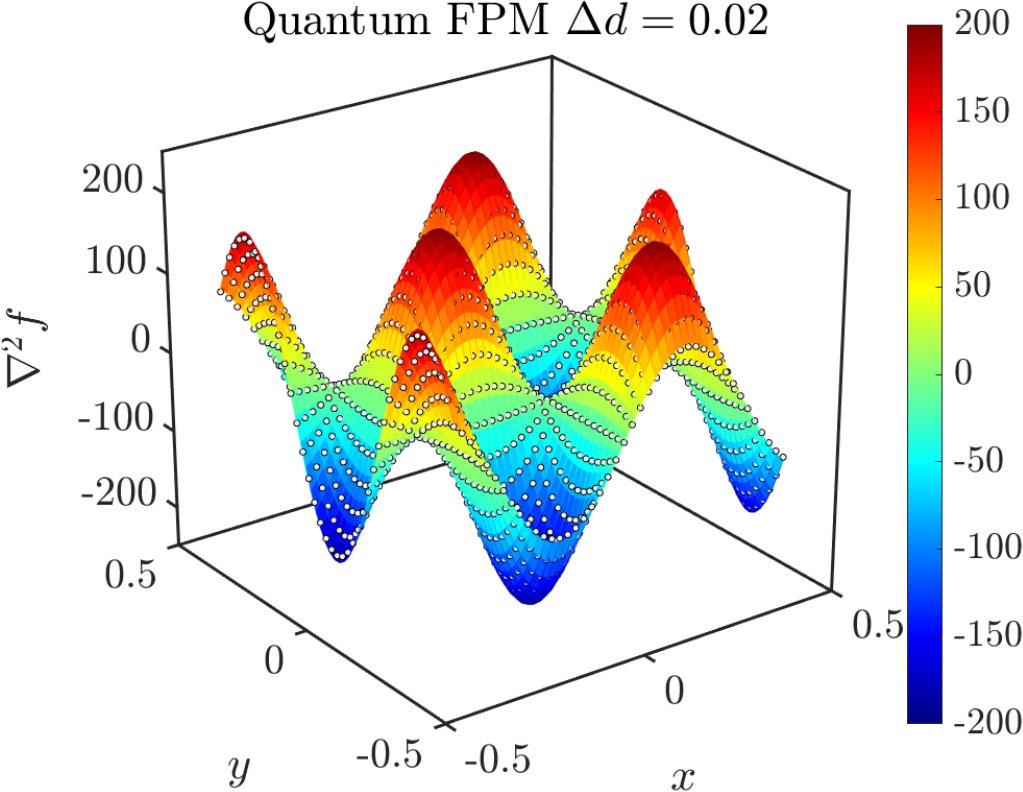}
			\caption{\centering\footnotesize Second-order partial derivative by nested sum (2601 nodes).}  
			\label{fig:6b}
		\end{subfigure}
		\begin{subfigure}[t]{0.35\textwidth}
			\centering
			\includegraphics[scale=0.3]{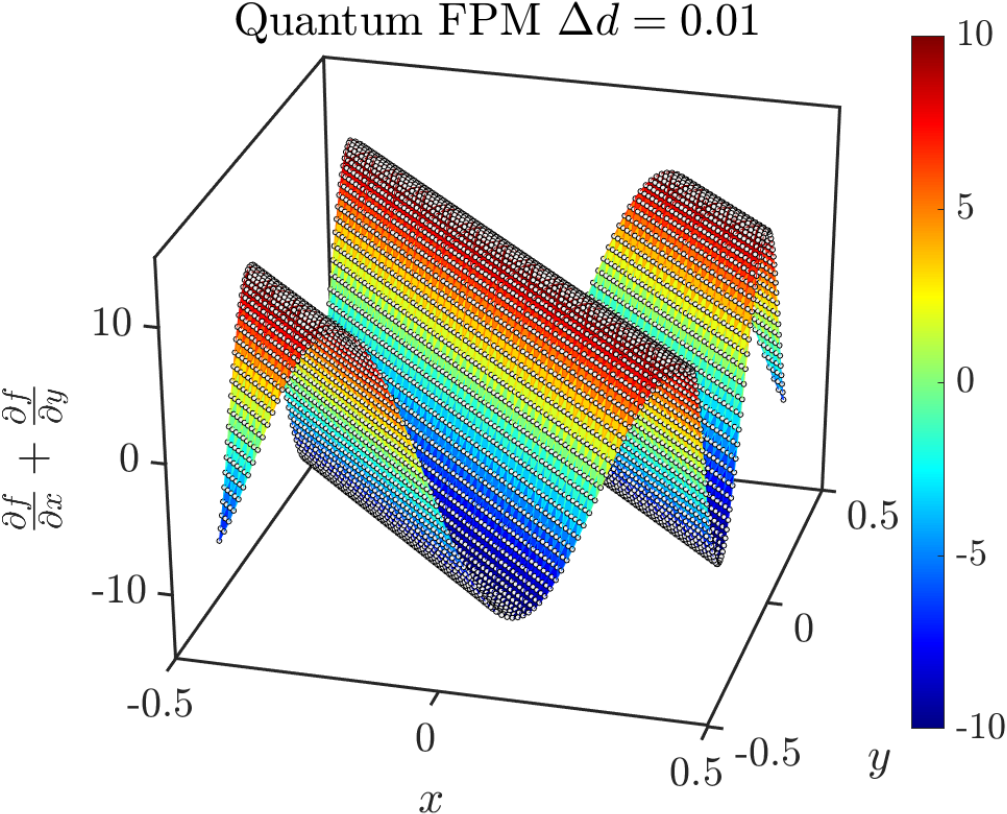}
			\caption{\centering\footnotesize First-order partial derivative (10201 nodes).}  
			\label{fig:6c}
		\end{subfigure}
		\begin{subfigure}[t]{0.35\textwidth}
			\centering
			\includegraphics[scale=0.3]{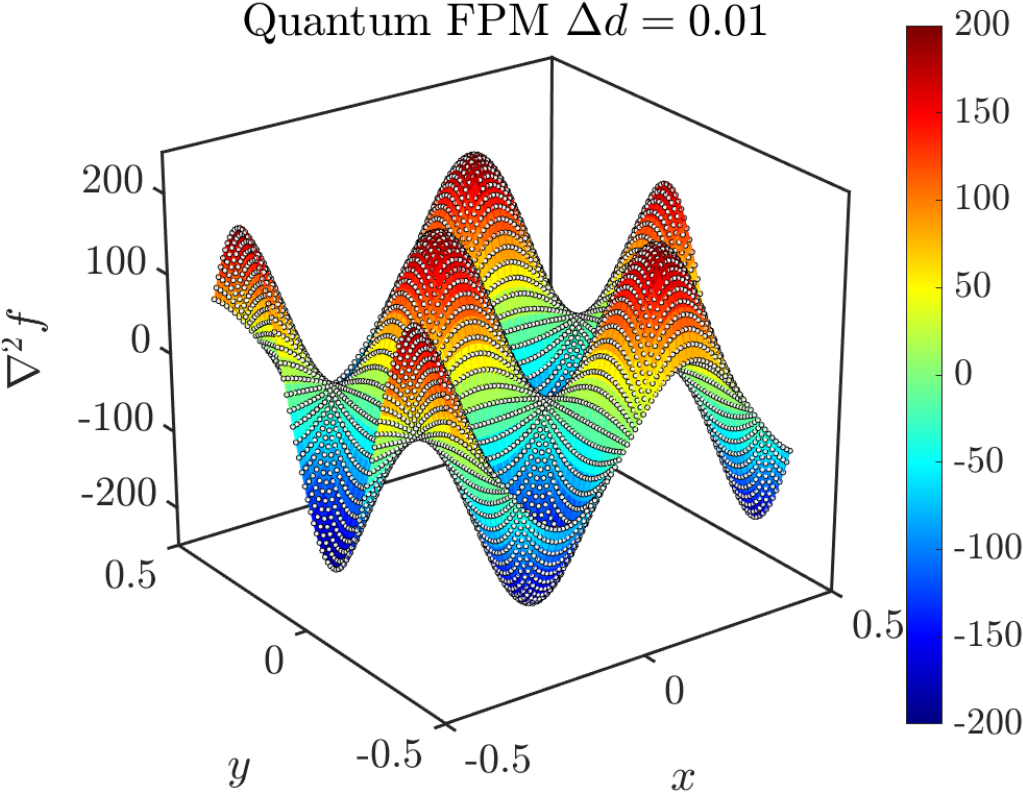}
			\caption{\centering\footnotesize Second-order partial derivative by nested sum (10201 nodes).}
			\label{fig:6d}
		\end{subfigure}
		\caption{\small Global quantum FPM approximation using different particle interval.}
		\label{fig_Case1:3}
	\end{figure}

\begin{figure}[H]
		\centering
		\begin{subfigure}[t]{0.32\textwidth}
			\centering
			\includegraphics[scale=0.3]{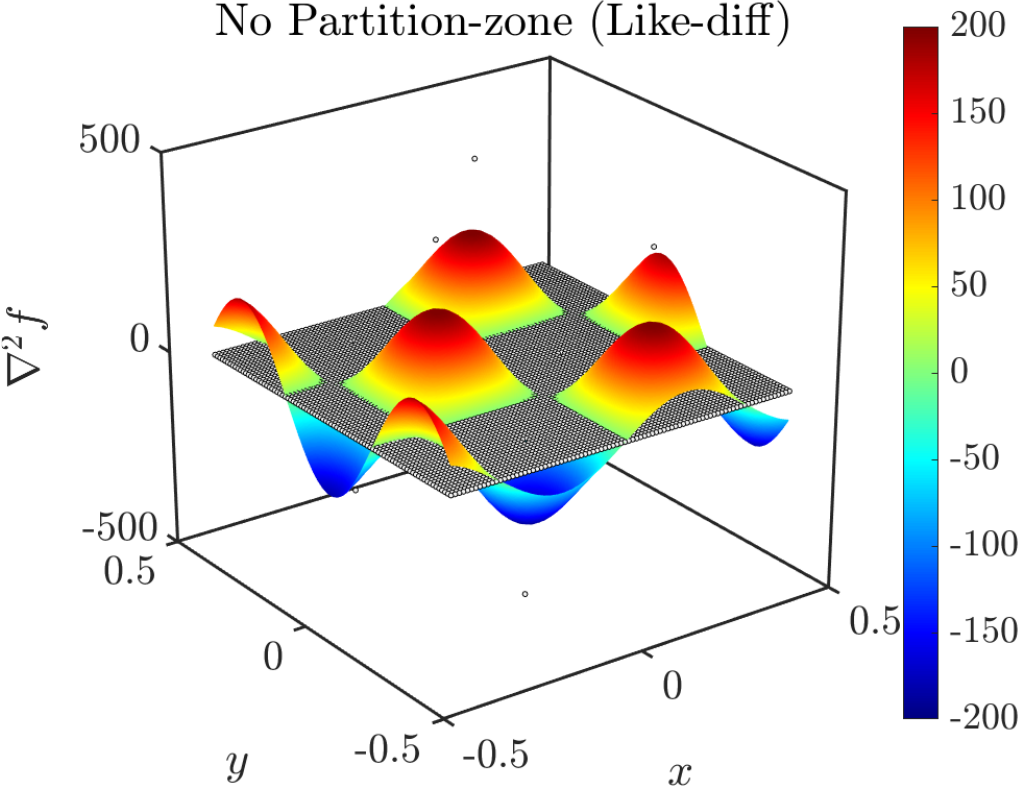}
			\caption{\centering\footnotesize Ancilla qubits ($t=4$) without partitioned zones.}
			\label{fig:6a}
		\end{subfigure}
		\begin{subfigure}[t]{0.32\textwidth}
			\centering
			\includegraphics[scale=0.3]{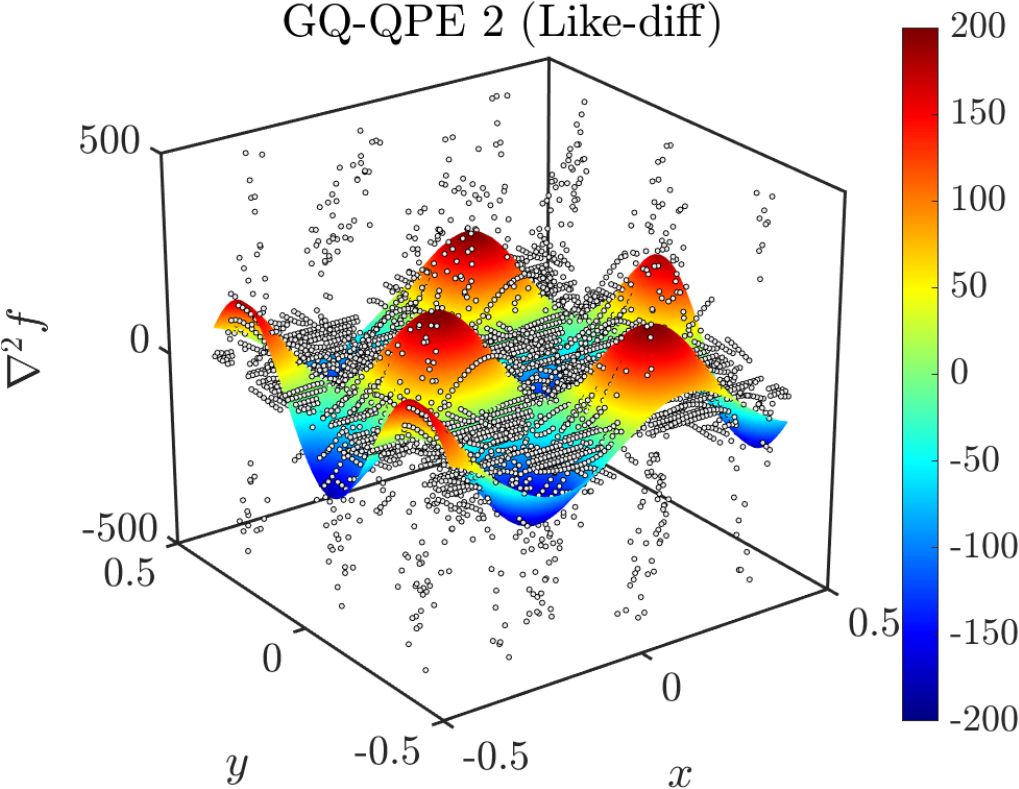}
			\caption{\centering\footnotesize Ancilla qubits ($t=2$).}  
			\label{fig:6b}
		\end{subfigure}
		\begin{subfigure}[t]{0.32\textwidth}
			\centering
			\includegraphics[scale=0.3]{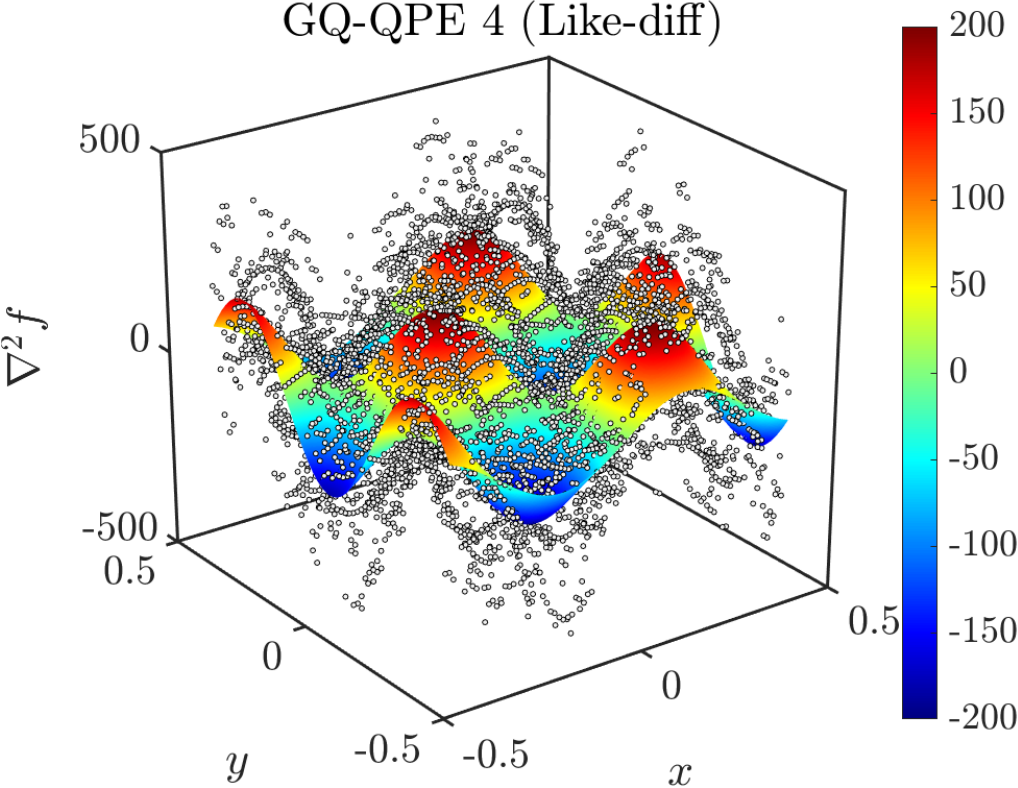}
			\caption{\centering\footnotesize Ancilla qubits ($t=4$).}  
			\label{fig:6c}
		\end{subfigure}
		\begin{subfigure}[t]{0.32\textwidth}
			\centering
			\includegraphics[scale=0.3]{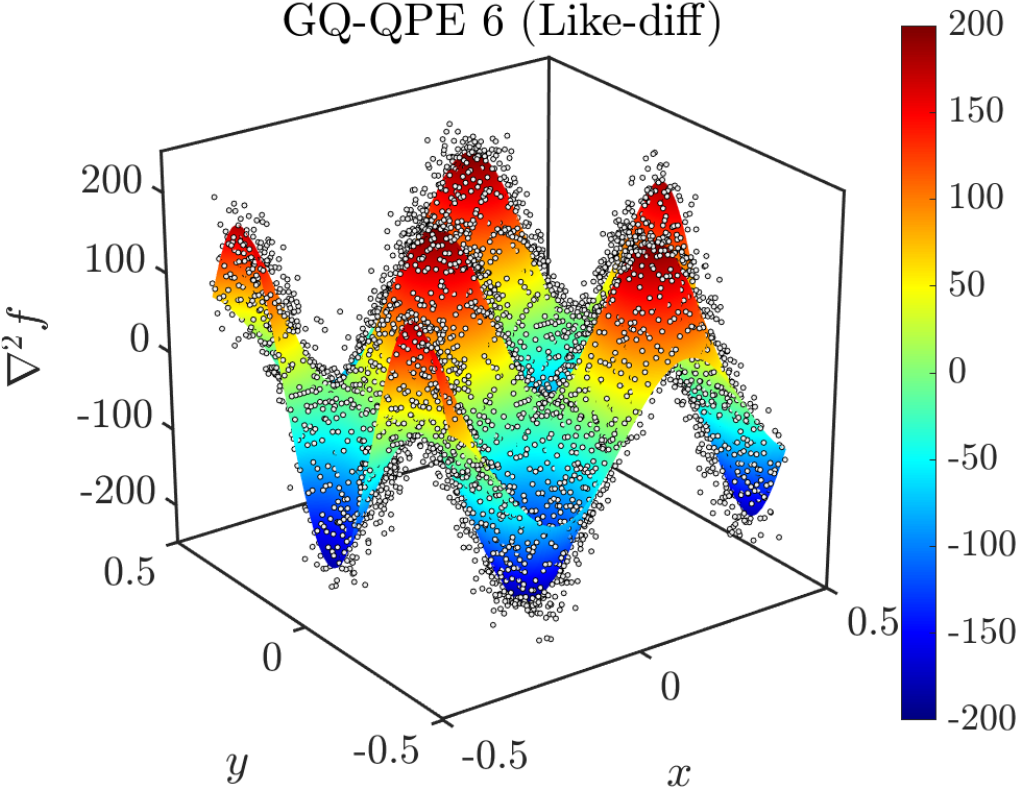}
			\caption{\centering\footnotesize Ancilla qubits ($t=6$).}
			\label{fig:6d}
		\end{subfigure}
		\begin{subfigure}[t]{0.32\textwidth}
			\centering
			\includegraphics[scale=0.3]{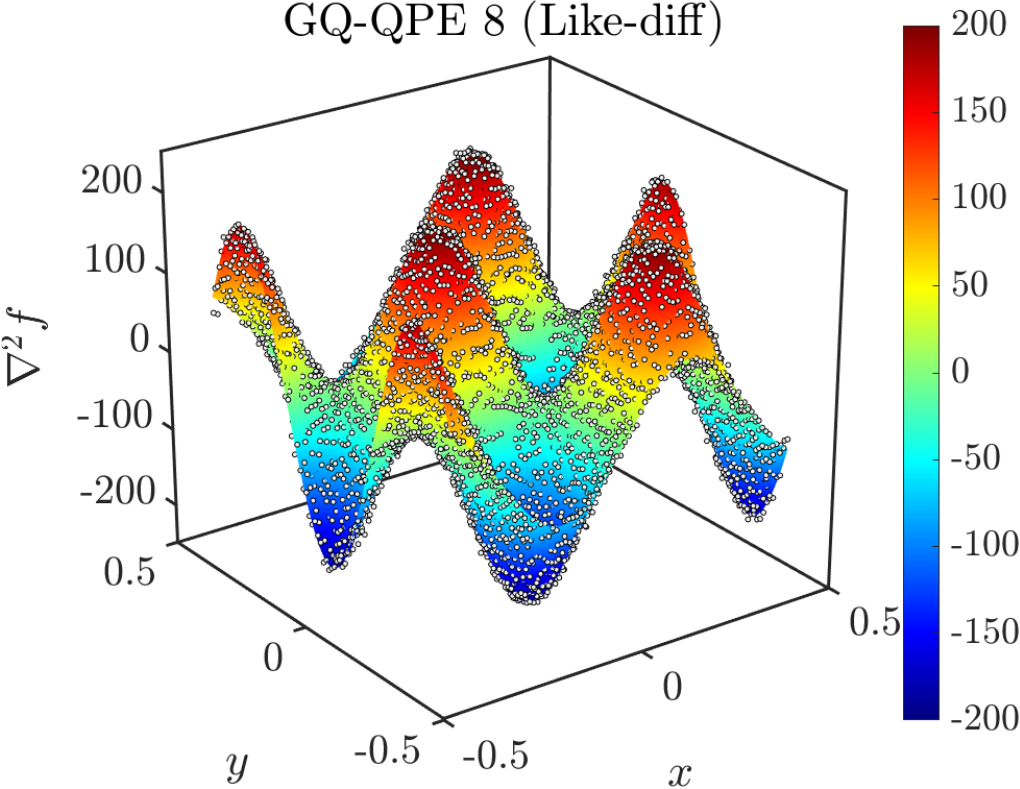}
			\caption{\centering\footnotesize Ancilla qubits ($t=8$).}  
			\label{fig:6e}
		\end{subfigure}
		\caption{\small Second-order partial derivative on approximated like-difference form.}
		\label{fig_Case1:4}
	\end{figure}

\begin{figure}[H]
		\centering
		\begin{subfigure}[t]{0.32\textwidth}
			\centering
			\includegraphics[scale=0.3]{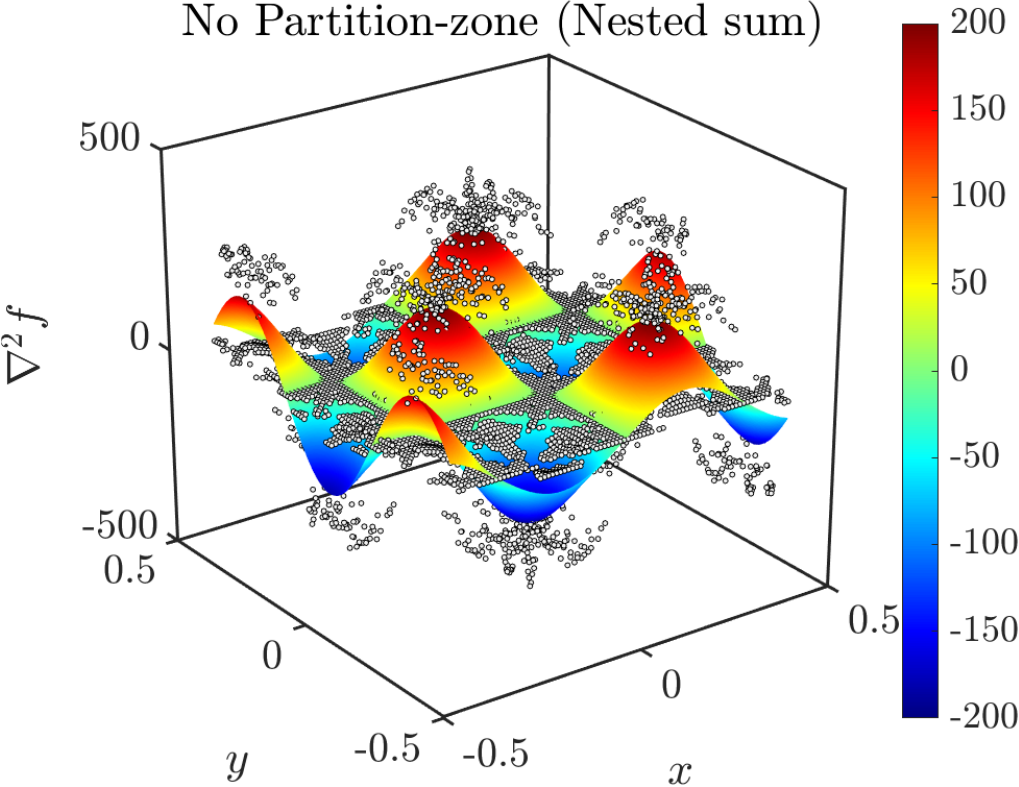}
			\caption{\centering\footnotesize Ancilla qubits ($t=4$) without partitioned zones.}
			\label{fig:6a}
		\end{subfigure}
		\begin{subfigure}[t]{0.32\textwidth}
			\centering
			\includegraphics[scale=0.3]{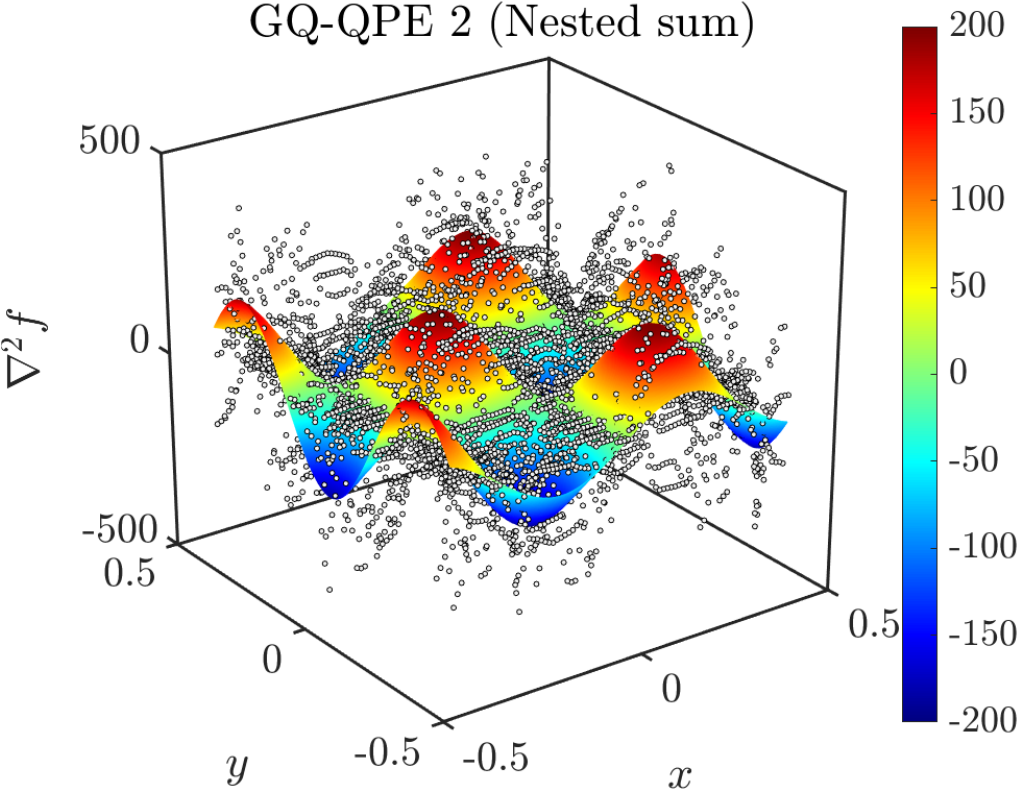}
			\caption{\centering\footnotesize Ancilla qubits ($t=2$).}  
			\label{fig:6b}
		\end{subfigure}
		\begin{subfigure}[t]{0.32\textwidth}
			\centering
			\includegraphics[scale=0.3]{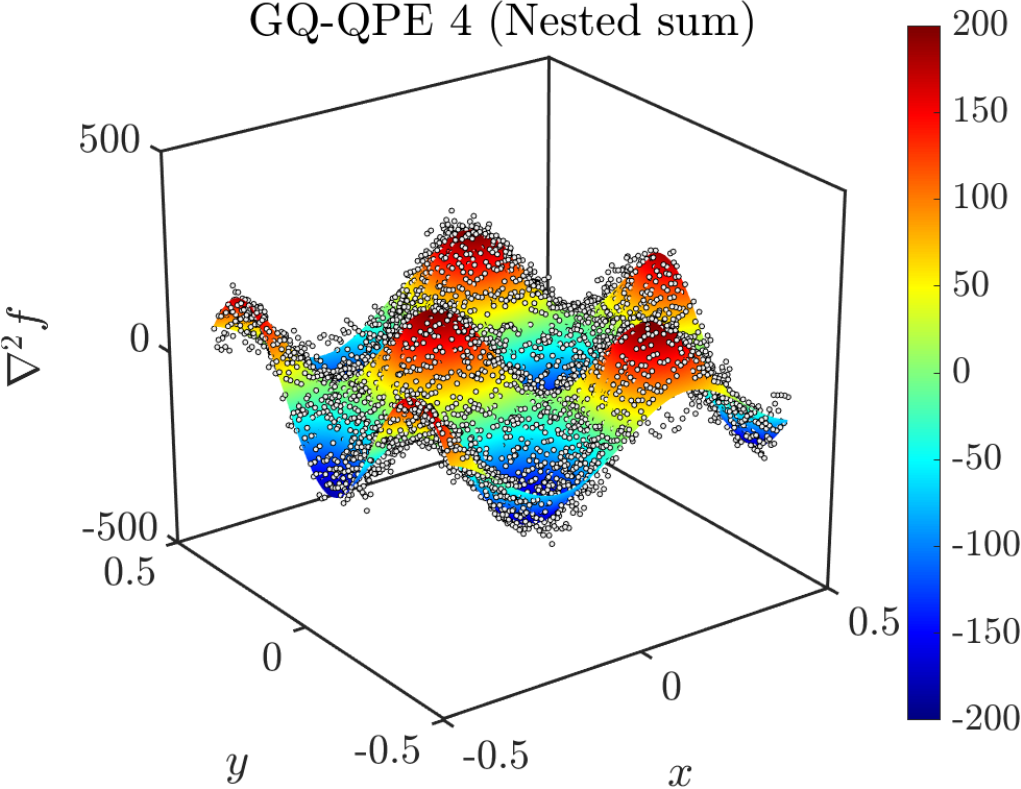}
			\caption{\centering\footnotesize Ancilla qubits ($t=4$).}  
			\label{fig:6c}
		\end{subfigure}
		\begin{subfigure}[t]{0.32\textwidth}
			\centering
			\includegraphics[scale=0.3]{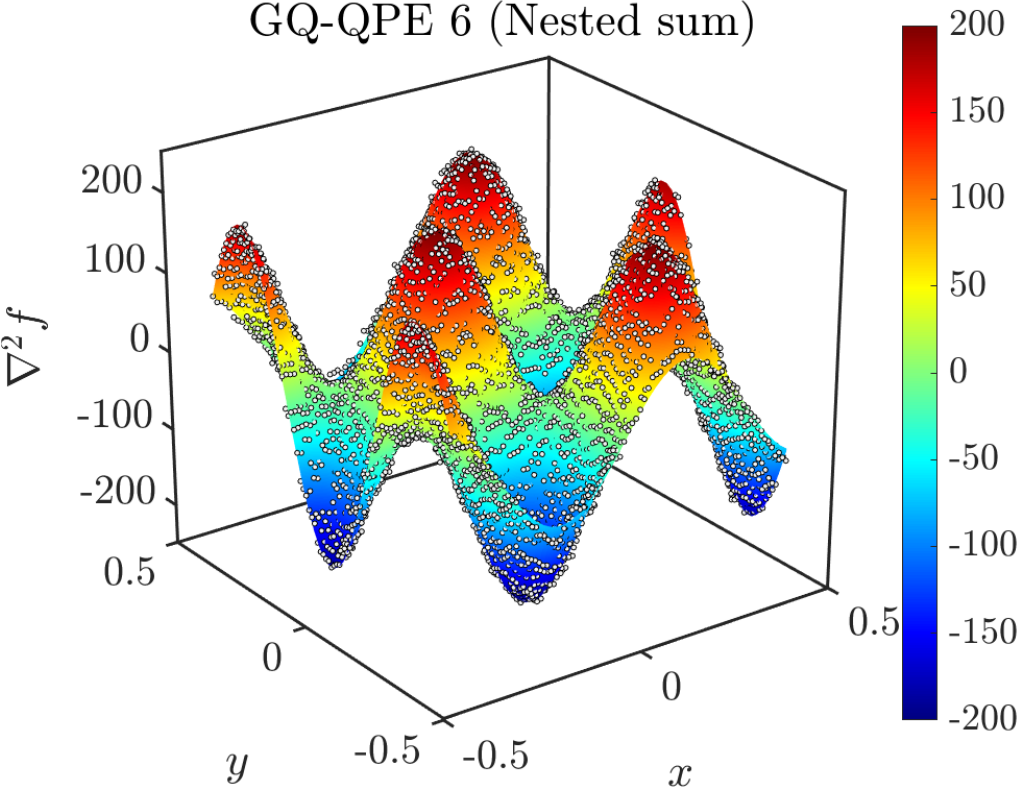}
			\caption{\centering\footnotesize Ancilla qubits ($t=6$).}
			\label{fig:6d}
		\end{subfigure}
		\begin{subfigure}[t]{0.32\textwidth}
			\centering
			\includegraphics[scale=0.3]{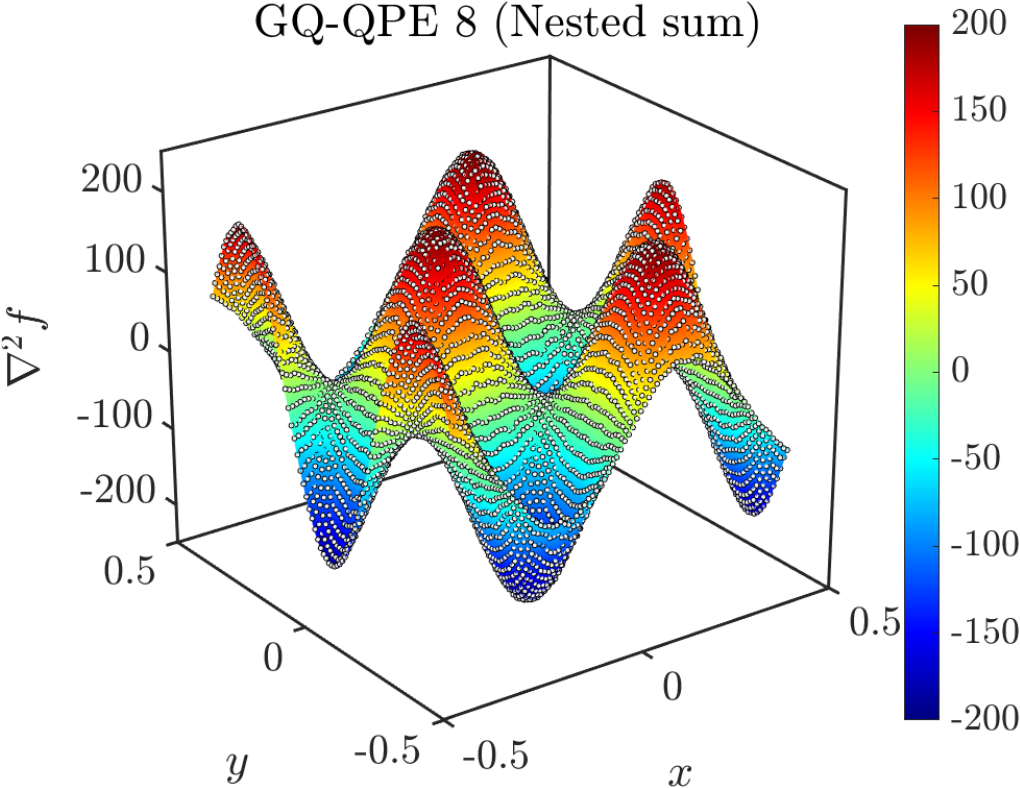}
			\caption{\centering\footnotesize Ancilla qubits ($t=8$).}  
			\label{fig:6e}
		\end{subfigure}
		\caption{\small Second-order partial derivative on approximated nested sum form.}
		\label{fig_Case1:5}
	\end{figure}

	Subsequently, considering more complicated variations on surface functions, it is validated based on the hybrid quantum computing of Eq.(\ref{eq_Case1:2}). As shown in Fig.(\ref{fig_Case1:3}), we adopted the aforementioned quantum FPM algorithm to capture the derivative changes of surface function using different particle interval $\Delta d$. Based on the complicated changes of second-order derivative, we captured the particle morphology in Figs.(\ref{fig_Case1:4},\ref{fig_Case1:5}) by using different discrete forms to further investigate their numerical performances on hybrid quantum procedures. Differently from conventional meshfree particle approximation, it is sensitive on the QPE estimation and the manifestation of denominator singularity ${1 \mathord{\left/
			{\vphantom {1 {r_{ij}^2}}} \right.
			\kern-\nulldelimiterspace} {r_{ij}^2}}$ in Eq.(\ref{eq_BASICdiscretization:3}). It is further concluded that the small size of computation possesses the better quantum fidelity with multi-partitioned zones and the discrete form on repetitive nested sum has the advantaged accuracy originally from high-precision approximated capability of finite particle method. Some details of conventional numerical performance for different mathematical methods can be referenced to an extent\supercite{WOS:001501337900001}.
	
	\begin{table}[H]
		\caption{\small\centering Time consuming on present validation procedures.}
		\label{tab_Case1:3}
		\centering \footnotesize 
		\begin{tabular}{
				>{\centering\arraybackslash}p{1.8cm} >{\centering\arraybackslash}p{1.3cm} >{\centering\arraybackslash}p{1.9cm} >{\centering\arraybackslash}p{1.9cm} >{\centering\arraybackslash}p{1.9cm} >{\centering\arraybackslash}p{1.9cm} >{\centering\arraybackslash}p{1.9cm}}
			\toprule
			\textbf{Interval $\Delta {d}$} & \textbf{Function} & \textbf{No Partition} & \textbf{QPE 2} & \textbf{QPE 4} & \textbf{QPE 6} & \textbf{QPE 8}\\
			\midrule
			\multirow{2}{*}
			{0.02} & (Fun1) & 151.192 s & 73.319 s & 82.174 s & 159.764 s & 536.649 s \\
			& (Fun2) & 152.786 s & 74.416 s & 88.356 s & 162.471 s & 553.983 s \\[2mm]
			\multirow{2}{*}
			{0.01} & (Fun1) & 680.269 s & 345.125 s & 410.021 s & 726.124 s & 2498.25 s \\
			& (Fun2) & 684.161 s & 349.456 s & 414.895 s & 737.772 s & 2531.63 s \\
			\bottomrule
		\end{tabular}
	\end{table}
	
	\begin{figure}[H]
		\centering
		\begin{subfigure}[t]{0.4\textwidth}
			\centering
			\includegraphics[scale=0.35]{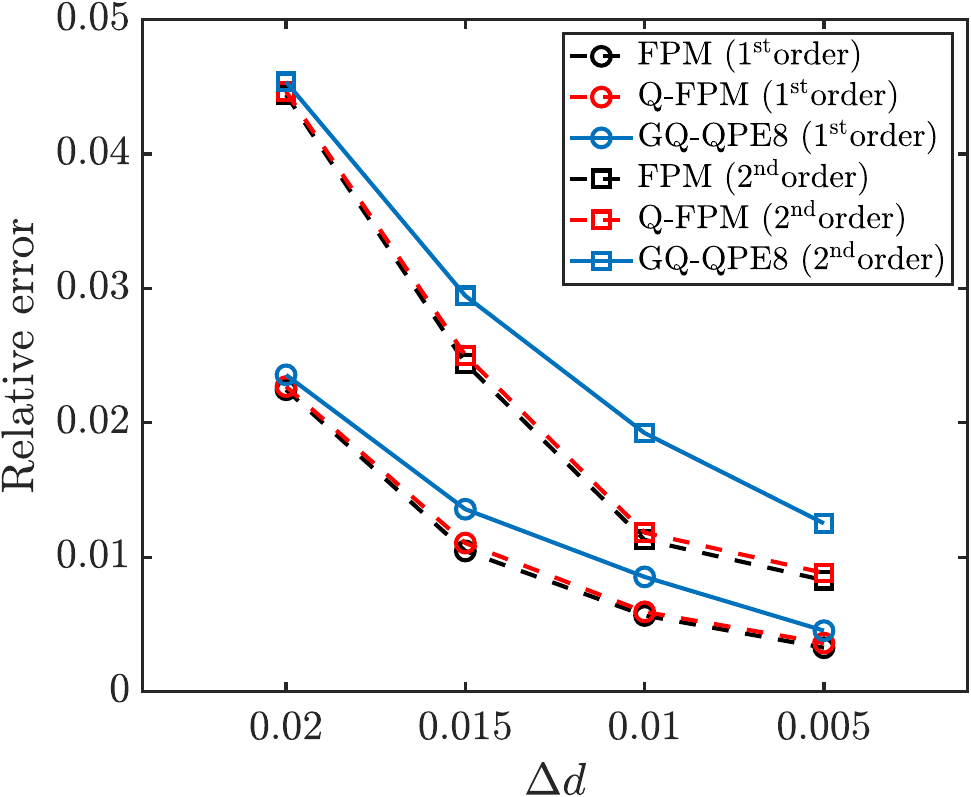}
			\caption{\centering\footnotesize Convergence on numerical partial derivative by nested sum form.}
			\label{fig:6d}
		\end{subfigure}
		\begin{subfigure}[t]{0.4\textwidth}
			\centering
			\includegraphics[scale=0.35]{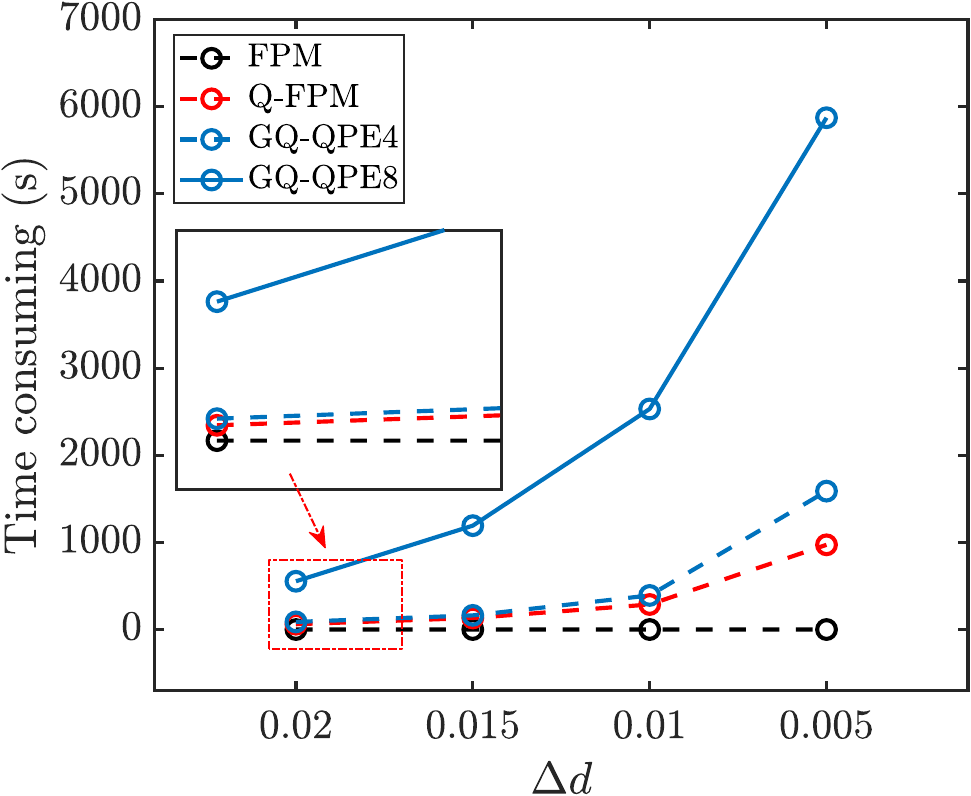}
			\caption{\centering\footnotesize Time consuming on hybrid quantum procedure.}  
			\label{fig:6e}
		\end{subfigure}
		\caption{\small Numerical approximation with the refinement of particle spacing.}
		\label{fig_Case1:6}
	\end{figure}

Additionally, the on-going time consumption and algorithmic convergence on present hybrid particle dynamics are also described detailedly in Table \ref{tab_Case1:3} and Fig.(\ref{fig_Case1:6}). It is observed that the present hybrid procedure has a stable convergent state reaching to error $0.5\%$ when $\Delta d=5$ mm. The inclusion of additional auxiliary qubits in QPE also leads to exponentially growing computational time requirements as particle spacing refinement increases. It notes that the accuracy degradation in hybrid quantum FPM implementations, while unavoidable given current NISQ-era device constraints (gate fidelity, coherence times), represents a necessary trade-off for pioneering quantum-classical computational synergies. \modifyFirstYellow{Its strategic importance persists as qubit scalability and error correction technologies maturely\supercite{WOS:001672321600001}.}
	
	\subsection*{Case 2: Dynamic viscoelastic Couette flow}\label{Case2}
\begin{figure}[H]
			\centering
			\includegraphics[scale=0.4]{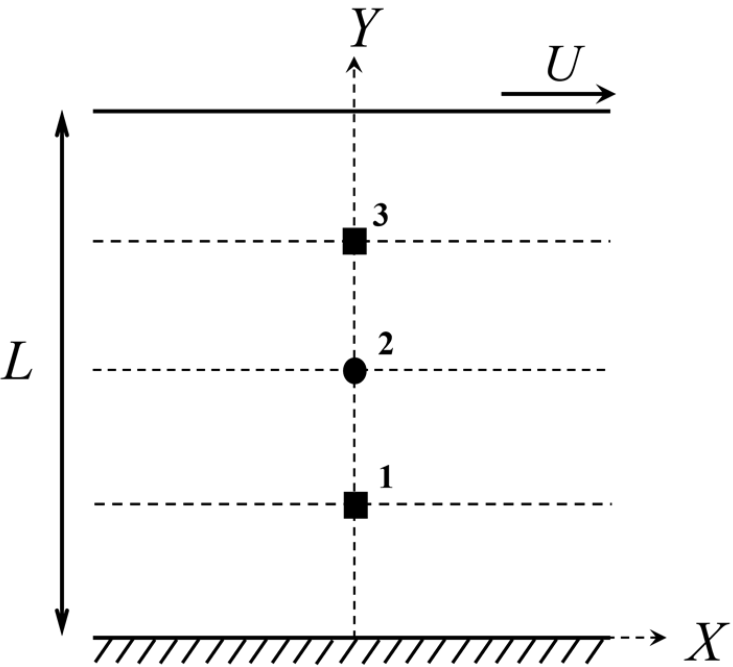}
			\caption{\small Physical model of Couette planar flow.}
			\label{fig_Case2:1}
		\end{figure}

	\normalsize \hspace{10pt}		
	This subsection presents a targeted verification of the proposed hybrid quantum FPM approach involving a time-dependent flow procedure. The unsteady Couette flow of Oldroyd-B viscoelastic fluids is validated. As shown in Fig.(\ref{fig_Case2:1}), the configuration consists of two infinite parallel plates and the bottom plate remains stationary while the top plate moves at a constant velocity $U = 0.01\ \rm{m/s}$. A transient shear flow driven periodically by a shear stress parallel to the X-axis is simulated. Analytical solutions for this Oldroyd-B flow are available in the literature\supercite{LI2023112213}. The characteristic velocity and length are set to $U = 1\ \rm{m/s}$ and $L = 1\ \rm{m}$, respectively. The Reynolds number is defined as ${\rm{Re}} = {{\rho UL} \mathord{\left/
			{\vphantom {{\rho UL} \mu }} \right.
			\kern-\nulldelimiterspace} \mu }$, the Weissenberg number as ${\rm{We}} = {{{\lambda _1}U} \mathord{\left/
			{\vphantom {{{\lambda _1}U} L}} \right.
			\kern-\nulldelimiterspace} L}$, and the ratio of viscoelasticity as ${\beta _0} = {\mu _s}/({\mu _s} + {\mu _p})$. Simulations are conducted across various Reynolds and Weissenberg numbers. Velocity profiles are probed at three fixed locations within the flow domain, as indicated in Fig.(\ref{fig_Case2:1}). The smoothing kernel length is set to $h = 1.4{d_0}$, where ${d_0}$ is the initial particle spacing, and the time step is $\Delta t = 1 \times {10^{ - 4}}\ \rm{s}$.
		
		\begin{figure}[H]
			\centering
			\begin{subfigure}[t]{0.4\textwidth}
				\centering
				\includegraphics[scale=0.35]{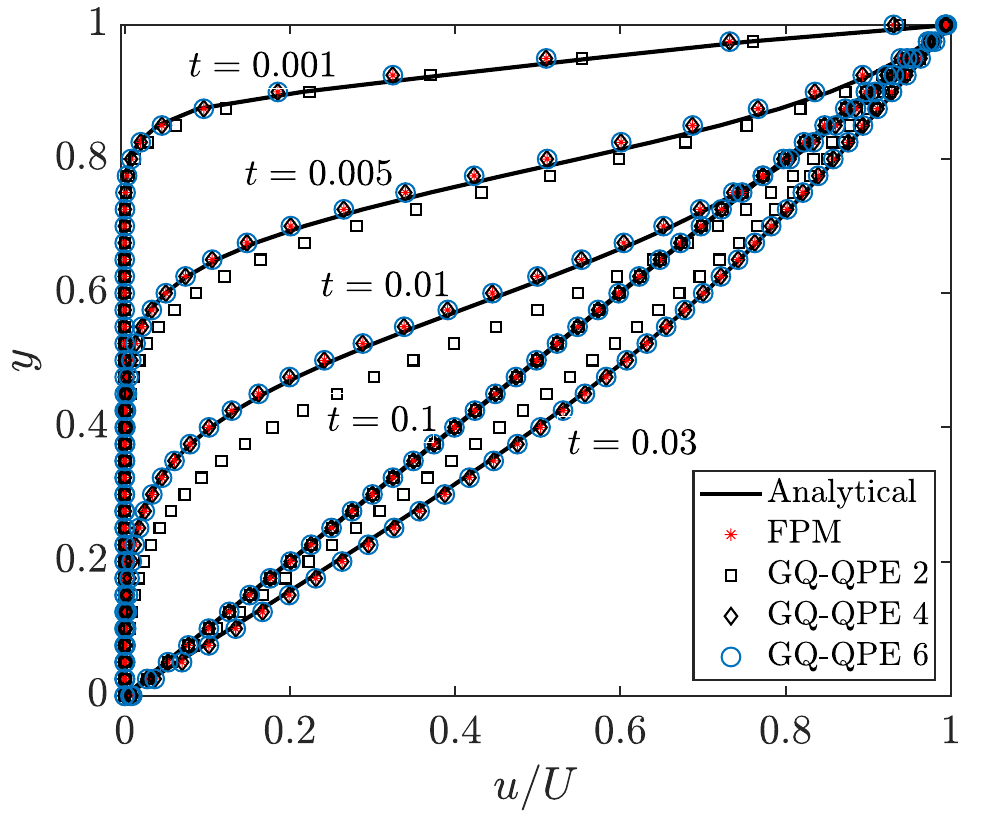}
				\caption{\centering\footnotesize Velocity profiles at different times $\textrm{We}=0.01$.}
				\label{fig_Case2:2a}
			\end{subfigure}
			\begin{subfigure}[t]{0.4\textwidth}
				\centering
				\includegraphics[scale=0.35]{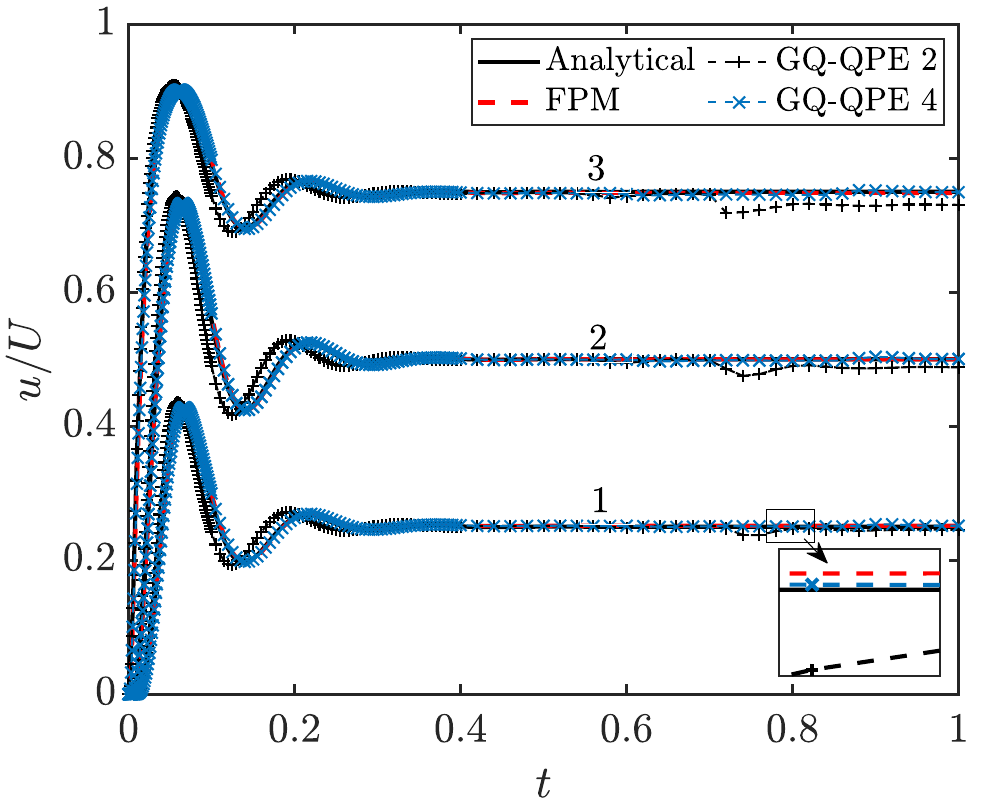}
				\caption{\centering\footnotesize Velocity profiles at different positions $\textrm{We}=0.1$.}  
				\label{fig_Case2:2b}
			\end{subfigure}
			\caption{\small Velocity overshoot of viscoelastic Couette flow. (Re $= 0.05$, ${\beta _0}= 0.1$)}
			\label{fig_Case2:2}
		\end{figure}
	
\begin{figure}[H]
		\centering
		\begin{subfigure}[t]{0.4\textwidth}
			\centering
			\includegraphics[scale=0.35]{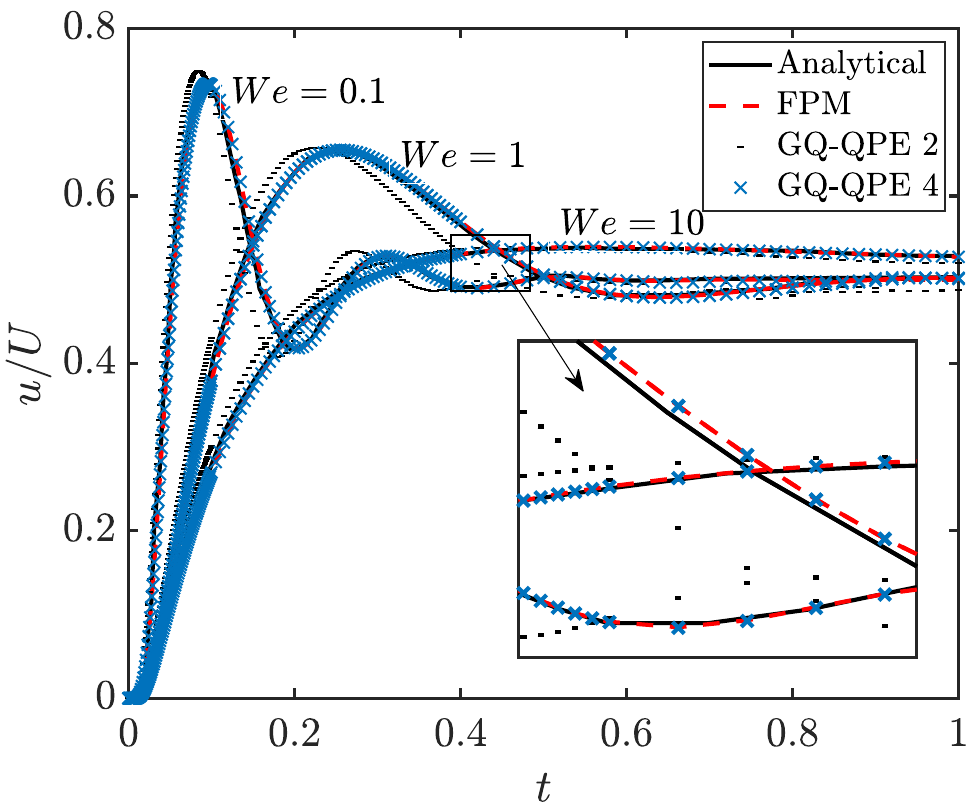}
			\caption{\centering\footnotesize Velocity profiles at position 2 and different Weissenberg numbers.}
			\label{fig:6d}
		\end{subfigure}
		\begin{subfigure}[t]{0.4\textwidth}
			\centering
			\includegraphics[scale=0.35]{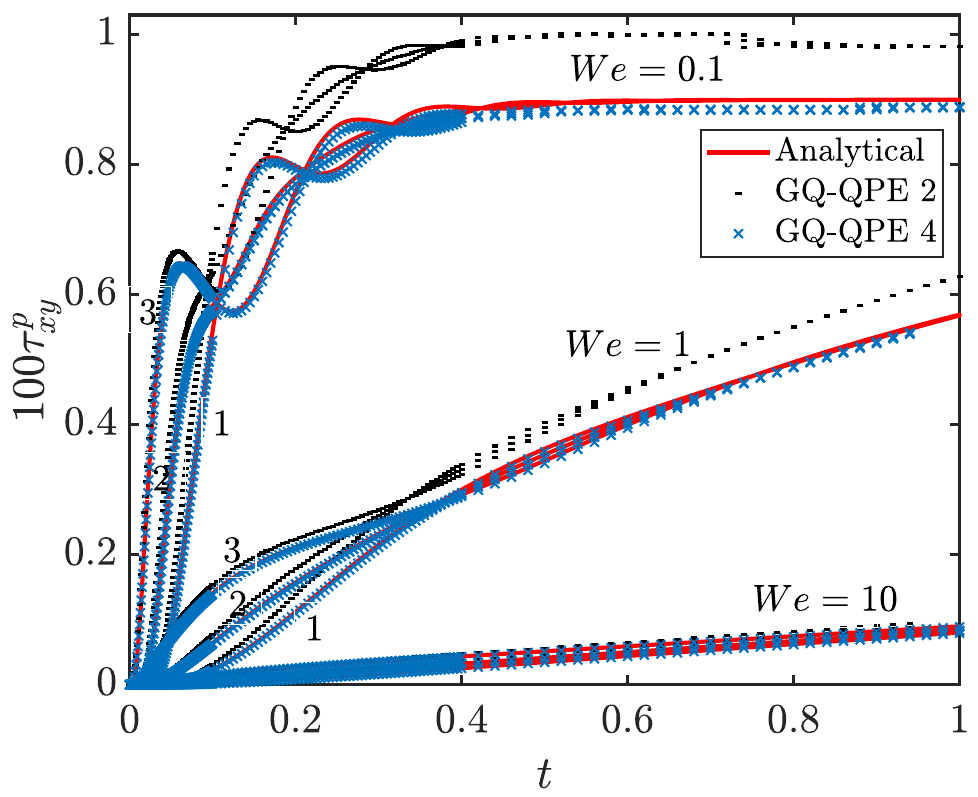}
			\caption{\centering\footnotesize Viscoelastic stress profiles at different positions and Weissenberg numbers.}  
			\label{fig:6e}
		\end{subfigure}
		\caption{\small Numerical results of viscoelastic Couette flow in hybrid quantum FPM. (Re $= 0.1$, ${\beta _0}= 0.1$)}
		\label{fig_Case2:3}
	\end{figure}

	Figs.(\ref{fig_Case2:2}-\ref{fig_Case2:3}) give different physical descriptions and numerical comparisons among these numerical results and analytical solutions. The numerical phenomenon reveals a time-dependent increase in the velocities of bottom fluid, attributed to the viscoelastic stress effects shown in Fig.(\ref{fig_Case2:2a}). For the viscoelastic fluid flow, particle velocity overshoot is observed, wherein fluid particles exhibit a transient peak velocity in Fig.(\ref{fig_Case2:2b}). Subsequently, the velocity fluctuations attenuate progressively until the system stabilizes into a linear particle motion. Fig.(\ref{fig_Case2:3}) presents the numerical results at varying spatial positions and Weissenberg numbers. Key findings indicate that both the particle velocity overshoot and fluctuations diminish with increasing Weissenberg numbers $\textrm{We}$, correlating with the weakening viscoelastic effects.
	
	Based on the advantaged multi-partitioned quantum computing with the frame on computational FPM method, the long-time evolutional quantum inner product process with 4 or 6 ancillary qubits achieves the favorable computational accuracy comparable to classical FPM method, while demonstrating excellent agreements with exact solutions. In contrast, the computational procedure employing only 2 ancillary qubits exhibits inferior performance, yielding significantly less accurate results in quantum phase estimation (QPE) measurements. As a result, the current results of designed dynamic viscoelastic Couette flow obviously confirm the validity of hybrid quantum-classical computational approach, while demonstrating the potential for implementing quantum computing in long-term evolution simulations of particle-laden fluid dynamics. This comprehensive methodology enables more extensive and in-depth integrated computations.
	
	Moreover, the computational accuracy of current hybrid quantum inner product process is critically dependent on the number of ancillary qubits in QPE shown in Fig.(\ref{fig_quantumFPM:partition}), particularly when accounting for quantum noise in noisy intermediate-scale quantum (NISQ) devices. Under such conditions, quantum measurement outcomes and state collapse behaviors exhibit significant variations. As emerging quantum-enhanced techniques, including variational quantum algorithms (VQAs)\supercite{Cerezo2021Variationalquantum} and quantum machine learning approaches, it demonstrated the improved capability for evaluating numerical solutions of quantum inner product and achieving fidelity. While the QPE-based inner product methodology remains fundamentally robust when considering the computational continuity and integrality requirements in fluid dynamics simulations\supercite{WOS:000477879500005,CHEN2024117428}.
	
\begin{figure}[H]
	\centering
	\begin{subfigure}[t]{0.4\textwidth}
		\centering
		\includegraphics[scale=0.44]{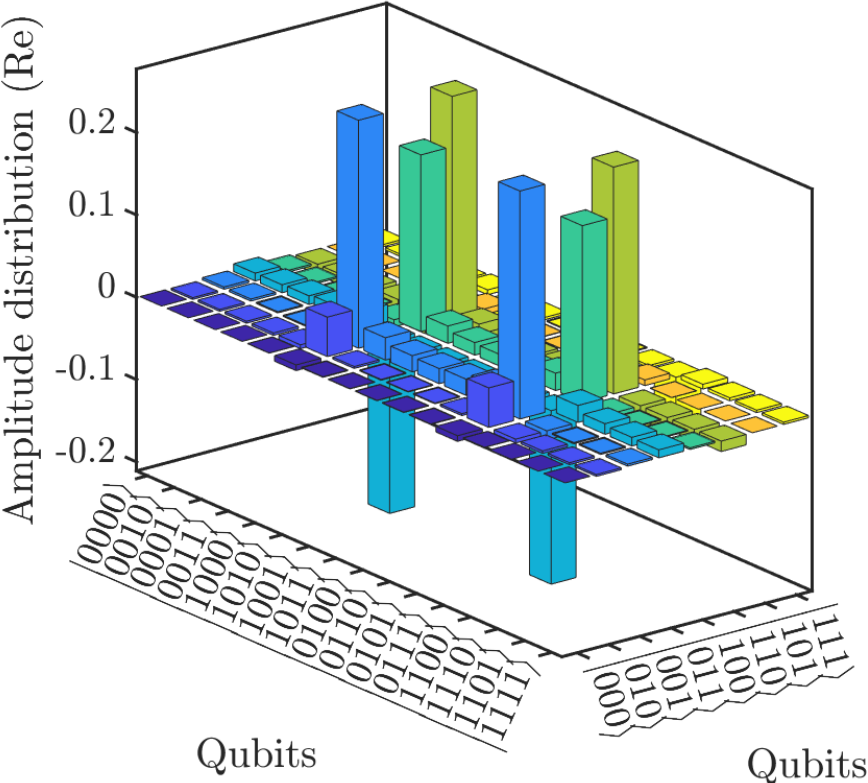}
		\caption{\centering\footnotesize Real part of quantum state amplitude.}
		\label{fig:6d}
	\end{subfigure}
	\begin{subfigure}[t]{0.4\textwidth}
		\centering
		\includegraphics[scale=0.44]{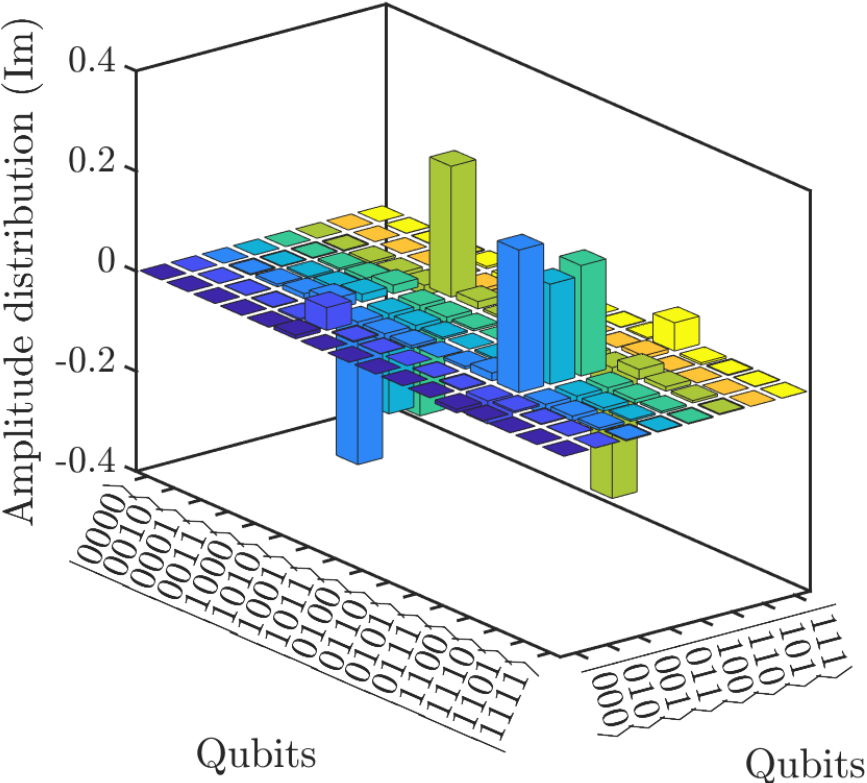}
		\caption{\centering\footnotesize Imaginary part of quantum state amplitude.}  
		\label{fig:6e}
	\end{subfigure}
	\caption{\small Quantum state amplitudes in high-dimensional Hilbert space.}
	\label{fig_Case2:4}
\end{figure}
	
\begin{figure}[H]
		\centering
		\begin{subfigure}[t]{0.4\textwidth}
			\centering
			\includegraphics[scale=0.35]{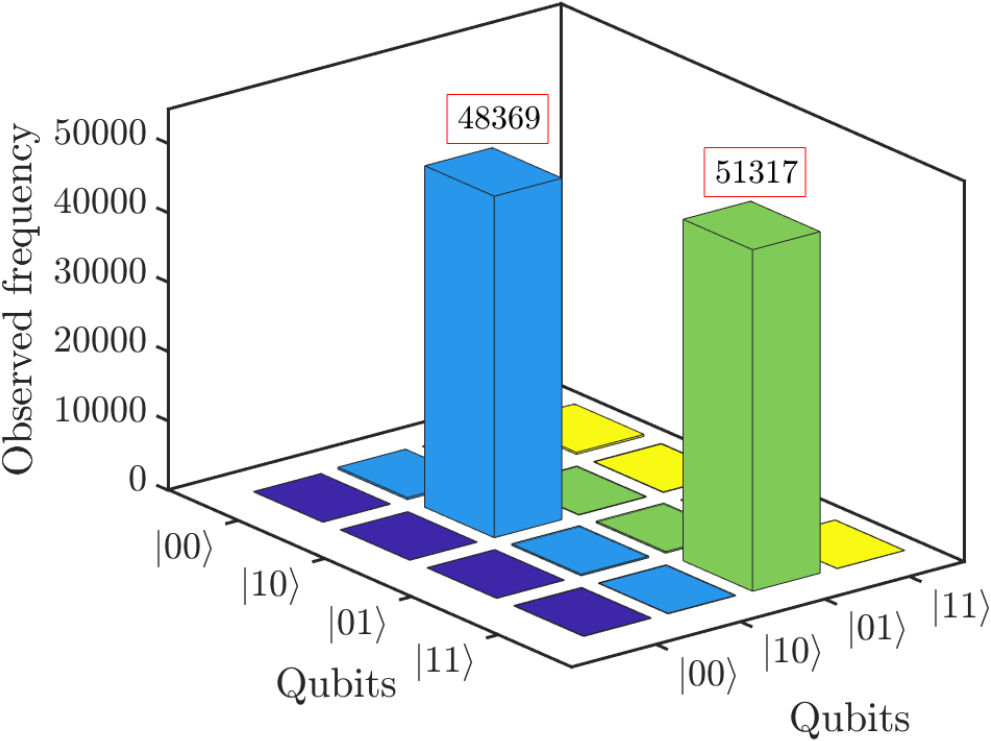}
			\caption{\centering\footnotesize Theoretical observation and measurement.}
			\label{fig:6d}
		\end{subfigure}
		\begin{subfigure}[t]{0.4\textwidth}
			\centering
			\includegraphics[scale=0.35]{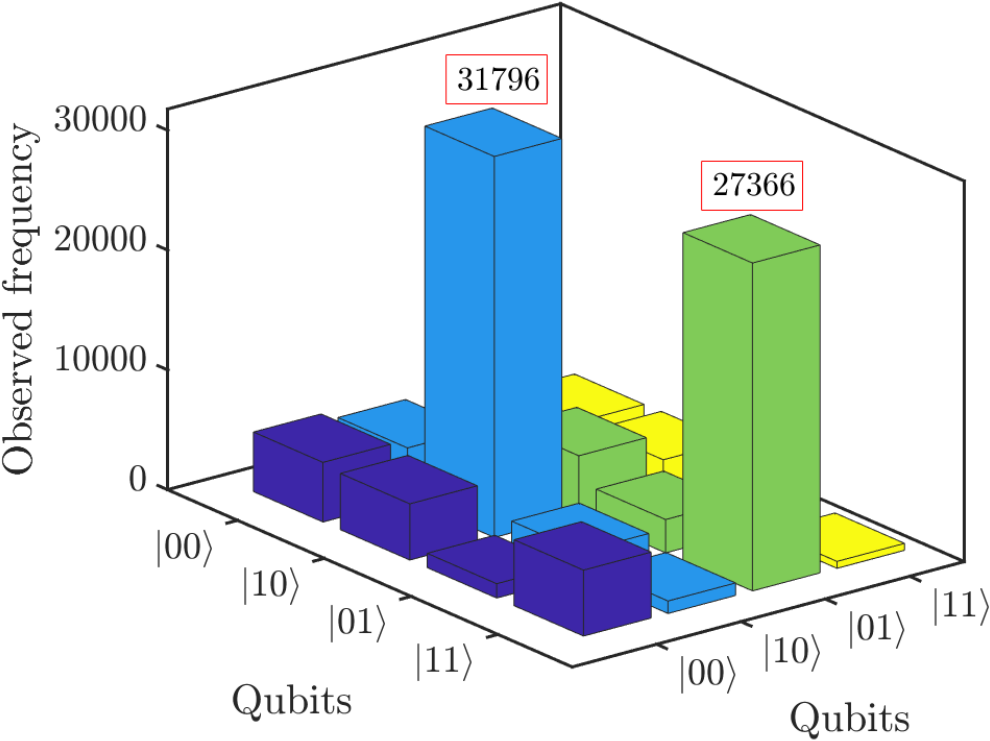}
			\caption{\centering\footnotesize Noisy quantum observation and measurement (Noise $\mathrm{I}$).}  
			\label{fig:6e}
		\end{subfigure}
		\begin{subfigure}[t]{0.4\textwidth}
			\centering
			\includegraphics[scale=0.35]{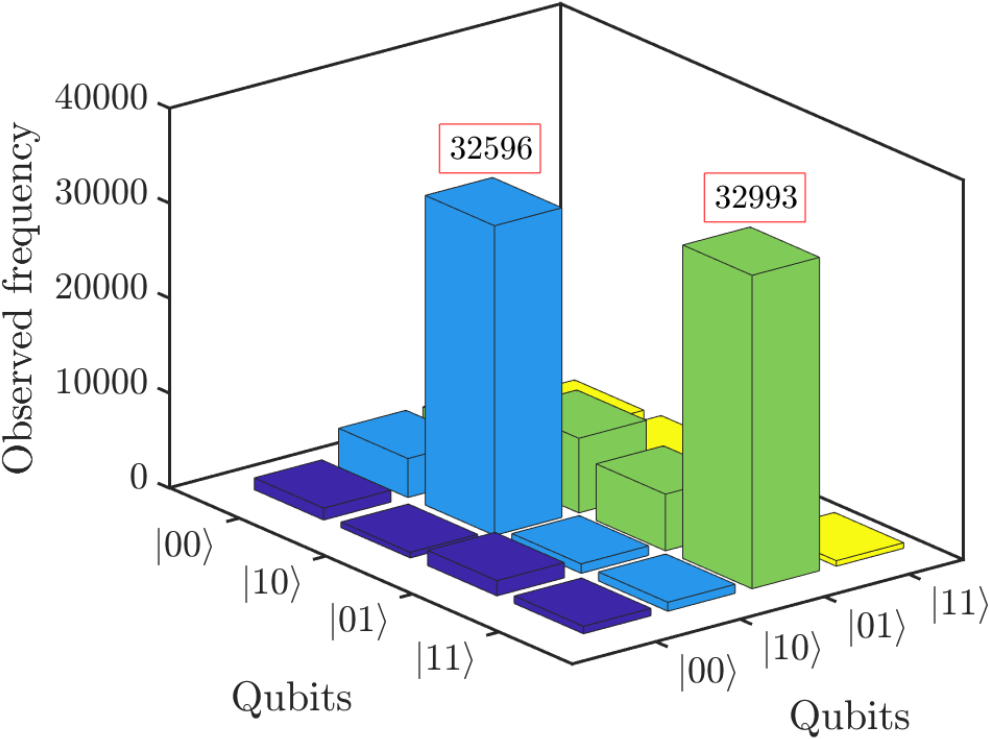}
			\caption{\centering\footnotesize Noisy quantum observation and measurement (Noise $\mathrm{II}$).}  
			\label{fig:6e}
		\end{subfigure}
		\begin{subfigure}[t]{0.4\textwidth}
			\centering
			\includegraphics[scale=0.35]{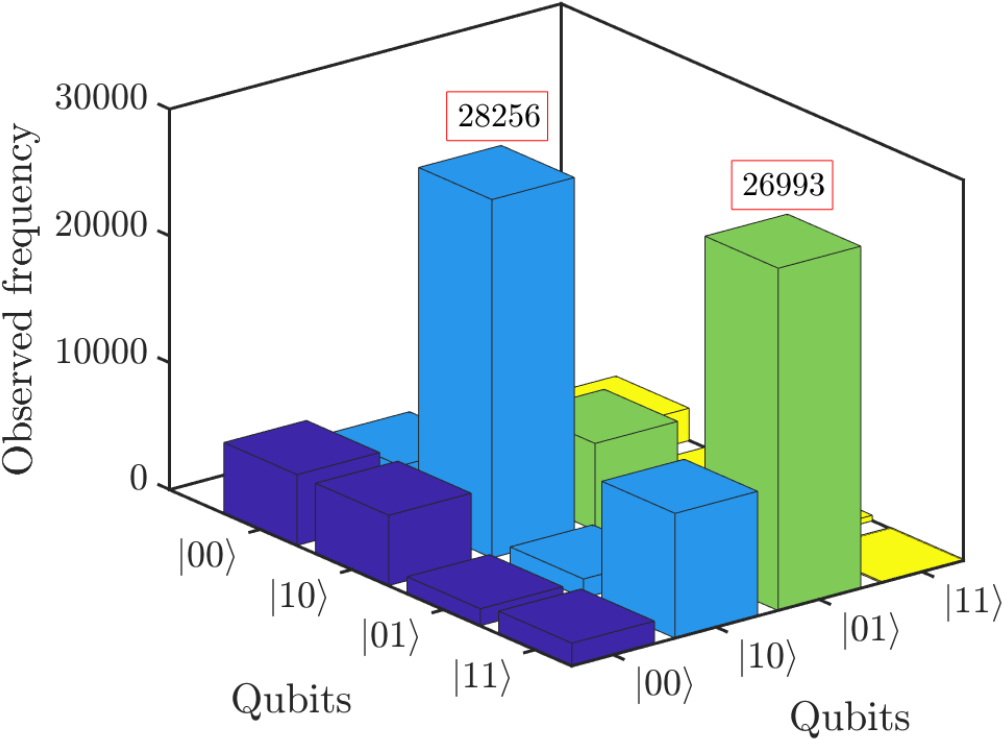}
			\caption{\centering\footnotesize Noisy quantum observation and measurement (Noise $\mathrm{III}$).}  
			\label{fig:6e}
		\end{subfigure}
		\caption{\small Quantum measurements and state collapse through 100,000 single-molecule observations.}
		\label{fig_Case2:5}
	\end{figure}

	Fig.(\ref{fig_Case2:4}) illustrates the composite Hilbert space formed by ancillary and measurement qubits in the present hybrid quantum FPM procedure. The observed probability amplitude distribution within this state space governs the measurement-induced wave function and collapse statistics, where the squared modulus of these amplitudes ${\left| {\left\langle {u\left| w \right.} \right\rangle } \right|^2}$ determines the projective measurement probabilities according to the Born rule. In this study, to rigorously evaluate the hybrid numerical performance, we conducted 100,000 effective experimental trials of noisy QPE computing, as demonstrated in Fig.(\ref{fig_Case2:5}). \modifyFirstYellow{The noisy quantum incorporated the variational decoherence noise models and distinct energy relaxation rates during the measurement processes. Noise I described the depolarizing channel only, whose symmetric depolarizing error is applied to all single-qubit gates with error probability $p_1 = 1 \times 10^{-3}$ and to all two-qubit gates with $p_2 = 1 \times 10^{-2}$. No additional decoherence or measurement error is introduced in this baseline configuration. Noise II described the depolarizing, amplitude damping and decoherence noises. In addition to the depolarizing channel of Noise I, qubit decoherence is modelled by the standard Kraus operator parameterised with the energy relaxation time $T_1 = 50\,\mu\text{s}$, the dephasing time $T_2 = 70\,\mu\text{s}$, and a characteristic gate time of $5\,\text{ns}$. Furthermore, an amplitude damping error during measurement is applied with a $3\%$ relaxation probability. This composite model reflects the dominant noise mechanisms on current superconducting qubit platforms. Noise III described the device-calibrated noise, including depolarizing rates, $T_1$ and $T_2$ times, gate durations, and readout error probabilities, which are directly extracted from the backend calibration data of the “OriginQ Wukong” superconducting quantum processor at the time of execution.}This approach ensures computational reliability and facilitates practical hybrid computation implementation on current NISQ devices. Despite inherent noise interference, the robust QPE protocol successfully captures accurate computational results through statistically significant sampling (100,000 quantum measurements), demonstrating the favorable hybrid computation performance.
	
	\modifyFirstYellow{Regarding the propagation of quantum computational errors within the time-stepping scheme, these errors from multi-qubit gate infidelity, multi-partition accumulation and finite ancilla qubits in QPE are evident in the baseline verification of Case 1. In principle, these errors could accumulate through the Symplectic predictor-corrector integration, as illustrated intuitively in the framework of Fig.(\ref{fig:quantum_frame}). However, the results in Figs.(\ref{fig_Case2:2}, \ref{fig_Case2:3}) indicate that the quantum inner product errors do not undergo such propagation. They remain bounded, primarily because the explicit integration scheme, the artificial viscosity formulation and the particle shifting technique embedded in the classical SPH framework can effectively absorb and constrain these perturbations. As noted in earlier studies\supercite{WOS:001501337900001}, local modifications to numerical kernels such as high-order FPM discretization, SPH summation primarily influence neighboring interactions, yet their impact on the global time-stepping solution is substantially moderated when integrated into the full multi-physics framework. Consequently, the conservation of physical invariants is maintained within a range similarly to that of conventional high-fidelity methods\supercite{2019Smoothed}. This observation confirms that, supported by a range of conventional discretization techniques, this non-end-to-end hybrid paradigm does not lead to a cumulative, amplifying propagation of local quantum computational errors.}

	\begin{figure}[H]
		\centering
		\begin{subfigure}[t]{0.4\textwidth}
			\centering
			\includegraphics[scale=0.355]{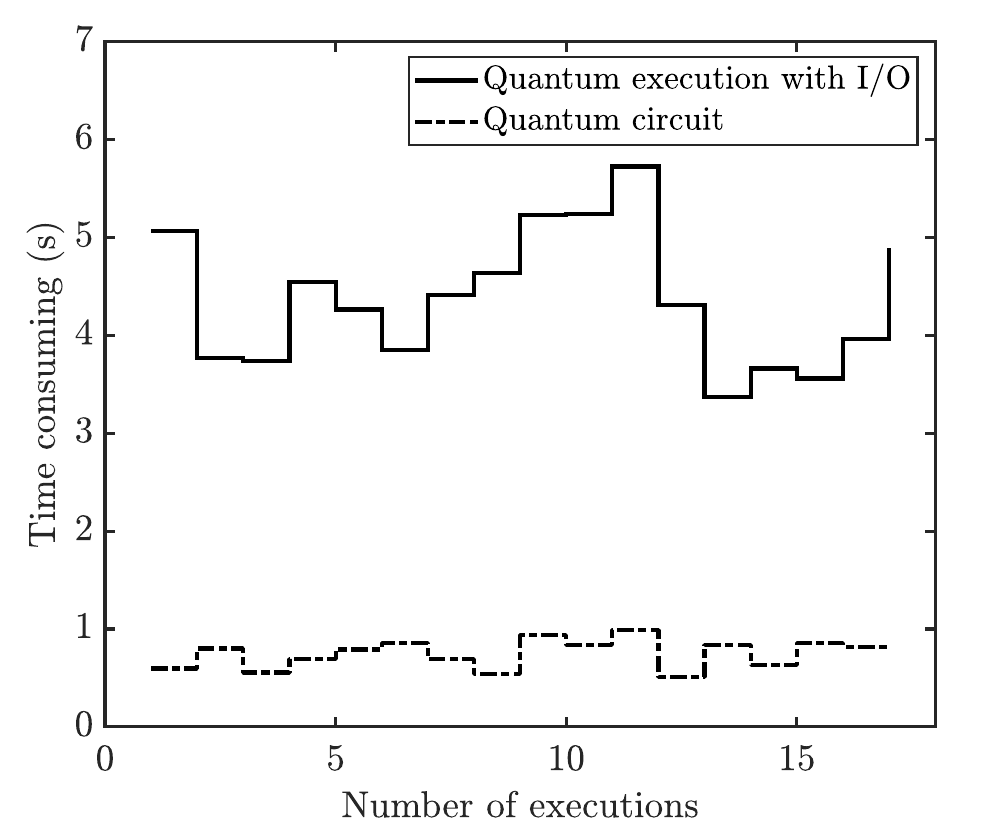}
			\caption{\centering\footnotesize QPE at $t=2$.}
			\label{fig:6d}
		\end{subfigure}
		\begin{subfigure}[t]{0.4\textwidth}
			\centering
			\includegraphics[scale=0.355]{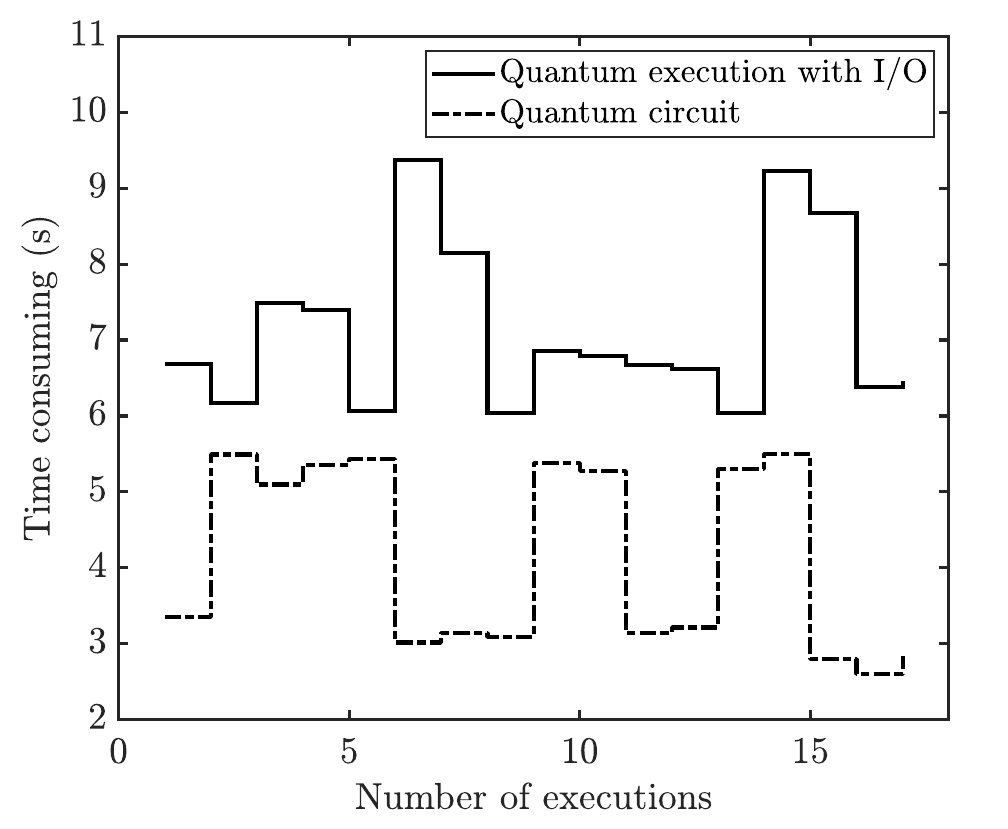}
			\caption{\centering\footnotesize QPE at $t=4$.}  
			\label{fig:6e}
		\end{subfigure}
		\caption{\small Extracted time consuming on quantum inner product and associated classical-quantum I/O overhead with specific measurements $M=100000$.}
		\label{fig_Case2:Revise}
	\end{figure}
	
	\begin{table}[H]
		\caption{\small\centering Operational statistics of 735 FPM particles multi-partially encoded in quantum registers.}
		\label{tab_Case2:1}
		\centering \footnotesize 
		\begin{tabular}{>{\centering\arraybackslash}p{0.6cm} >{\centering\arraybackslash}p{2.6cm} >{\centering\arraybackslash}p{1.5cm} >{\centering\arraybackslash}p{1.2cm} 
				>{\centering\arraybackslash}p{1.8cm} >{\centering\arraybackslash}p{2.5cm} >{\centering\arraybackslash}p{3.4cm}}
			\toprule
			\textbf{QPE}& \textbf{\parbox{2.6cm}{Total steps on\\quantum register}} & \textbf{\parbox{1.5cm}{Observed frequency}} & \textbf{\parbox{1.2cm}{Total\\time}}& \textbf{\parbox{1.8cm}{Time at\\per particle}} & \textbf{\parbox{2.5cm}{Time at per\\particle and step}} & \textbf{\parbox{3.4cm}{Time at per particle \\step and measurement}}\\
			\midrule
			2& 20,002 &100,000&241,273 s & 328.26 s& 0.01641 s & $1.641\times 10^{-7}$ s\\[1mm]
			4& 20,002 &100,000&306,740 s & 417.33 s& 0.02086 s & $2.086\times 10^{-7}$ s\\
			\bottomrule
		\end{tabular}
	\end{table}

\modifyFirstYellow{Finally, to quantify the temporal overhead associated with the classical-quantum interface as discussed in Section \ref{sec:complexity}, Fig.(\ref{fig_Case2:Revise}) extracted the time consuming characteristic on the quantum inner product evaluation with multi-partitioned circuits, including three constituent phases: classical pre-processing and amplitude encoding; quantum circuit execution, comprising the swap test core and quantum phase estimation; classical post-processing and measurement retrieval, where the collapsed bit-strings from repeated shots are collected and statistically processed to extract the inner product estimate. Some corresponding descriptions of theoretical complexity have been elaborated in the previous contents. It can be observed that the average time consuming of single-running quantum circuit is approximately 0.9 s at QPE $t=2$ ($9 \times 10^{-6}$ s/measurement), and 4.3 s at QPE $t=4$ ($4.3 \times 10^{-5}$ s/measurement). The total time including I/O procedure is approximately 4.5 s at QPE $t=2$, and 7.8 s at QPE $t=4$. The time solely used for I/O operations is both approximately 3.5 s. It should be noted that the time measurement of each component encompasses the function call overhead, i.e., the stack-level input/output latency. Theoretically, a single quantum measurement can reach a time scale of the order of $10^{-7}$ s, as described in Table \ref{tab_Case2:1}.}

\modifyFirstYellow{Overall, Table \ref{tab_Case2:1} summarizes the operational statistics of 735 FPM particles encoded in multi-partitioned quantum registers.}The number of ancillary qubits emerges as the dominant factor governing total computation time, particularly for large-scale models requiring multiple iteration steps. Under these conditions, the temporal cost significantly exceeds classical computational scales (by $2-3$ orders of magnitude), reflecting fundamental limitations in current quantum phase estimation implementations and quantum device architectures generally. \modifyFirstYellow{Notably, quantum computing demonstrates the inherent advantages for small-scale, single-step computations, achieving latencies as low as $10^{-7}$ s.}However, the analysis reveals substantial developmental requirements for complex, large-scale dynamical computations, particularly regarding temporal efficiency and precision maintenance in noisy environments. \modifyFirstYellow{As a result, the present hybrid quantum-classical approaches remain essential for transitional-stage implementations.}We anticipate that the continued development of quantum hardware will further enhance the reliability and robustness of these quantum computational outcomes.
	
	\subsection*{Case 3: Dynamic viscoelastic Poiseuille flow with HWN discussion}
	\normalsize \hspace{10pt}
	In this subsection, numerical analysis on high Weissenberg viscoelastic fluid flow is further investigated based on present hybrid quantum computing procedure. Notably, given the inherent limitations of current quantum simulation algorithms and the non-trivial computational overhead of hybrid quantum-particle operations quantified in Table \ref{tab_Case2:1}, this work employs viscoelastic Poiseuille flow as a benchmark system to rigorously evaluate the HWN behaviors and improved meshfree numerical performance on accuracy and convergence. \modifyFirstYellow{Through this paradigm, we demonstrated both the methodological novelty and the computational potential of developed FPM framework and quantum-hybridized implementation.}
	
	\begin{figure}[H]
		\centering
		\includegraphics[scale=0.25]{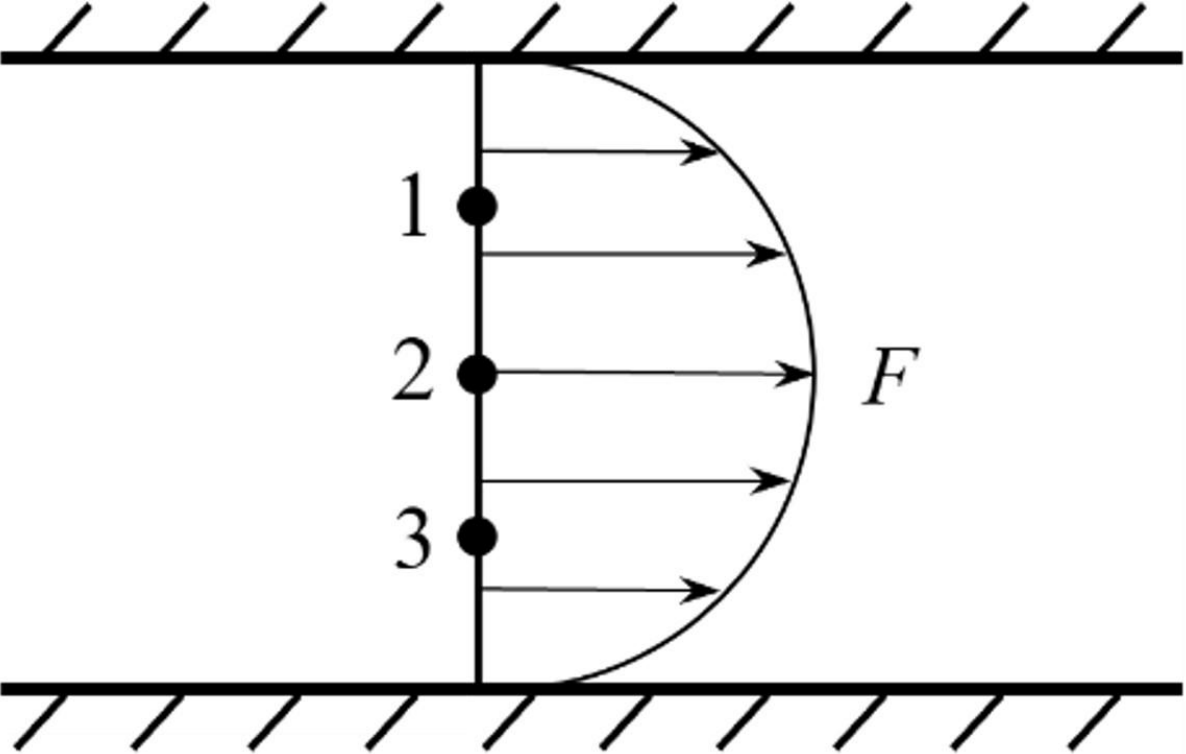}
		\caption{\small Sketch of viscoelastic Poiseuille flow.}
		\label{fig_Case3:model}
	\end{figure}
	
	\begin{figure}[H]
		\centering
		\includegraphics[scale=0.4]{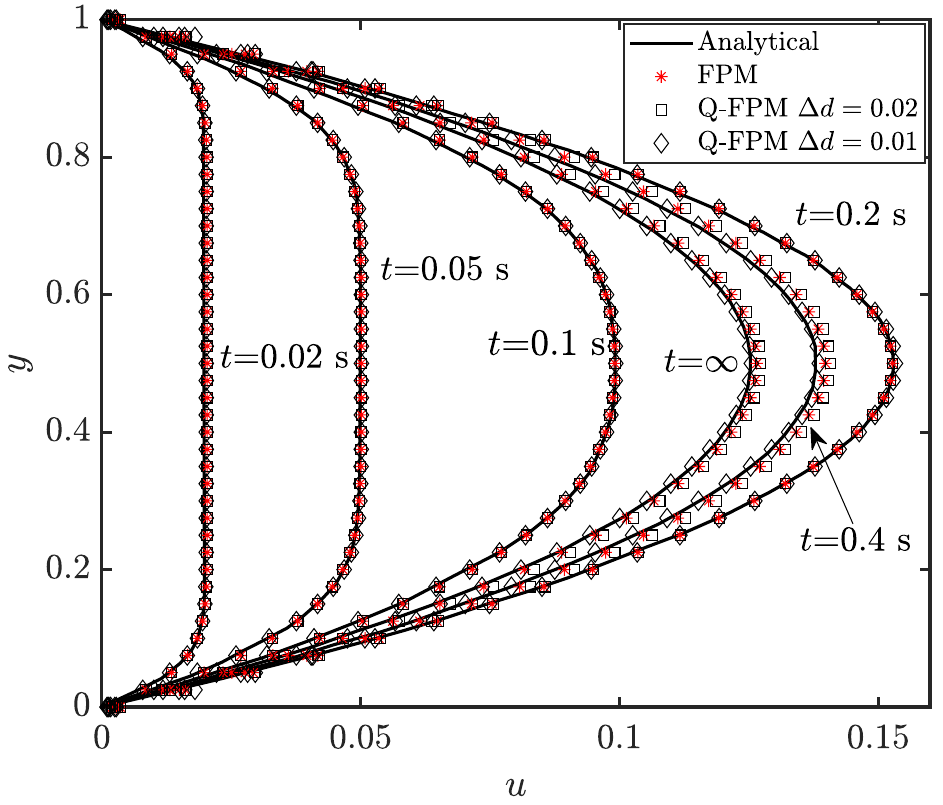}
		\caption{\small Profile on measured velocity by present quantum FPM procedure.}
		\label{fig_Case3:1}
	\end{figure}

\begin{figure}[H]
	\centering
	\begin{subfigure}[t]{0.45\textwidth}
		\centering
		\includegraphics[scale=0.4]{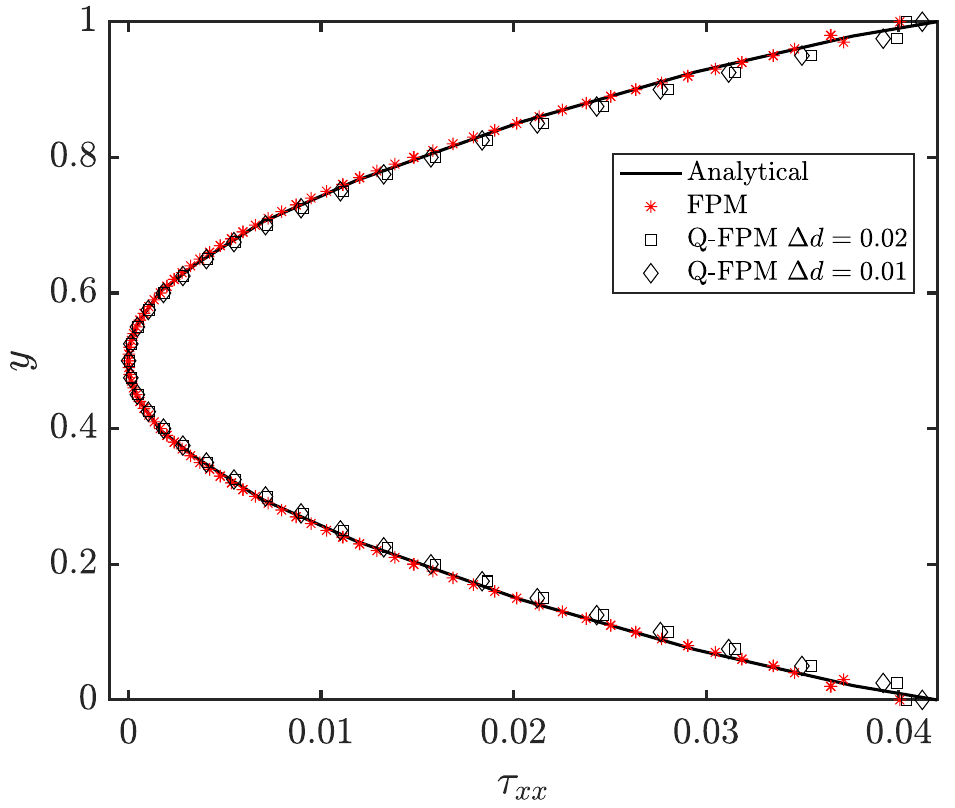}
		\caption{\centering\footnotesize Viscoelastic stress components ${\tau _{xx}}$.}
		\label{fig:6d}
	\end{subfigure}
	\begin{subfigure}[t]{0.45\textwidth}
		\centering
		\includegraphics[scale=0.4]{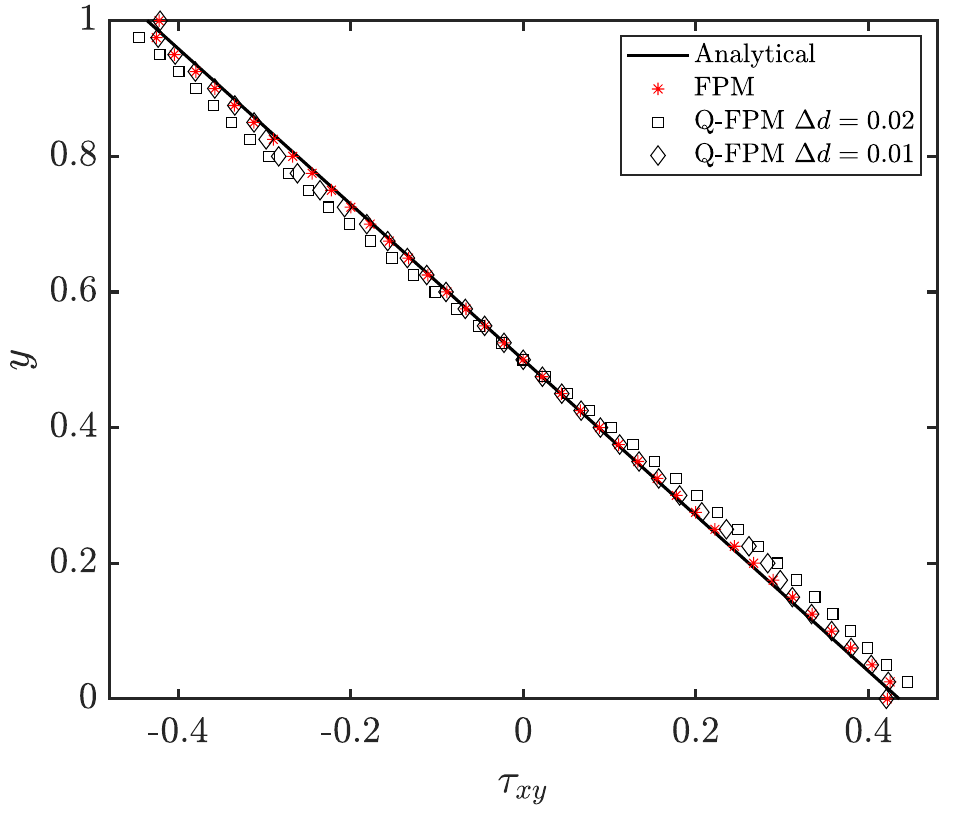}
		\caption{\centering\footnotesize Viscoelastic stress components ${\tau _{xy}}$.}  
		\label{fig:6e}
	\end{subfigure}
	\caption{\small Profile on measured viscoelastic stress components at steady flow state.}
	\label{fig_Case3:2}
\end{figure}
	
	The schematic representation of present Poiseuille flow system is depicted in Fig.(\ref{fig_Case3:model}), specifying the unit-length spacing between two infinite parallel plates maintained in a stationary state. The computational setup employs periodic boundary conditions in horizontal direction, where a constant external force $F=1$ N applied in the positive x-direction drives the viscoelastic fluid flows through the resultant shear stress until achieving balanced laminar motion. Three equidistant sampling probes are strategically placed to quantify the evolution on physical field variables. The dimensionless parameters are characterized by Reynolds number ${\rm{Re}} = {{\rho UL} \mathord{\left/
			{\vphantom {{\rho UL} \mu }} \right.
			\kern-\nulldelimiterspace} \mu }$, Weissenberg number ${\rm{We}} = {{{\lambda _1}U} \mathord{\left/
			{\vphantom {{{\lambda _1}U} L}} \right.
			\kern-\nulldelimiterspace} L}$, and ratio of viscoelasticity ${\beta} = {\mu _s}/({\mu _s} + {\mu _p})$. The characteristic velocity is $U = 0.1\ \rm{m/s}$, the characteristic length is $L = 1\ \rm{m}$ and ${\Delta d}$ denotes the initial spatial distance.
	
	Fig.(\ref{fig_Case3:1}) captured the evolutional velocity with analytical solutions\supercite{XU2022782} at different time, which can demonstrate the computational accuracy of present numerical scheme and show excellent agreements on quantum FPM. During the transient flow development, the characteristic velocity overshoot phenomenon caused by viscoelastic effects is clearly captured before reaching steady-state pipe flow. Comparative velocity profiles under steady-state conditions reveal the relative errors of 1.37\% (FPM), 1.91\% (present Q-FPM with ${\Delta d=0.02}$) and 0.50\% (present Q-FPM with ${\Delta d=0.01}$), respectively. In Fig.(\ref{fig_Case3:2}), the viscoelastic stress components are also computed across varying initial particle spacings, demonstrating satisfactory convergence behavior. With progressive particle refinements, numerical discrepancies near boundaries systematically diminish, asymptotically approaching the analytical solution\supercite{XU2022782}. Therefore, the effectiveness of current hybrid quantum-particle dynamics approach can be demonstrated through the quantum-classical coupling accuracy and convergence.	
	
	\begin{figure}[H]
		\centering
		\begin{subfigure}[t]{0.45\textwidth}
			\centering
			\includegraphics[scale=0.4]{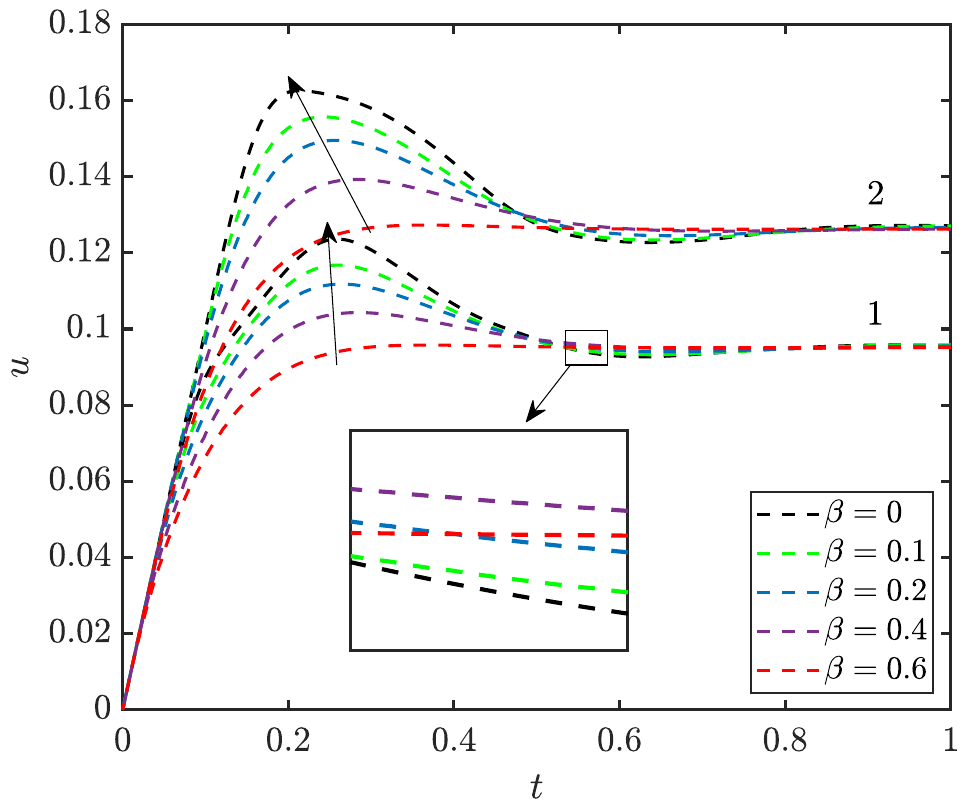}
			\caption{\centering\footnotesize The evolution on velocity at fixed position.}
			\label{fig:6d}
		\end{subfigure}
		\begin{subfigure}[t]{0.45\textwidth}
			\centering
			\includegraphics[scale=0.4]{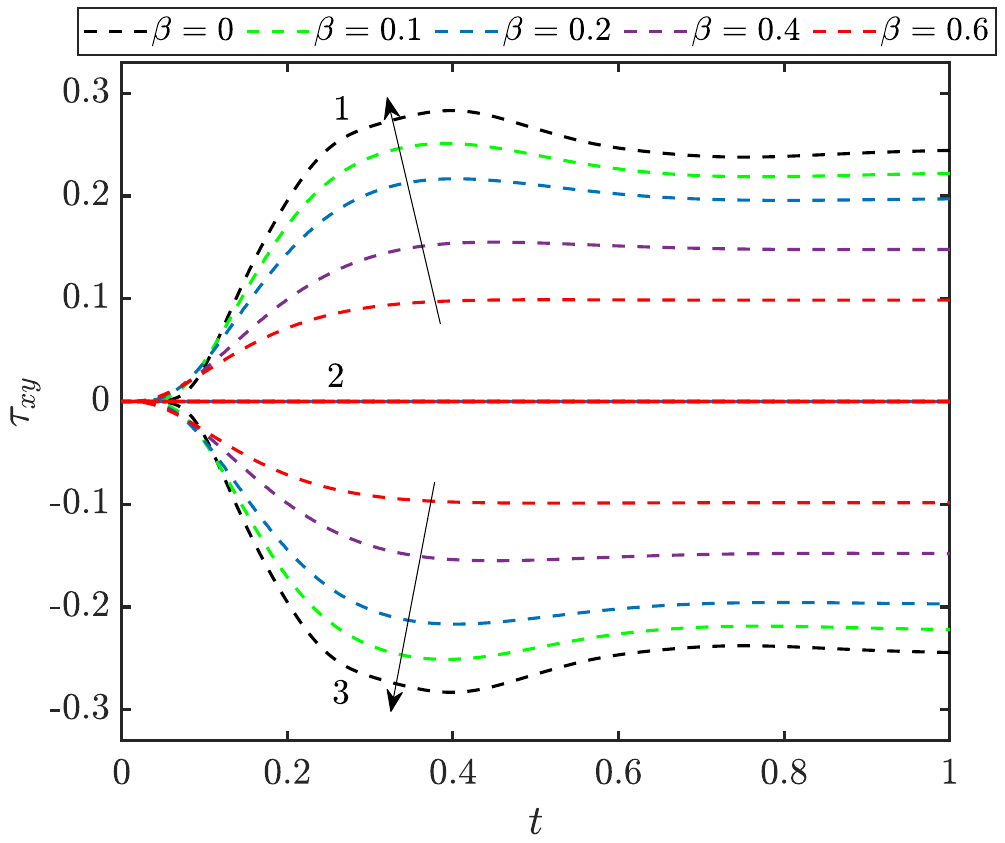}
			\caption{\centering\footnotesize The evolution on stress components ${\tau _{xy}}$ at fixed position.}  
			\label{fig:6e}
		\end{subfigure}
		\caption{\small Numerical evolution on present hybrid quantum FPM with the ranging of ratio of viscoelasticity ${\beta}$. (Re $= 0.1$, We $= 0.01$)}
		\label{fig_Case3:3}
	\end{figure}

\begin{figure}[H]
	\centering
	\begin{subfigure}[t]{0.45\textwidth}
		\centering
		\includegraphics[scale=0.4]{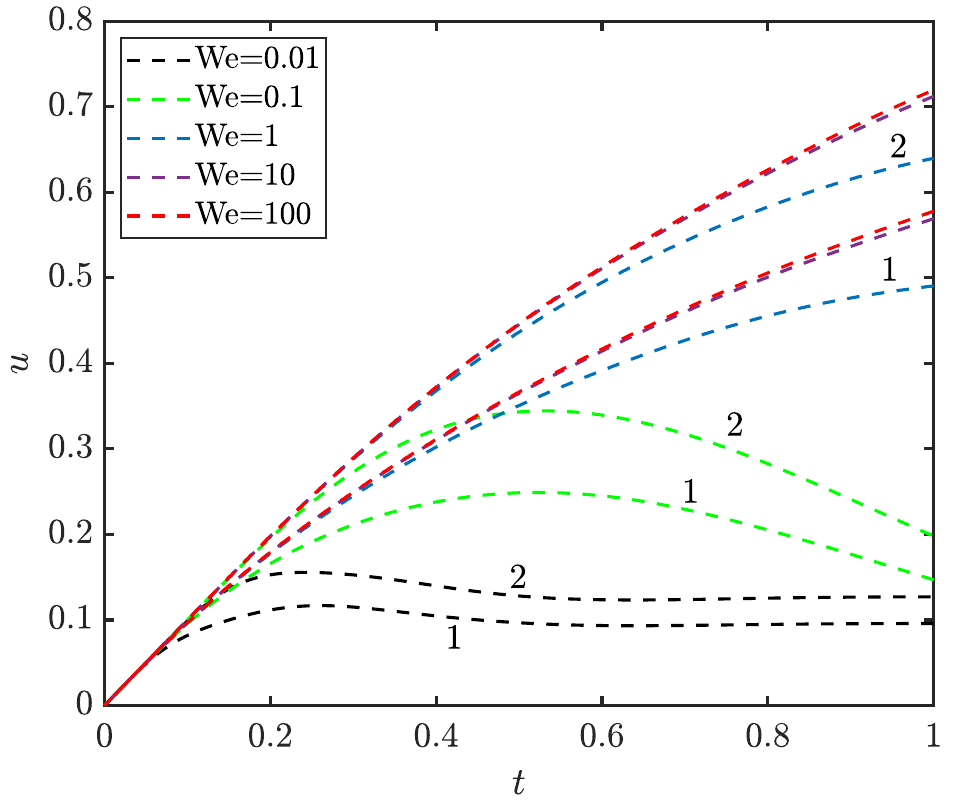}
		\caption{\centering\footnotesize The evolution on velocity at fixed position. (${\beta} = 0.1$)}
		\label{fig:6d}
	\end{subfigure}
	\begin{subfigure}[t]{0.45\textwidth}
		\centering
		\includegraphics[scale=0.4]{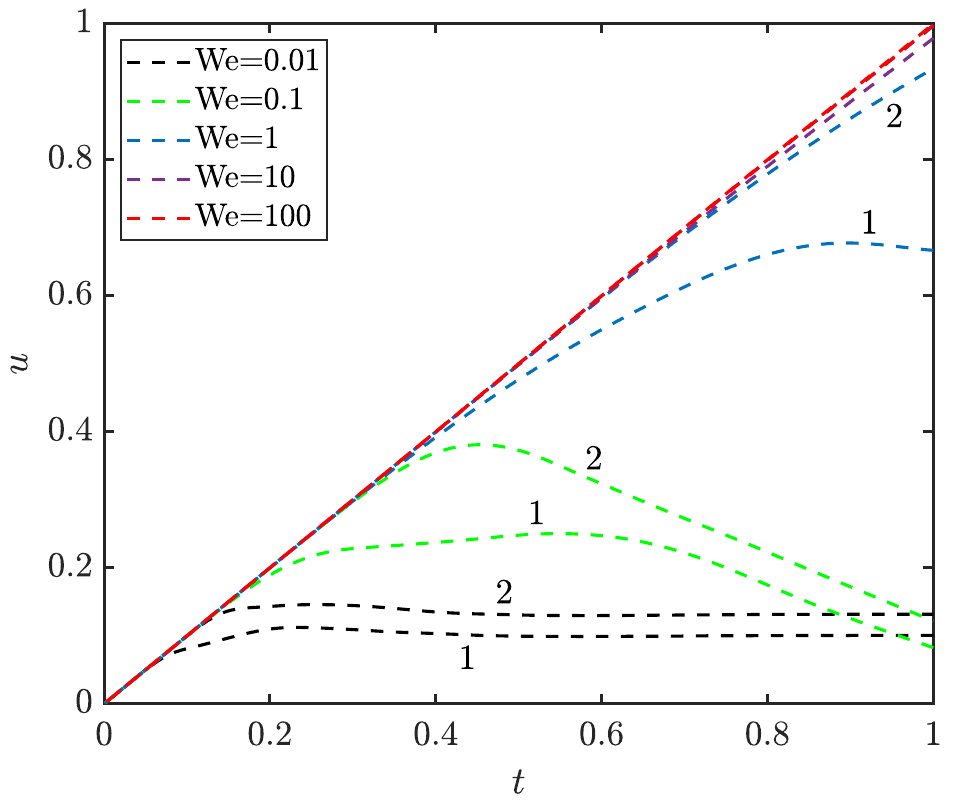}
		\caption{\centering\footnotesize The evolution on velocity at fixed position. (${\beta} = 0$)}  
		\label{fig:6e}
	\end{subfigure}
	\caption{\small Numerical evolution on present hybrid quantum FPM with the increasing of Weissenberg number ${\rm{We}}$. (Re $= 0.1$)}
	\label{fig_Case3:4}
\end{figure}
	
	Subsequently, based on the developed quantum FPM approach and Log conformation evolution, it is further focus on the high Weissenberg viscoelastic fluid flow and classical formulations of Oldroyd-B and UCM constitutive models. As shown in Fig.(\ref{fig_Case3:3}), the velocity evolution and dominant shear stress development in Poiseuille flow are computed across varying ratio of viscoelasticity ${\beta} = {\mu _s}/({\mu _s} + {\mu _p})$ under specified Reynolds number and relatively low Weissenberg number. It is observed that with increasing ratios ${\beta}$ (enhanced viscous dominance), the flow characteristics asymptotically approach Newtonian fluid behaviors. The spatial positions of fluid probes 1 and 2 are explicitly defined in Fig.(\ref{fig_Case3:model}). Notably, with increasing elastic contributions, the flow procedure exhibits progressively pronounced velocity overshoot phenomena accompanied with fluctuating velocity distributions. When the ratio parameter ${\beta} = 0$, the fluid model is transformed to the constitutive UCM model, which possibly exists inherent numerical instabilities and singularities in peak velocity variations. To address these computational challenges, the artificial viscosity method in Eq.(\ref{eq_NS_AVdiscretization:1}) is typically employed similarly as the elasto-viscous stress splitting (EVSS) technique in finite element analysis\supercite{RAJAGOPALAN1990159,KING2021104556}, where controlled introduction of numerical dissipation stabilizes the numerical solutions while maintaining essential physical fidelity.
	
	The absence of viscous terms transforms the momentum equation from a parabolic form (as in Newtonian or Oldroyd-B fluids) to a hyperbolic partial differential equation (as in purely elastic UCM fluids). This loss of explicit dissipation often results in numerical instability or divergence of the solution. Stabilization approaches such as BSD, EVSS, and AVSS, mitigate these issues by reintroducing elliptic operators into the momentum equation, thereby effectively restoring its parabolic character in a manner analogous to artificial viscosity. For instance, the techniques like EVSS augment viscous dissipation to preserve numerical stability. Nevertheless, as the Weissenberg number rising, these instabilities become more pronounced until a critical threshold is reached, the value of which depends on both the constitutive model and the numerical scheme adopted.
	
	\begin{figure}[H]
		\centering
		\begin{subfigure}[t]{0.45\textwidth}
			\centering
			\includegraphics[scale=0.4]{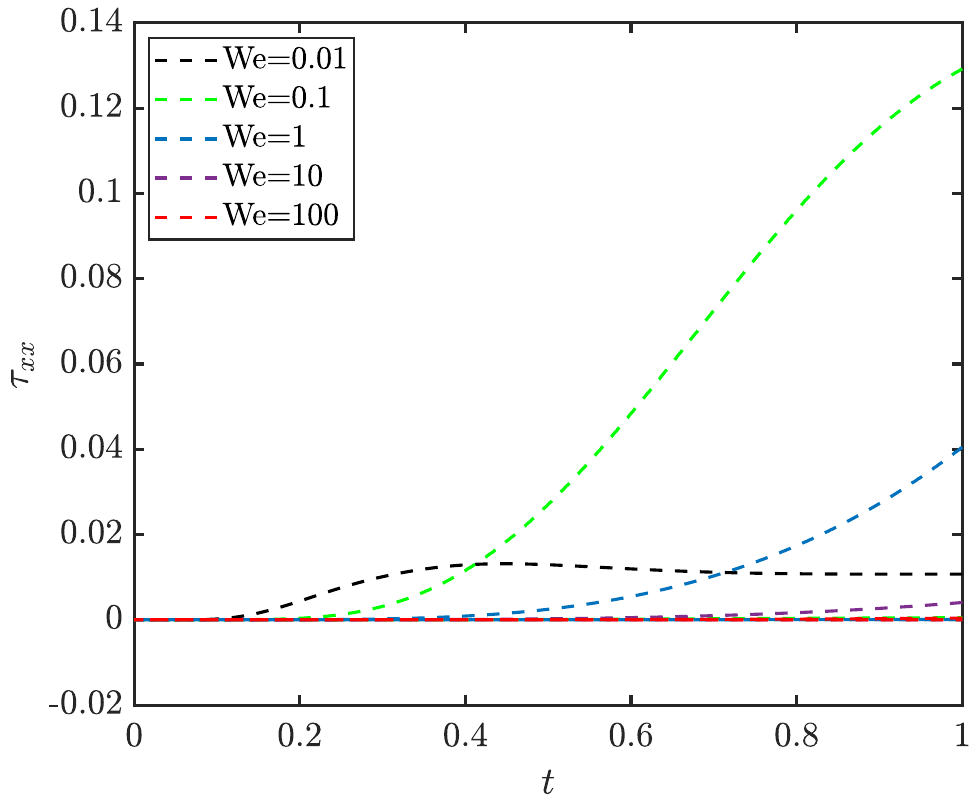}
			\caption{\centering\footnotesize The evolution on stress components ${\tau _{xx}}$ at fixed position. (${\beta} = 0.1$)}
			\label{fig_Case3:51}
		\end{subfigure}
		\begin{subfigure}[t]{0.45\textwidth}
			\centering
			\includegraphics[scale=0.4]{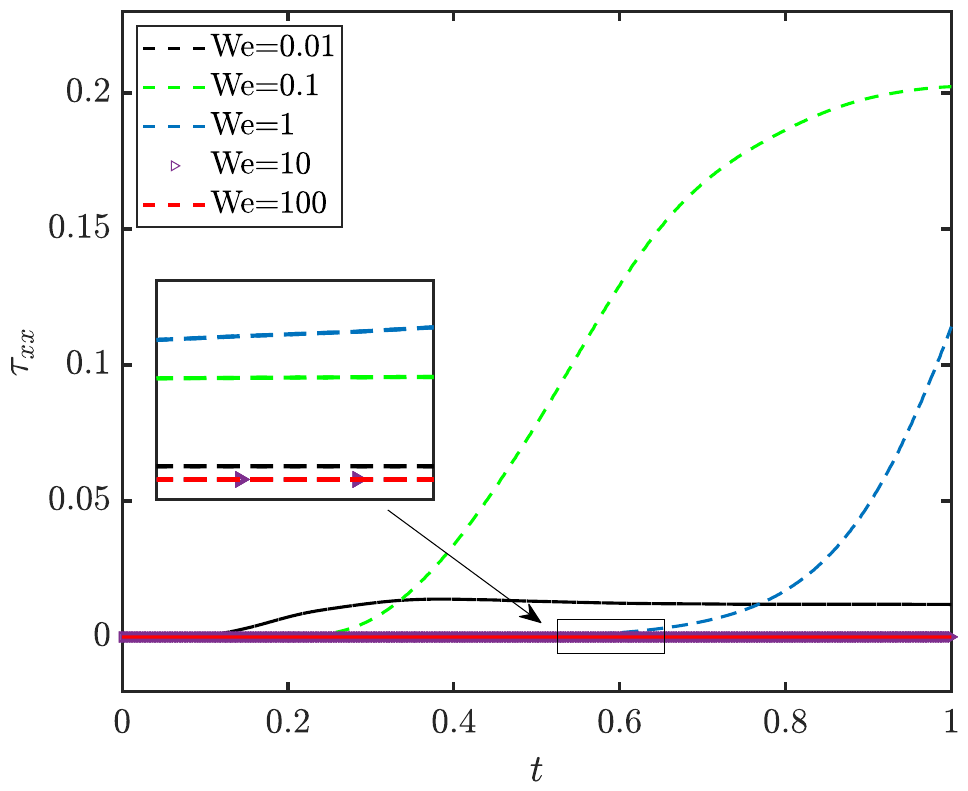}
			\caption{\centering\footnotesize The evolution on stress components ${\tau _{xx}}$ at fixed position. (${\beta} = 0$)}  
			\label{fig_Case3:52}
		\end{subfigure}
		\begin{subfigure}[t]{0.45\textwidth}
			\centering
			\includegraphics[scale=0.4]{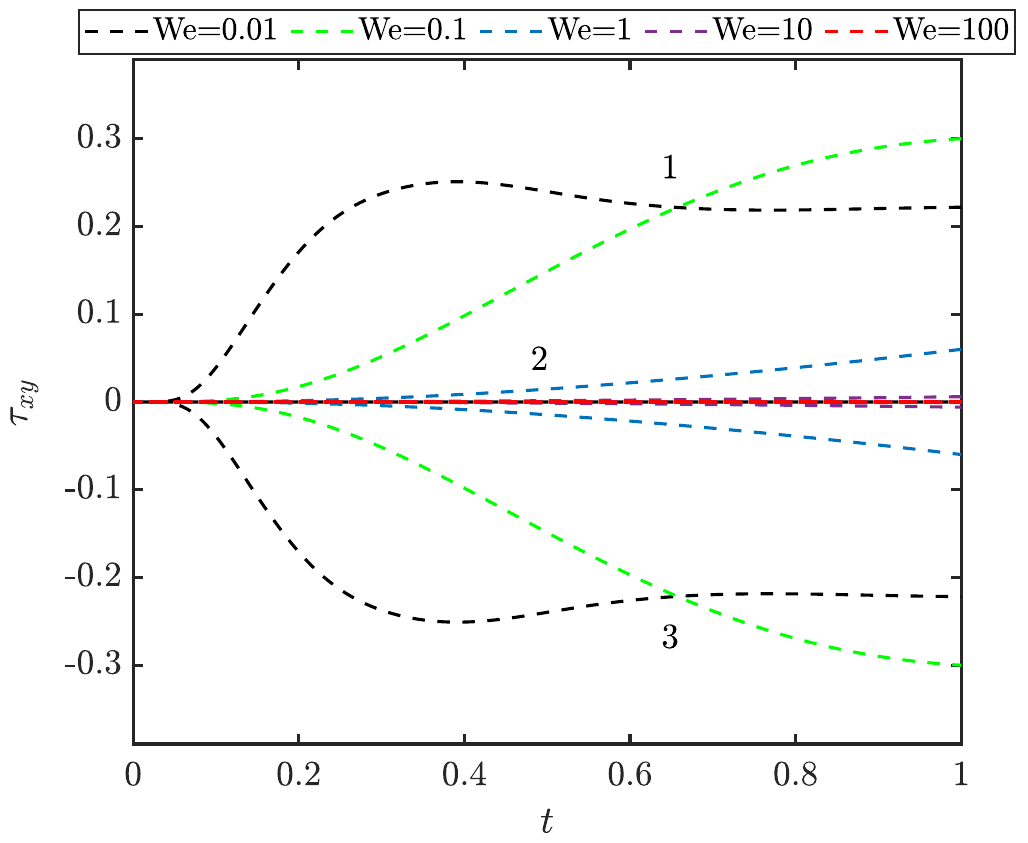}
			\caption{\centering\footnotesize The evolution on stress components ${\tau _{xy}}$ at fixed position. (${\beta} = 0.1$)}
			\label{fig_Case3:53}
		\end{subfigure}
		\begin{subfigure}[t]{0.45\textwidth}
			\centering
			\includegraphics[scale=0.4]{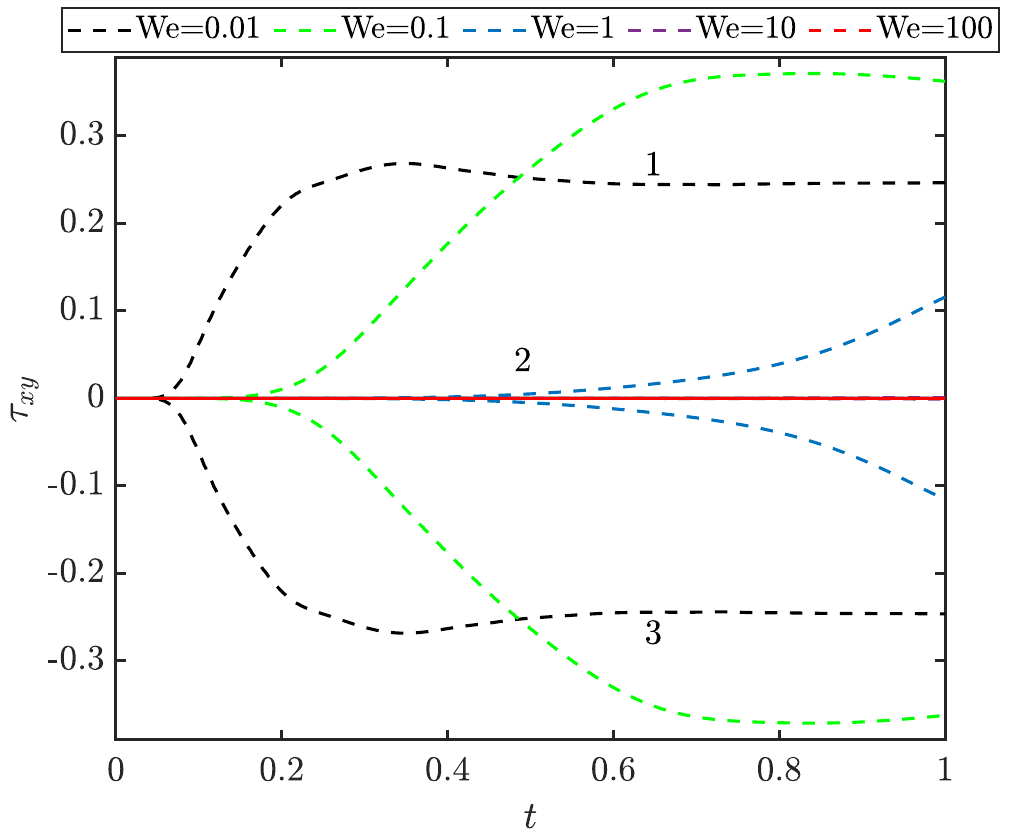}
			\caption{\centering\footnotesize The evolution on stress components ${\tau _{xy}}$ at fixed position. (${\beta} = 0$)}  
			\label{fig_Case3:54}
		\end{subfigure}
		\caption{\small Numerical evolution on present hybrid quantum FPM with the increasing of Weissenberg number ${\rm{We}}$. (Re $= 0.1$)}
		\label{fig_Case3:5}
	\end{figure}
	
	Further, we investigated the viscoelastic flow behaviors with the increasing of Weissenberg number based on the present Oldroyd-B and typical UCM constitutive models. Through the entire quantum-hybrid computing and HWN-resolving Log conformation approach, numerical instability can be effectively suppressed combing with additional artificial viscosity. As illustrated in Fig.(\ref{fig_Case3:4}), the velocity evolution of viscoelastic Poiseuille flow computed by the current numerical framework is presented, with particular emphasis on the progression at locations of Probe 1 and Probe 2. It can be observed that with the increasing of Weissenberg number, the velocity profile exhibits a more linearized trend. The purely elastic upper convected Maxwell (UCM) model demonstrates a pronounced tendency for sharp velocity gradients at extreme points, a behavior consistent with the manifestations detailed in Fig.(\ref{fig_Case3:3}). Additionally, as the Weissenberg number increasing, the discrepancy of velocity between Positions 1 and 2 diminishes significantly. At $\rm{We} = 100$, the velocity variations across different heights become nearly identical, indicating that the interlayer shear stress inducing nonlinear velocity changes, converges to the same distribution regardless of vertical positions.
	
	The specific manifestations underlying this trend can be observed from the evolution of elastic stress components in Fig.(\ref{fig_Case3:5}). In conjunction with the time-dependent viscoelastic constitutive equations, explicit form of Eq.(\ref{eq_NSdiscretization:14}) and logarithmic form of Eq.(\ref{eq_HWNLog:12}), the qualitative analysis confirms that as the relaxation time ${\lambda _1}$ increasing in corresponding Weissenberg number, the temporal variations of elastic stress are primarily governed by their velocity gradients and existing stress fields. The viscous shear stress term of constitutive model serves as the main driver for interlayer flow. A gradual reduction in this viscous component leads to a significant decrease in interlayer shear, thereby resulting in the velocity trends described in above figures. These mechanistic insights are corroborated by numerical changes in the elastic stress components depicted in Figs.(\ref{fig_Case3:52}, \ref{fig_Case3:54}). The results also demonstrate that introducing artificial viscosity effectively suppresses numerical instabilities as the rationale underpinning techniques such as aforementioned EVSS approach, which explicitly decouples and stabilizes the numerical discretization on Navier Stokes equations. \modifyFirstYellow{It is worth clarifying that the resolution of HWN challenge and the effective computation of purely elastic flows are primarily achieved through the Log-conformation representation and the EVSS technique, with the artificial viscosity formulation providing additional numerical stability\supercite{LI2023112213,KING2021104556}. The quantum inner product itself does not offer specific stabilization for this problem, for which this observation aligns with the earlier discussion in Case 2 concerning propagation of quantum computational errors within the time-stepping scheme and conservation of invariants. Furthermore, the presence of quantum noise does not compromise the SPD properties of conformation tensor ${\boldsymbol{\hat A}}$. As shown in Eq.(\ref{eq_HWNLog:6}), the quantum inner product contributes only to the velocity gradient term through a non-end-to-end local evaluation, while the eigen decomposition of ${\boldsymbol{\hat A}}$ and the logarithmic transformation are performed entirely via classical arithmetic. Consequently, the eigenvalue-based operations and integration on Eqs.(\ref{eq_HWNLog:10}, \ref{eq_HWNLog:11}) rigorously preserve the SPD property of conformation tensor throughout the HWN computational process.}
	
	\begin{figure}[H]
		\centering
		\includegraphics[scale=0.5]{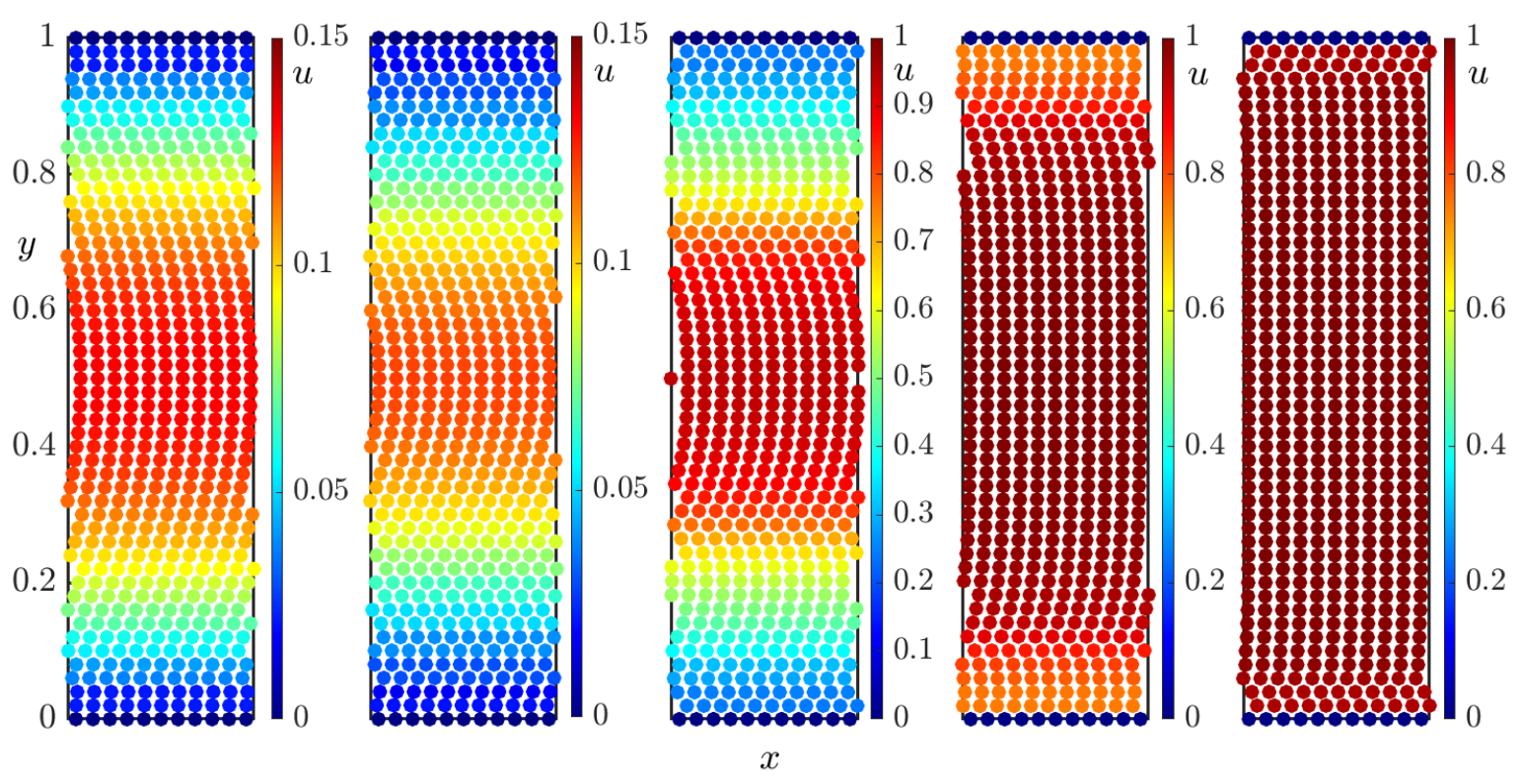}
		\caption{\small Profile on calculated velocities by present hybrid quantum computing, from left to right: We $= 0.01, 0.1, 1, 10, 100$. (Re $= 0.1$, ${\beta} = 0$)}
		\label{fig_Case3:7}
	\end{figure}
	
	Finally, we extracted the velocity profiles on the pipe flow at time $t = 1$ for Weissenberg numbers ranging sequentially from 0.01 to 100, as shown in Fig.(\ref{fig_Case3:7}). Consistent with the preceding analysis, at $\rm{We} = 10, 100$, the viscous shear contribution becomes nearly negligible. The viscoelastic flow is predominantly driven by purely elastic stresses and the external force $F$, exhibiting a steady linear increase and distinct wall slip phenomena. This behavior aligns with the interfacial dynamics typical of highly elastic, purely elastic fluids under shear flows. Meanwhile, these advances provide critical insights for transitioning quantum-enhanced non-Newtonian fluid simulations toward practical engineering applications.
	
	\section{Concluding remarks}
	\normalsize \hspace{10pt}
	In this study, we built a hybrid quantum framework for fluid flow, integrating multi-partitioned quantum zones with meshfree finite particle method. Through a purposefully designed sequence of numerical experiments, progressing from fundamental to complicated scenarios, the proposed methodology was thoroughly validated and analyzed. The results demonstrate that integrating quantum computing to hybridize conventional linear combinations in common computational particle dynamics serves as an effective computing strategy. Furthermore, by extending the numerical investigation into viscoelastic, highly elastic, and purely elastic fluids under high Weissenberg number conditions, the applicability of quantum-hybrid algorithm is significantly broadened.
	
	As represented by smoothed particle hydrodynamics (SPH)\supercite{AUYEUNG2024108909,10.1063/5.0268240}, material point method (MPM), and current finite particle method (FPM) in meshfree computational particle dynamics, these numerical techniques such as corrected kernels for linear combinations and approximated difference operators for physical quantity smoothing can be implemented by adopting quantum bits to construct quantum states. Quantum processors’ natural parallelism might enable the real-time modeling of massive particle systems, something that remains computationally prohibitive on traditional HPC architectures. Therefore, this approach establishes a foundation for the development of quantum-hybrid computing. Considering the limitations of current quantum devices (particularly in qubit counts) and associated computational costs, the multi-partitioned zones proposed in this study help to provide a holistic view in more complicated fluid simulations to an extent.
	
	Similarly, while the established quantum algorithms such as the swap test for inner product and quantum phase estimation providing accurate inner product results, along with emerging variational quantum algorithms, this study still adopts the classical swap test and multi-ancilla QPE process for integration into the simulations. This decision is resulted from the current readiness of these methods for physics-based simulations and high performance quantum processing unit (QPU) computations, and it represents an original attempt at such integration within the meshfree/particle computational framework firstly.
	
	\modifyFirstYellow{The results indicate that, while the on-chip quantum execution time occupies only a small fraction of total wall-clock time, the classical-quantum data I/O transferring and the readout of measurement outcomes across multiple partitions, constitute the dominant bottlenecks. This observation aligns with the asymptotic analysis of Section \ref{sec:complexity}. The multi-partitioned strategy mitigates this bottleneck by restricting the encoding dimension $N$ to a fixed block size, thereby capping the per-partition I/O latency at a hardware-tractable constant. Nevertheless, the cumulative I/O cost accumulated over all partitions remains a primary constraint on the overall throughput of hybrid framework, and its reduction will be contingent on advances in quantum memory and high-speed classical-quantum interconnects.}
	
	It is undeniable that several limitations remain, including quantum hardware constraints, fidelity issues, and computational overhead in three-dimensional simulations. This work should therefore be regarded as a preliminary effort and foundational study in quantum-classical hybrid computing. In alignment with the perspective\supercite{RevModPhys.94.015004,Madsen2022Quantum,PhysRevA.109.052423,Kenton2023Optimising}, quantum computers are unlikely to entirely replace classical systems but rather function through hybrid mechanisms. This research expands the scope of hybrid high performance computing that integrates quantum co-processors and QPU architectures with traditional particle-based methods. Future work will focus on addressing the computational efficiency challenges associated with preparing a large number of quantum circuits in a nested loop manner. Despite this limitation, the primary contribution of this study lies in the proposal of a novel computational paradigm. It contributes a meaningful basis for future scientific and technological advances in quantum hardware and computational fluid dynamics, and will continue to evolve with ongoing developments.
	\\
	\\
	\noindent  \textbf{Authorship contribution statement}\quad\small
	{Yudong Li: Writing – review \& editing, Writing – original draft, Visualization, Validation, Conceptualization, Methodology, Investigation, Data curation. 
		Wenkui Shi: Writing – review \& editing, Investigation, Formal analysis. 
		Zhihao Qian: Investigation, Formal analysis. 
		Yan Li: Resources, Investigation, Funding acquisition. 
		Chunfa Wang: Investigation, Formal analysis. 
		Ling Tao: Investigation, Formal analysis. 
		Zhuojia Fu: Resources, Investigation, Funding acquisition. 
		Moubin Liu: Writing – review \& editing, Supervision, Resources, Methodology, Investigation, Formal analysis, Conceptualization. 
		Zhiqiang Feng: Supervision, Software, Resources, Project administration, Investigation, Funding acquisition, Formal analysis.}
	\\
	\\
	\noindent  \textbf{Declaration of competing interest}\quad\small
	{The authors declare that they have no known competing financial
		interests or personal relationships that could have appeared to influence
		the work reported in this paper.}
	\\
	\\
	\noindent  \textbf{Acknowledgements}\quad\small
	{This work has been partially supported by the National Natural Science Foundation of China [Grant Nos. 12602300, 12032002 and U22A20256] and the Sino-German Mobility Programme [No. M-0210]. The experiments were executed on the Origin Quantum Cloud using the “OriginQ Wukong” superconducting quantum computer and quantum simulators.}
	
	\bibliographystyle{elsarticle-num} 
	\begin{footnotesize }
		\bibliography{reference}
	\end{footnotesize }
	
\end{document}